\documentclass{aastex63}

\usepackage{comment}
\usepackage{todonotes}
\usepackage{multirow}
\usepackage{threeparttable}
\usepackage{color}

\newcommand{\PreserveBackslash}[1]{\let\temp=\\#1\let\\=\temp}
\def\l{\left}
\def\r{\right}

\def\be{\begin{equation}}
\def\ee{\end{equation}}
\def\nn{\nonumber}
\shorttitle{Comprehensive analysis of the tidal effect in gravitational waves and cosmology}
\shortauthors{B. Wang, Z. Zhu, A. Li, W. Zhao.}

\begin{document}

\title{Comprehensive analysis of the tidal effect in gravitational waves and implication for cosmology}

\correspondingauthor{Bo Wang}
\email{ymwangbo@ustc.edu.cn}
\correspondingauthor{Wen Zhao}
\email{wzhao7@ustc.edu.cn}

\author[0000-0002-3784-8684]{Bo Wang}
\affiliation{CAS Key Laboratory for Research in Galaxies and Cosmology, Department of Astronomy,
\\
University of Science and Technology of China, Hefei 230026, China}
\affiliation{School of Astronomy and Space Sciences, University of Science and Technology of China, Hefei, 230026, China}
\author[0000-0001-9189-860X]{Zhenyu Zhu}
\affiliation{Department of Astronomy, Xiamen University, Xiamen 361005, China}
\affiliation{Institut f{\"u}r Theoretische Physik, Max-von-Laue-Stra{\ss}e 1, 60438 Frankfurt, Germany}
\author[0000-0001-9849-3656]{Ang Li}
\affiliation{Department of Astronomy, Xiamen University, Xiamen 361005, China}
\author[0000-0002-1330-2329]{Wen Zhao}
\affiliation{CAS Key Laboratory for Research in Galaxies and Cosmology, Department of Astronomy,
\\
University of Science and Technology of China, Hefei 230026, China}
\affiliation{School of Astronomy and Space Sciences, University of Science and Technology of China, Hefei, 230026, China}



\begin{abstract}

Detection of gravitational waves (GWs) produced by coalescence of compact binaries provides a novel way to measure the luminosity distance of GW events. Combining their redshift, they can act as standard sirens to constrain cosmological parameters. For various GW detector networks in 2nd-generation (2G), 2.5G and 3G, we comprehensively analyze the method to constrain the equation-of-state (EOS) of binary neutron-stars (BNSs) and extract their redshifts through the imprints of tidal effects in GW waveforms. We find for these events, the observations of electromagnetic counterparts in low-redshift range $z < 0.1$ 
are important for constraining the tidal effects. Considering 17 different EOSs of NSs or quark-stars, we find GW observations have strong capability to determine the EOS. Applying the events as standard sirens, and considering the constraints of NS's EOS derived from low-redshift observations as prior, we can constrain the dark-energy EOS parameters $w_0$ and $w_a$. In 3G era, the potential constraints are $\Delta w_0\in (0.0006,0.004)$ and $\Delta w_a\in(0.004,0.02)$, which are 1-3 orders smaller than those from traditional methods, including Type Ia supernovas and baryon acoustic oscillations. The constraints are also 1 order smaller than the method of GW standard siren by fixing the redshifts through short-hard $\gamma$-ray bursts, due to more available GW events in this method. Therefore, GW standard sirens, based on the tidal effect measurement, provide a realizable and much more powerful tool in cosmology.

\end{abstract}

\keywords{gravitational waves --- instrumentation: detectors --- stars: neutron --- cosmology: dark energy}


\section{Introduction} 
\label{sec:intro}

The discovery of first gravitational-wave (GW) event GW150914 \citep{Abbott2016GW1stA,Abbott2016GW1stB,Abbott2016GW1stC,Abbott2016GW1stD, Abbott20162016e,Abbott20162016f, Abbott20162016abbott}, produced by the coalescence of two massive black holes (BHs), marks the new era of GW astronomy. Slightly later, the first binary neutron-star (BNS) coalescence GW event, GW170817,
together with its electromagnetic (EM) counterparts,
opens the window of multi-messenger GW astronomy \citep{GW170817ligo}.
As the primary targets of ground-based GW detectors, such as LIGO and Virgo, the GW events of BNSs  and/or NSBHs provide a new
approach to explore the internal structure of NSs \citep{GW170817ligo,GW190425ligo}. In addition, these events can be treated as the standard sirens to measure various cosmological parameters \citep{Schutz1986}, in particular the Hubble constant \citep{GW170817ligoHubbleNature} and the equation-of-state (EOS) of dark energy.

The internal structure of NSs remains a challenge for current astrophysics.
The ultra-high density state inside NS is an unknown field with new scientific potential.
This extreme condition is inaccessible for terrestrial collider experiments and is theoretically challenging.
In the EM observations of astrophysics,
the constraints on EOS of NS mainly come from the measurement of radius and mass of NSs.
Several NSs with a mass greater than 2$M_ \odot $ have been observed (\cite{AntoniadisFreireWex2013,FonsecaPennucciEllis2016,Arzoumanian2018,Cromartie2020}; https://stellarcollapse.org/nsmasses),
which set the maximum of allowable NS mass larger than 2$M_ \odot $
and can exclude some EOS models.
{However, the constraint from the mass observations is not a strong one.}
Lindblom proposed that the EOS of NSs can be further constrained
by observing the radius-mass relation \citep{Lindblom1992}.
{Related} studies have been carried out and
provided some constraints on the EOS \citep{Riley2019,MillerLamb2019}.
{However, the uncertainties are still large in deriving constraints from current radius measurements.}
With the increasing sensitivities of GW detectors,
and the detection of GW170817 and GW190425,
the GW signal together with its EM counterpart
provides a new method to explore the EOS of NSs.
In the final stage of the inspiral BNS or NSBH,
the tidal deformation which contains imprints of the internal structure of NS can affect the orbital motion as well as the waveform of the emitted GWs
\citep{AbdelsalhinGualtieriPani2018,BanihashemiVines2020,BiniDamourFaye2012,Landry2018,VinesFlanaganHinderer2011,VinesFlanagan2013}.
Thus, the EOS of NS can be determined by analyzing the GW signals from these coalescence \citep{ReadMarkakisShibata2009, HindererLackey2010}.
GW170817  set a constraint on tidal deformability $ \lambda $,
whose relation with  gravitational mass alone
$\lambda(m)$ depends on the EOS \citep{love1911,Hinderer2008,HindererLackey2010}, with the observation of GW for the first time \citep{GW170817ligo},
and excluded some of the EOS models \citep{AbbottAbbottAbbott2018}.

In order to obtain a better measurement of the EOS of NSs, it is necessary to combine the information from an ensemble of BNSs.
The first idea of combining multiple sources with Fisher matrix method was reported in Ref.\citep{MarkakisReadShibata2012}.
Del Pozzo and Agathos {\it et al}. proposed the first fully Bayesian study that combined information to constrain the EOS by using the tidal effects \citep{PozzoAgathos2013}.
Their results showed that when taking the NS mass as $ 1.4 M_ \odot $,
tens of GW observations can constrain $ \lambda $ to an accuracy of $\sim10\%$.
Lackey and Wade showed that in the range of 1--2$M_ \odot$,
$ \lambda $ can be measured to an accuracy of 10--50\%,
and the overwhelming majority of this information comes from the loudest $\sim$ 5 GW events \citep{LackeyWade2015}.
In order to reduce the computational cost of the study,
these authors adopted a Fisher matrix approach.
Agathos {\it et al.}
improved this analysis by including
the effects of NS spin in the GW waveforms,
and obtained the consistent results.
Vivanco {\it et al.} carried out Bayesian inference and optimized the previous data processing method.
They found that
LIGO and Virgo are likely to place constraints
on the EOS of NS
similar as Ref.\citep{LackeyWade2015}
by combining the first
forty BNS detections \citep{VivancoSmithThrane2019}.

In all these works, the authors only investigated the potential constraint of EOS of NSs by focusing on the 2G GW detectors, including LIGO \citep{AasiAbbott2015},
Virgo \citep{AcerneseAgathos2014}, KAGRA \citep{AsoMichimura2013, KAGRA2018}, and  LIGO-India \citep{LIGOindia2011}. More importantly, these works did not consider the important role of EM counterparts in the analysis. From the observations of GW170817, it is known that BNS mergers have the EM counterparts \citep{gw170817-kilonova} called ``kilonova" or ``macronova", which has the nearly isotropic optical emission \citep{kilonova}. For the events at low-redshift range, {\it i.e.} $z\lesssim 0.1$, they can be detected by various optical telescopes including Zwicky Transient Factory (ZTF), Large Synoptic Survey Telescope (LSST) \citep{lsst}, Wide Field Survey Telescope  (WFST) and so on. Once the EM counterparts are detected, the position and redshift of these GW events can be fixed by their host galaxies, and the degeneracy between these parameters and the EOS parameters of NS can be broken. For this reason, with the help of EM-counterpart observations for the low-redshift GW events, the constraint of NS's EOS is expected to be significantly improved. In this article,
we consider a sample of representative EOS models \citep{Ozel2016,Zhu2018,Zhou2018,Xia2019} 
under
the 2-solar-mass constraint.
We linearly expand the relationship of NS $ \lambda $-$ m $ at the average mass of NSs $ 1.35M_ \odot $ as $ \lambda = Bm + C $,
which is similar to the treatment in Ref.\citep{PozzoAgathos2013}.
Based on the latest BNS merger rate \citep{AbbottAbbottAbbott2019},
we simulated a bunch of BNS sources.
Using the networks of 2G, 2.5G LIGO A+ \citep{LIGOaPlusNoise} and 3G  Einstein telescope (ET) (\cite{PunturoAbernathy2010},https://www.et-gw.eu/) and  Cosmic Explorer (CE) \citep{Abbott2017ETCE, Dwyer2015ETCE} GW detectors,
we assume that every detected GW signal at low redshift $z<0.1$ or $z<0.05$ has an detectable EM counterpart,
which gives the precise redshift and location of the GW source.
We adopt Fisher matrix approach to calculate the covariance matrix of $ B $ and $ C $, which quantifies the capabilities of GW detectors for the determination of NS's EOS.
To highlight the importance of detectable EM counterparts,
we also compare the results with those in the case without EM counterparts.




As an important issue in GW astronomy, the GW events can be treated as the standard sirens to probe the expansion history of the Universe, since the observations on the GW waveform can measure the luminosity distance $d_{\rm L}$ of the events, without having to rely on a cosmic distance ladder \citep{Schutz1986}. For the standard sirens in low-redshift range, for instance in 2G and/or 2.5G era, one can measure the Hubble constant $H_0$ \citep{GW170817ligoHubbleNature}, and hope to solve the $H_0$-tension problem in the near future \citep{chen2018}. While, in 3G era, a large number of high-redshift events will be detected, and the standard sirens will provide a novel way to detect the cosmic dark energy \citep{SathyaprakashSchutzBroeck2010,ZhaoBroeckBaskaran2011}. The key task of this issue is to determine the redshift of these GW events. The most popular way is from the observations of their EM counterparts and/or host galaxies \citep{GW170817ligoHubbleNature,SathyaprakashSchutz2009,SathyaprakashSchutzBroeck2010,ZhaoBroeckBaskaran2011,cairg,Yan2020,Yu2020,LISA}. As mentioned above, for the low-redshift BNS mergers, the kilonovae as the primary EM counterparts could emit the nearly isotropic EM radiations \citep{kilonova}, which are expected to be detected by various optical telescopes. However, for the high-redshift events, the expected EM counterparts are the short-hard $\gamma$-ray bursts (shGRBs) and the afterglows \citep{grb}. This method of redshift measurement has three defects: First, the optical radiation of afterglow is faint, and only the nearby sources can be detected. Second, the radiation of shGRBs is believed to be emitted in a narrow cone more or less perpendicular to the plane of the inspiral. Therefore, these EM counterparts are expected to be detected only if they are nearly face-on. Third, in this method, the inclination angle $\iota$ of the events should be well determined by EM observations, which is needed to break the degeneracy between $\iota$ and $d_{\rm L}$ \citep{zhao2019,schutz2019}. For these reasons, only a small
fraction of the total number of BNSs are expected to be observed with shGRBs, and the application of GW sirens in cosmology is limited \citep{SathyaprakashSchutzBroeck2010,ZhaoBroeckBaskaran2011,cairg,ZhaoWen2018}.

The alternative method to measure the redshift of GW events is by counting the tidal effects in GW waveform, which was firstly proposed in \citep{MessengerRead2012}. The basic idea is that, the contributions of tidal effects in the GW phase evolution break a degeneracy between the mass parameters and redshift and thereby allow the simultaneous measurement of both the effective distance and the redshift for individual sources. Since this method depends only on the GW observations, all the detectable GW events can be used as standard sirens. Thus, in particular in 3G era, the number of useful events will be dramatically large.  In Ref.\citep{PozzoLiMessenger2017},
Del Pozzo {\it et al.} proposed for the first time to use Bayesian method to investigate the measurement errors of cosmic parameters
with the GW events detected by ET.
However, they directly assumed that the EOS of NS is known, did not discuss how to determine from the observations. In this article, 
with the EOS model determined at low redshift as a prior,
in light of the  tidal effect in GW,
we calculate the errors of redshift $\Delta z$ and luminous distance $\Delta d_{\rm L}$ of GW sources with the GW observation from the high-redshift range $ 0.1<z<2 $ observed by the assumed 3G detector networks.
Then, with the help of the property of Fisher matrix,
we convert $\Delta z$ and $\Delta d_{\rm L}$
into the errors of parameters of dark energy.
For comparison,
we also calculate the measurement errors of the dark energy parameters
without EM counterparts observed at low redshift.

This paper is organized as follows.
In Sec. \ref{sec:GWandDetectors},
we review the mathematical preliminaries of GW waveforms from BNS coalescence and the response of GW detectors to these signals.
In Sec. \ref{sec:EOS}, we briefly introduce the EOS models adopted in this paper.
In Sec. \ref{sec:Fiser}, we describe our method to calculate the Fisher matrix for a set of parameters required to fully describe a GW signal from BNS coalescence.
In Sec. \ref{sec:EOSdetermination}, we show the results of distinguishing different EOS models by 2G, 2.5G, 3G detector networks,
and compare the results with and without the presence of detectable EM counterparts.
In Sec. \ref{sec:DEdetermination}, we discuss the determination of the parameters for dark energy by using 3G detectors.
At the end, in Sec. \ref{sec:conclusion}, we summarize and discuss our main results in this article.

Throughout this paper, we choose the units in which $G = c = 1$, where $G$ is the Newtonian gravitational constant,
and $c$ is the speed of light in vacuum.

\section{GW waveforms and the detector response}
\label{sec:GWandDetectors}

In this section,
we briefly review the detection of GWs by a ground-based network,
which includes $N_d$ GW detectors.
The sizes of these detectors are all much smaller than the wavelength of GWs.
We use the vector $\mathbf r_I$ with $I=1,2,...,N_d$
to denote the locations of the detectors,
which is given by
\be
\mathbf r_I=R_{\oplus}  (\sin\varphi_I\cos\alpha_I,
    \sin\varphi_I\sin\alpha_I, \cos\varphi_I),
\ee
in the celestial coordinate system.
Here $\varphi_I$ is the latitude of the detector,
$\alpha_I\equiv\lambda_I+\Omega_r t$,
where $\lambda_I$ is the east longitude of the detector,
$\Omega_r$ is the rotational angular velocity of the Earth,
and $R_{\oplus}$ is the radius of the Earth.
In this paper,
we take the zero Greenwich sidereal time at $t=0$.

\begin{table}[htbp]
\caption{The coordinates of the interferometers used in this study.
Orientation is the smallest angle made by any of the arms
and the local north direction \citep{ChuWenBlair2012,VitaleEvans2017}.
}
\begin{center}
\label{positionDetectors}
\begin{tabular}{|c|c |c| c| c| }
 \hline
  &
$\varphi$ &
$\lambda$ &
$\gamma$ &
$\zeta$
\\ \hline
LIGO Livingston &
$30.56^\circ$ &
$-90.77^\circ$ &
$243^\circ$ &
$90^\circ $
\\ \hline
LIGO Hanford &
$46.45^\circ$ &
$-119.41^\circ$ &
$171.8^\circ$ &
$90^\circ $
\\ \hline
Virgo &
$43.63^\circ$ &
$10.5^\circ$ &
$116.5^\circ$ &
$90^\circ $
\\ \hline
KAGRA &
$36.25^\circ$ &
$137.18^\circ$ &
$0^\circ$ &
$90^\circ $
\\ \hline
LIGO India &
$19.09^\circ$ &
$74.05^\circ$ &
$0^\circ$ &
$90^\circ $
\\ \hline
Einstein Telescope in Europe &
$43.54^\circ$ &
$10.42^\circ$ &
$19.48^\circ$ &
$60^\circ $
\\ \hline
Cosmic Explorer in the U.S. &
$30.54^\circ$ &
$-90.53^\circ$ &
$162.15^\circ$ &
$90^\circ $
\\ \hline
Assumed detector in Australia  &
$-31.51^\circ$ &
$115.74^\circ$ &
$0^\circ$ &
---
---
\\ \hline
\end{tabular}
\end{center}
\end{table}
\begin{figure*}[htbp]
\centering
\includegraphics[width = .6\linewidth]{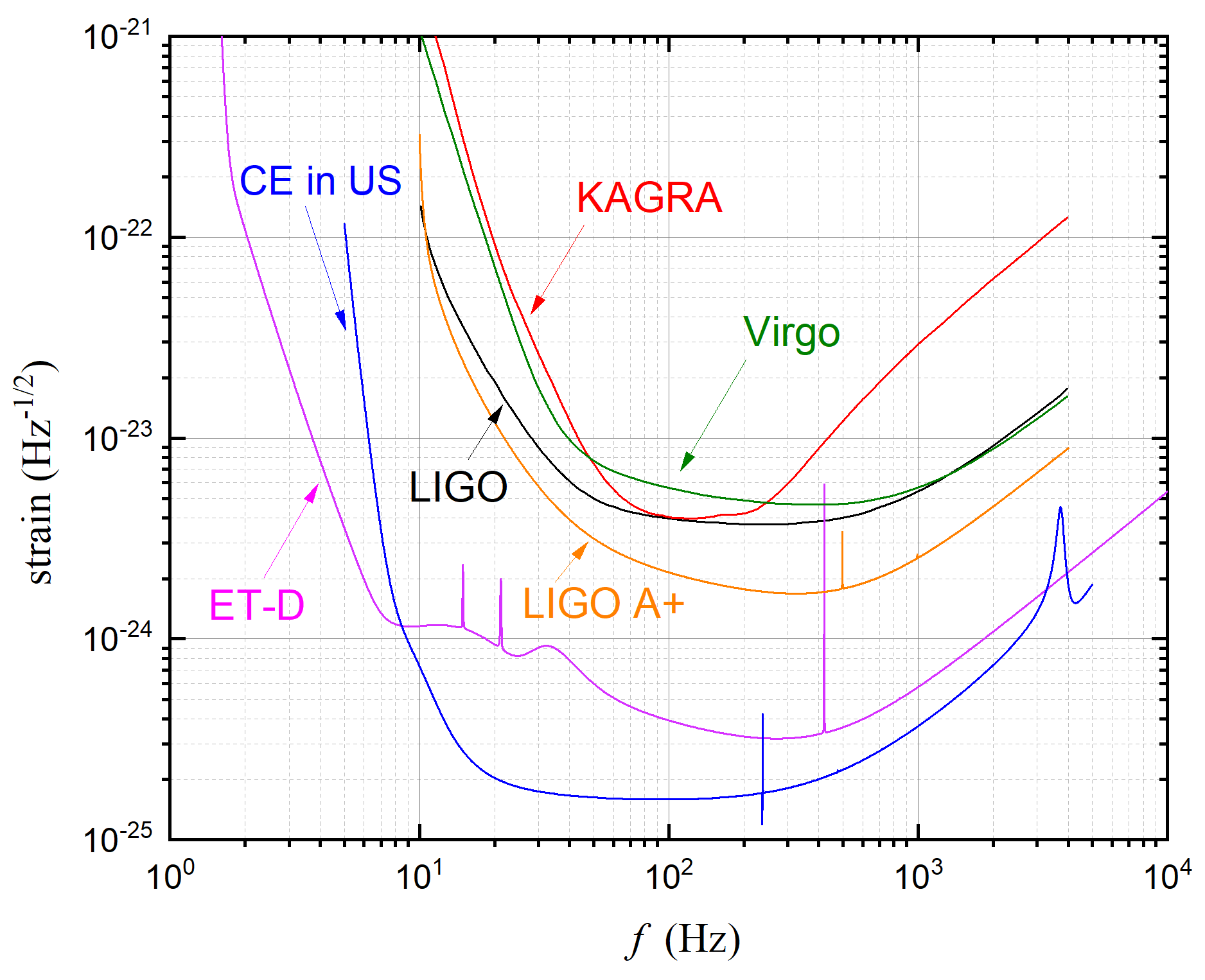}
\caption{ The design amplitude spectral density of
ET-D, CE,
LIGO A$+$, and LIGO, Virgo, KAGRA  noise level (\cite{Dwyer2015ETCE,Abbottetal2016LIGOdesignnoise,KAGRA2018,Abbott2017ETCE,LIGOaPlusNoise},
https://www.et-gw.eu/).
We take the noise level of the assumed detector in Australia to be the same as CE in the U.S.,
and take LIGO-India to be the same as LIGO-H.
    }
     \label{NoiseCurveLIGOaPlus}
\end{figure*}

For a transverse-traceless GW signal detected by
a single detector labeled by $I$,
the response is a linear combination of
the plus mode $h_+$ and cross modes $h_\times$ as the following
\be
d_I(t_0+\tau_I+t)
=
F_I^+ h_+(t) + F_I^\times h_\times(t), ~~~
0<t<T,
\ee
where $t_0$ is the arrival time of the wave at the coordinate origin and
$\tau_I=\mathbf n\cdot \mathbf r_I(t)$
is the time required for the wave to travel from the origin
to reach the $I$-th detector at time $t$,
with $t\in [0,T]$ being the time label of the wave and
$T$ being the signal duration,
and $\mathbf n$ is the propagation direction of a GW.
In the above,
the antenna pattern functions $F_I^+$ and $F_I^\times$ \citep{JaranowskiKrolakSchutz1998,Maggiore2007}
depend on the location of the source $(\theta_s,\phi_s)$,
the polarization angle $\psi_s$,
the latitude $\varphi$ and the longitude $\lambda$ of the detector
on the Earth,
the orientation angle $\gamma$
of the detector's arms
which is measured counter-clockwise from East of the Earth
to the bisector of the interferometer arms,
and the angle between the interferometer arms $\zeta$.
In Table \ref{NoiseCurveLIGOaPlus},
we list the parameters for LIGO, Virgo, KAGRA, and the
LIGO-India, as well as the potential
ET in Europe,
CE experiment in the U.S.,
and the assumed detector in Australia respectively.
The design amplitude spectral density for these detectors' noise (\cite{Dwyer2015ETCE,Abbottetal2016LIGOdesignnoise,KAGRA2018,Abbott2017ETCE,LIGOaPlusNoise},
https://www.et-gw.eu/)
is plotted in Fig.\ref{NoiseCurveLIGOaPlus}.
It is observed that the noises in 3G detectors
are lower than the 2G detectors by about one order of magnitude,
and the 2.5G detector lies in the middle.
The change in orbital frequency over a single GW cycle is negligible
during the inspiral phase of the binary merger.
Therefore,
under the stationary phase approximation (SPA) \citep{ZhangLiu2017,ZhangZhaoEtAl,ZhaoWen2018},
the Fourier transform of the time-series data from
the $I$-th GW detector can be obtained as follows,
\be
d_I(f)=\int^T_0 d_I(t) e^{2\pi i f t}dt.
\ee

We define a whitened data set in the frequency domain
in terms of the one-side noise spectral density  $S_I(f)$
as the following
\citep{WenChen2010},
\be
\hat d_I(f)\equiv S_I^{-1/2}(f)d_I(f).
\ee
For a detector network,
this can be rewritten as \citep{WenChen2010},
\be\label{NetResponse}
\hat{\bf d}(f)=e^{-i\Phi}\hat{\bf A}{\bf h}(f),
\ee
where $\Phi$ is the $N_d\times N_d$ diagonal matrix with
$\Phi_{IJ}=2\pi f \delta_{IJ}(\mathbf n \cdot \mathbf r_I(f))$,
and
\be
\hat{\bf A}{\bf h}(f)
=\left[
\frac{F_1^+h_+(f)+F_1^\times h_\times(f)}{\sqrt{S_1(f)}},
\frac{F_2^+h_+(f)+F_2^\times h_\times(f)}{\sqrt{S_2(f)}},
...,
\frac{F_{N_d}^+h_+(f)+F_{N_d}^\times h_\times(f)}{\sqrt{S_{N_d}(f)}}
\right]^T .
\ee
Note that, $F_I^+$, $F_I^\times$, $\Phi_{IJ}$ are all functions
with respective to frequency in general,
and are taken as the values at
$t_f=t_c-(5/256)\mathcal M_c^{-5/3}(\pi f)^{-8/3}$
\citep{Maggiore2007}
under SPA,
where $t_c$ is the binary merger time,
$\mathcal M_c\equiv M \eta^{3/5}$ is the chirp mass of binary system with component masses $m_1$ and $m_2$,
and $M\equiv(1+z)(m_1+m_2)$,
$\eta\equiv m_1 m_2/(m_1+m_2)^2$.
We also note that
the masses in this paper are defined as the physical (intrinsic) mass,
and the observed mass $m_{\rm obs}$ is related to
the physical mass through
\be
m_{\mathrm{obs}}=(1+z)m,
\ee
where $z$ is the redshift of the GW source.

In general,
in order to estimate parameters,
one needs to use the spinning inspiral-merge-ringdown
merger waveform template of the compact binaries.
For simplify,
similar to previous works \citep{SathyaprakashSchutzBroeck2010,ZhaoBroeckBaskaran2011,TaylorGair2012,ZhaoWen2018},
we only consider the waveforms in the inspiralling stage
and adopt the restricted post-Newtonian (PN)
approximation of the waveform for the non-spinning systems
\citep{SathyaprakashSchutz2009,Cutler1993,BlanchetEtAl2002etAl},
which is expected to have no significant differences
in our result with the full waveforms.
The SPA Fourier transform of the GW waveform
from a coalescing BNS is given by \citep{SathyaprakashSchutz2009},
\be\label{Fhtrans}
F_I^+ h_+(f) + F_I^\times h_\times(f)
=
\mathcal A_I f^{-7/6}{\rm exp}
\{
i\l[2\pi f t_c-\pi/4+2\psi(f/2)
+ \Phi_{\rm { tidal }}(f)\r]
-\varphi_{I,(2,0)}
\},
\ee
with the Fourier amplitude $\mathcal A_I$ given by
\be
\mathcal A_I=\frac{1}{d_{\rm L}}\sqrt{(F_I^+(1+\cos^2\iota))^2
    +(2F_I^\times\cos\iota)^2}\sqrt{5\pi/96}\,\pi^{-7/6}
    \mathcal M_c^{5/6},
\ee
where $d_{\rm L}$ is the luminosity distance of the GW source,
and $\iota$ is the inclination angle between the binary's orbital angular momentum and the line of sight.
In the above,
the phase $\psi$ is contributed by the  point-particle approximation
of the NSs,
$\varphi$ is related to the detection capability of a GW detector,
and
$\Phi_{\rm { tidal }}$ is the contribution
of the finite-size deformation effects
of the BNS.
Under the 3.5 PN approximation for the phase,
$\psi$
and $\varphi_{I,(2,0)}$ are given by
\citep{SathyaprakashSchutz2009,BlanchetEtAl2002etAl}
\be\label{phasePointGW}
\psi=-\psi_c+\frac{3}{256\eta}\sum^7_{i=0}\psi_i(2\pi M f)^{i/3},
\ee
\be
\varphi_{I,(2,0)}=\tan^{-1}
\l(
-\frac{2\cos\iota F_I^\times}{(1+\cos^2\iota)F_I^+}
\r)
,
\ee
and $\Phi_{\rm { tidal }}$ is given by \citep{MessengerRead2012},
\begin{eqnarray}\label{phiTidal}
 \Phi_{\rm { tidal }}(f)
&=&
 \sum_{a=1}^{2} \frac{3 \lambda_{a}(1+z)^{5}}{128 \eta M^{5}}\Big[
-\frac{24}{\chi_{a}}\left(1+\frac{11 \eta}{\chi_{a}}\right)(\pi M f)^{5/3}
\nn\\
&&
-\frac{5}{28 \chi_{a}}(3179-919 \chi_{a}-2286 \chi_{a}^{2}
+260 \chi_{a}^{3} )(\pi M f)^{7/3} \Big],
\end{eqnarray}
where the sum is over the components of the binary,
$\chi_{a}\equiv m_{a} / (m_1+m_2)$,
$\lambda_{a}\equiv \lambda\left(m_{a}\right) $.
The parameter  $\lambda$ characterizes
changes of the quadruple of a NS
given an external gravitational field,
and is comparable in magnitude with the
3PN and 3.5PN phasing terms for NSs \citep{MessengerRead2012}.
The $\lambda$-$m$ relation depends on the EOS models of NSs,
thus different EOS models shall lead to different
functions of $\lambda(m)$.
Since the NS masses are approximately taken to be a Gaussian
distribution with a mean of $1.35 M_{\odot}$
and a standard deviation of $0.15 M_{\odot}$
\citep{Thorsett1999,Stairs2004,PozzoLiMessenger2017},
to parameterize EOS models,
we shall express the $\lambda$-$m$ relation as a linear function
within $1$-$\sigma$ range of the NS mass-distribution as following
\be\label{lambdaMlinear}
\lambda=B\,m+C,
\ee
with $B$ and $C$ as two tidal-effect parameters,
which are expected to be determined through GW detection.
Each set of values of $B$ and $C$ represents one EOS model of NS.
EOS can be determined by fixing $B$ and $C$ through observations.

\section{EOS models and tidal deformability}
\label{sec:EOS}

\begin{figure*}[htbp]
\gridline{\fig{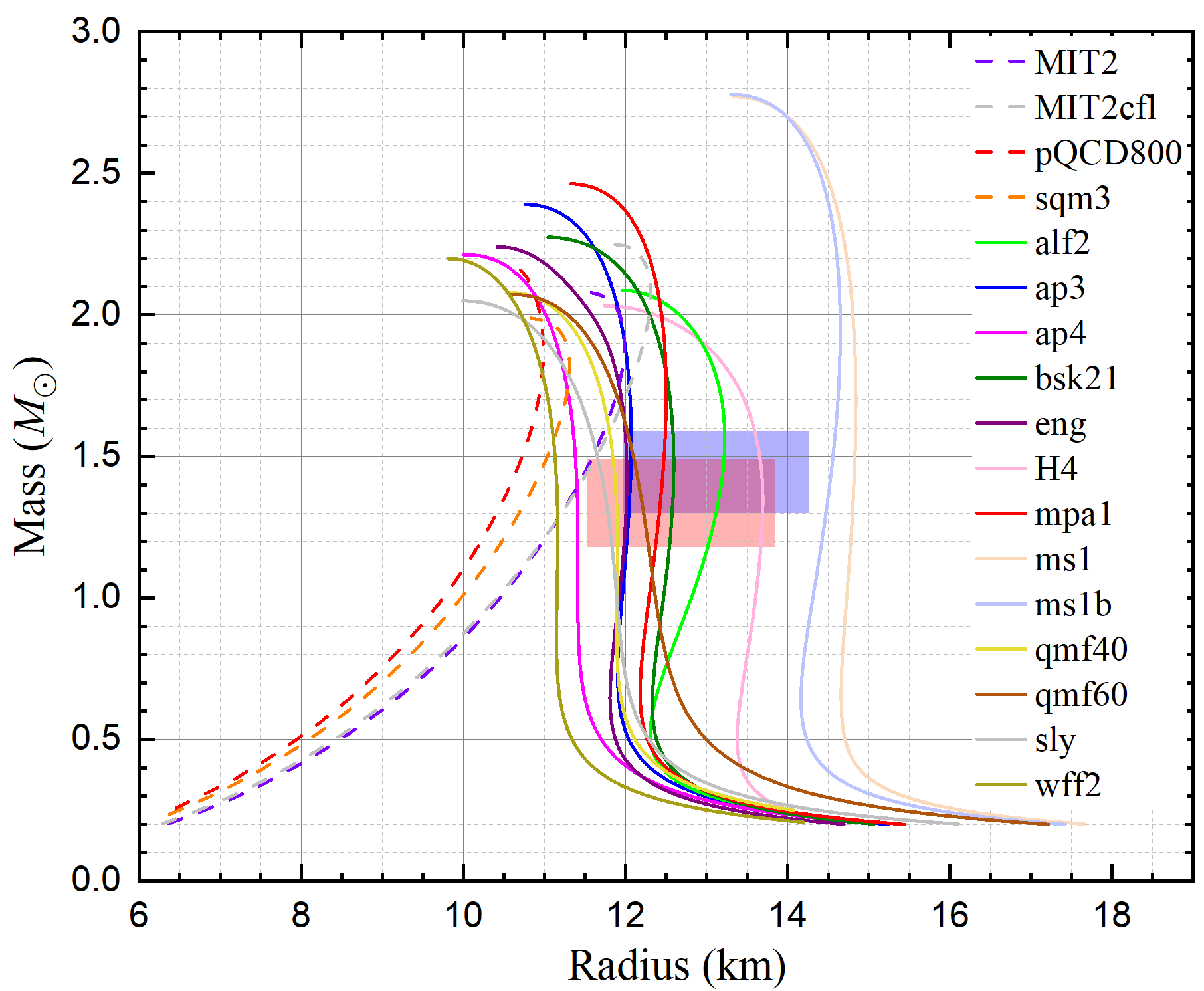}{0.48\textwidth}{(a)}
          \fig{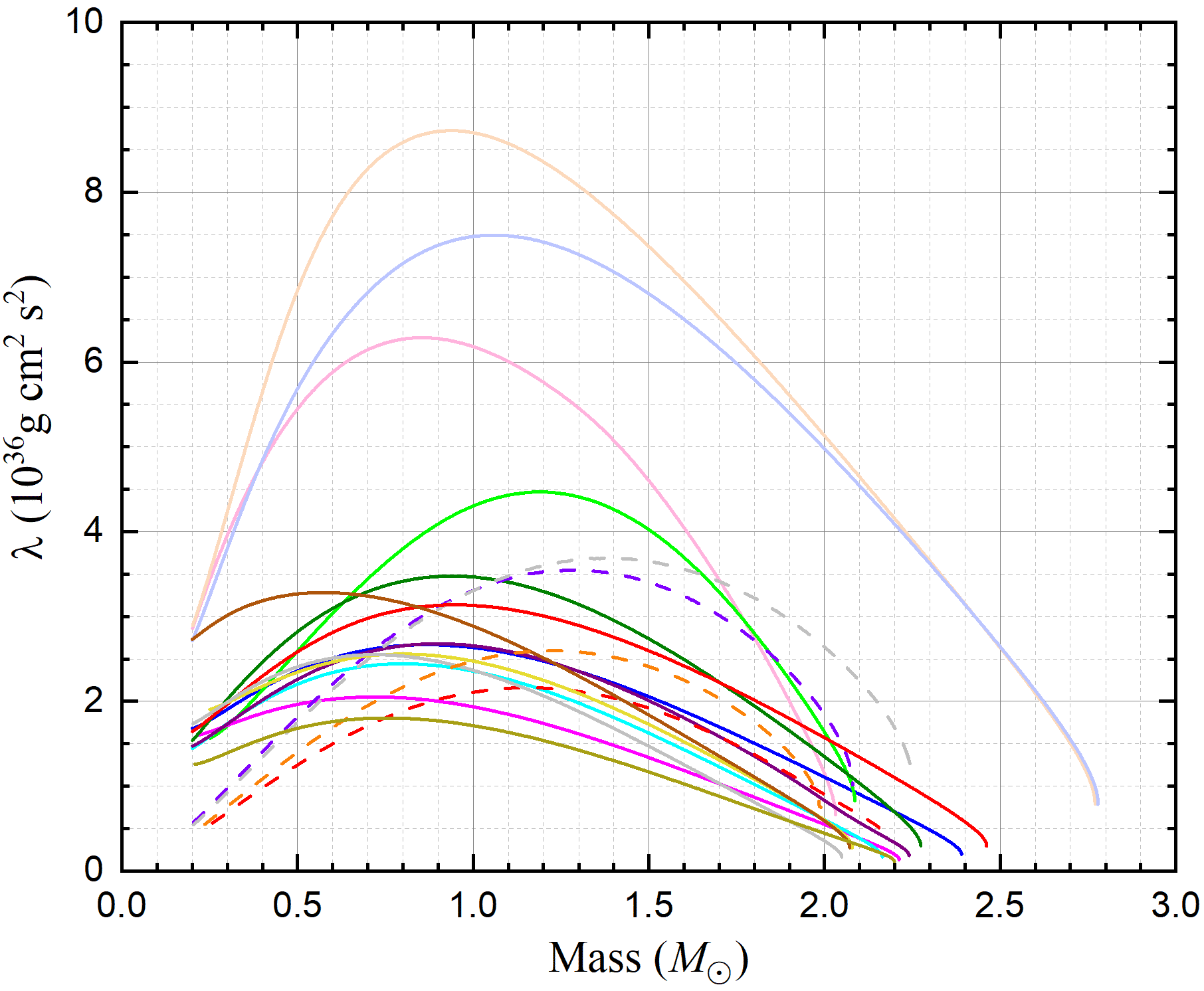}{0.48\textwidth}{(b)}
          }
\caption{
(a)
$m$-$r$ relations for various EOS models.
{The blue and red boxes are the $m$-$r$ error ranges to the 68.3\% credibility interval of the millisecond pulsar PSR J0030+0451 inferred  by Ref.\citep{MillerLamb2019} and Ref.\citep{Riley2019} using the NICER data, respectively.}
(b)
$\lambda$-$m$ relations for the EOS models,
which are selected under $m_{\rm max}>2M_\odot$.
{The maximum mass of each curve is the maximum mass for the corresponding EOS.}
The dashed lines are theoretical EOSs for quark stars.
    }
     \label{lambdaM}
\end{figure*}

Since the astrophysical observations indicate
the existence of NSs with masses larger than $\sim2M_\odot$
\citep{AntoniadisFreireWex2013,DemorestPennucciRansom2010, FonsecaPennucciEllis2016,Arzoumanian2018,Cromartie2020},
in this paper we consider a sample of candidate NS EOSs \citep{Ozel2016,Zhu2018,Zhou2018,Xia2019} with varying stiffness under the 2-solar-mass constraint.
The $m$-$r$ relations are plotted in Fig.\ref{lambdaM} (a),
and the corresponding $\lambda$-$m$ relations are shown in Fig.\ref{lambdaM} (b).
Note that we supplement 4 EOSs of quark stars, as illustrated by the dashed lines in Fig.\ref{lambdaM}, which are self-bound compact stars discussed in the literature.
The corresponding fitted values of $B$ and $C$
are listed in Table \ref{lambdaMfit}, together with the maximum masses and the radii of $1.35M_{\odot}$ stars.
We mention here that ms1, ms1b are incompatible, to the $90\%$ credibility interval, with the tidal deformability constraint from GW170817 \citep{GW170817ligo},
while H4 is marginally consistent.
Among neutron star EOS models, ap4, ms1, ms1b, wff2 are outside the boundaries, to the $68.3\%$ credibility interval,  of the mass and radius constraints obtained for PSR J0030+0451 \citep{MillerLamb2019,Riley2019}.
We make the reasonable assumption that all neutron stars are described
by the same EOS \citep{Forbes2019},
{\it i.e.},
all NSs have the same values of $B$ and $C$.
By the numerical simulation of GW from BNSs
with observable electromagnetic counterparts
at low redshift,
we shall study the discrimination of these models
by networks of GW detectors .

For most BNS mergers at high redshift,
the electromagnetic counterparts are not expected to be observed.
However, it is found from (\ref{phiTidal}) that when considering the tidal effects,
$z$ and $m$ do not bound with each other,
in contrary to the point-particle approximation (\ref{phasePointGW}).
This property enables us to estimate redshifts
by only using information coming only from the
GW observations
without electromagnetic counterparts,
which shall be discussed in latter sections when determining the parameters of dark energy.

\begin{table*}[htbp]
\centering
\caption{The  values of $B$ and $C$
fitted according to Eq.(\ref{lambdaMlinear})
for the EOS models in Fig.\ref{lambdaM}.
For each EOS,
the maximum allowable mass and the radius for a NS with $1.35M_\odot$ are also listed.}
\label{lambdaMfit}
\begin{threeparttable}
\begin{tabular}{l  c c c c}
\hline
\hline
\multirow{2}*{EOS}
& \multirow{2}*{$B$($10^{36}$g\,cm$^2$s$^2$$/M_{\odot}$) }
& \multirow{2}*{$C$($10^{36}$g\,cm$^2$s$^2$) }
& \multirow{2}*{$M_{\rm max} (M_\odot) $}
& \multirow{2}*{$R_{1.35}$(km)}
\\
\\
\hline
alf2    & $  -1.47 $ & $ 6.30 $ & 2.09 & 13.15
\\
\hline
ap3   & $  -1.40 $ & $ 4.17 $  & 2.39  &  12.05
\\
\hline
ap4     & $ -1.34 $ & $ 3.36 $ & 2.21 & 11.40
\\
\hline
bsk21  & $ -1.87 $ & $ 5.56  $ &  2.27 & 12.58
\\
\hline
eng   & $ -1.60 $ & $ 4.43 $  & 2.24 &  12.01
\\
\hline
H4    & $ -3.85 $ & $ 10.44  $   & 2.03  & 13.69
\\
\hline
mpa1    & $ -1.37  $ & $ 4.67$ & 2.46  &  12.43
\\
\hline
ms1    & $  -3.35 $ & $12.41 $ & 2.77  &   14.81
\\
\hline
ms1b  & $   -2.03  $ & $  9.89$ & 2.78   &  14.49
\\
\hline
qmf40  & $ -1.74 $ & $ 4.35 $ & 2.08  & 11.88
\\
\hline
qmf60  & $  -2.25  $ & $5.22$ & 2.07 &  12.19
\\
\hline
sly  & $ -1.99  $ & $ 4.46$ & 2.05  &  11.75
\\
\hline
wff2   & $ -1.24 $ & $ 3.04$ & 2.20  &  11.16
\\
\hline
MIT2\tnote{*} & $  -0.44 $ & $ 4.10  $  & 2.08  &  11.32
\\
\hline
MIT2cfl\tnote{*}  & $0.14 $ & $ 3.47 $  & 2.25  &  11.33
\\
\hline
pQCD800\tnote{*}  & $ -0.77  $ & $  3.10$ &  2.17 &  10.50
\\
\hline
sqm3\tnote{*}   & $ -0.60 $ & $ 3.35$ & 1.99  &  10.78
\\
\hline\hline
\end{tabular}
 \begin{tablenotes}
        \footnotesize
        \item[*] These are the EOS models of quark stars.
      \end{tablenotes}
    \end{threeparttable}
\end{table*}

\section{Fisher matrix and analysis method}
\label{sec:Fiser}
The match-filter method
is commonly used to estimate the parameters
from detected waveforms against certain theoretical templates.
For an unbiased estimation
(the ensemble average of which is the true value),
according to the Cramer-Rao bound \citep{Cramer1946},
the lower bound for the covariance matrix
of estimated parameters is the inverse of the Fisher matrix \citep{Kay}
when considering statistical errors,
which is method-independent.
Thus, the errors of the estimated parameters $p_i$ can be calculated by using Fisher matrix $F_{ij}$
through the expression $\langle\delta p_j\delta p_k\rangle
=(F^{-1})_{jk}$ in many fields \citep{MartinezSaar2010}. 
The Fisher matrix
of a network including $N_d$ independent GW detectors is given by \citep{Finn1992, Finn1993},
\be
F_{ij}=\langle\partial_i{\bf\hat d}|\partial_j{\bf\hat d}\rangle,
\ee
where $\partial_i{\bf\hat d}\equiv\partial{\bf\hat d}(f)/\partial p_i$,
and $\langle a|b \rangle
\equiv
2\int_{f_{\rm low}}^{f_{\rm up}}
\l\{a(f) {b}^*(f)+{a}^*(f) b(f)\r\} df$.
For a given BNS system,
$\hat{\bf d}(f)$ in Eq.(\ref{NetResponse})
depends on twelve system parameters
$(\mathcal M_c,$ $\eta,$ $t_c,$ $\psi_c,$ $\iota,$ $\theta_s,$ $\phi_s,$
$\psi_s,$ $d_{\rm L},$ $z,$ $B,$ $C)$,
where $\psi_c$ is defined in Eq.(\ref{phasePointGW}),
$B$ and $C$ are defined in Eq.(\ref{phiTidal}),
and the other parameters are all defined previously.
By taking the noise in detectors as stationary and Gaussian,
the optimal squared signal-to-noise ratio (SNR) is given by
\citep{Finn1992, Finn1993}
\be
\rho^2=\langle{\bf\hat d}|{\bf\hat d}\rangle.
\ee
Once the total Fisher matrix $F_{ij}$ is calculated,
an estimate of the root mean square (RMS) error,
$\Delta p_i$, in measuring the parameter $p_i$ can then be calculated,
\be
\Delta p_i=(F^{-1})_{ii}^{1/2}.
\ee
The correlation coefficient between two
parameters will be
$r_{ij}\equiv(F^{-1})_{ij}/[(F^{-1})_{ii}(F^{-1})_{jj}]^{1/2}$,
Note that $r_{ij}=0$ indicates the independency of
$p_i$ and $p_j$,
and $|r_{ij}|=1$ indicates the complete correlation of
$p_i$ and $p_j$,
i.e., $p_i$ and $p_j$ are degenerate in data analysis,
and the Fisher matrix is not invertible.

For practical computation,
the analytical expression of the solution for
$\partial{\bf \hat d}(f)/\partial p_i$
is usually not available,
and one needs to use numerical methods.
We adopt an approximation
$\partial{\bf \hat d}(f)/\partial p_i
\simeq ({\bf \hat d}(f;p_i+\delta p_i)-{\bf \hat d}(f;p_i))/\delta p_i$
which was used by Ref.\citep{ZhaoWen2018},
and calculate the Fisher matrix numerically.
Note that the Fisher matrix yields only the lower limit
of the covariance matrix, which might be slightly different from the corresponding results derived from Monte Carlo simulations.

\section{Determination of EOS of the neutron stars}
\label{sec:EOSdetermination}

To extract EOS information from the GW detection,
great detector sensitivity is required. In this paper,
we investigate four types of detector networks,
including 2G, 2.5G and 3G detectors,
for GW detection,
and compare their abilities for the EOS determination.

{\bf $\bullet$ LHV} one LIGO-Livingston detector,
one LIGO-Hanford detector, and one Virgo detector.

{\bf $\bullet$ LHVIK} LHV and additionally one LIGO-India detector
and one KAGRA.

{\bf $\bullet$ 3LA+} Three assumed LIGO A$+$ detectors locate in the sites of LHV respectively.

{\bf $\bullet$ ET2CE} One ET-D detector in Europe and two CE detectors in the U.S. and in Australia respectively.
The optimal site localizations for
this kind of  network with planning 3G detectors
 have been studied in Ref.\citep{RaffaiGondanHengKelecsenyi2013}.

The coordinates, orientations, and open angles of the detectors are listed in Table \ref{positionDetectors},
the design amplitude spectral densities of the detectors' noise are shown in Fig.\ref{NoiseCurveLIGOaPlus}.
According to the detection rates in Refs.\citep{GW190425ligo,AbbottAbbottAbbott2019},
we shall generate the random BNS samples simulatively with
$m_{1}=m_{2}=1.35M_{\odot}$
at low redshift $z<0.1$
to investigate the resolution of
$B$ and $C$ for the tidal effects in Eq.(\ref{lambdaMlinear}).

\subsection{The LHV network}
\label{subsec:LHV}

\begin{figure*}[htbp]
\gridline{\fig{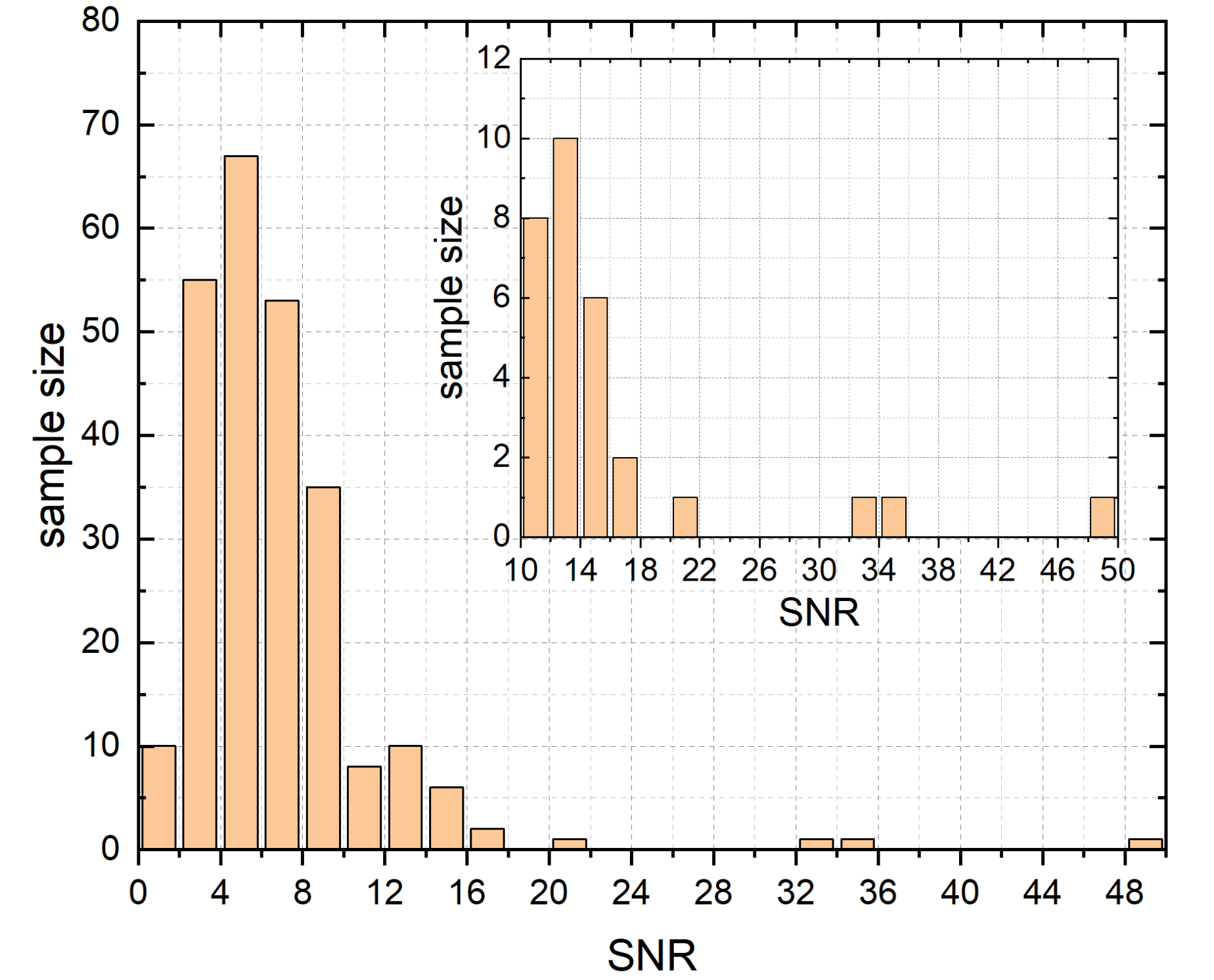}{0.48\textwidth}{(a)}
          \fig{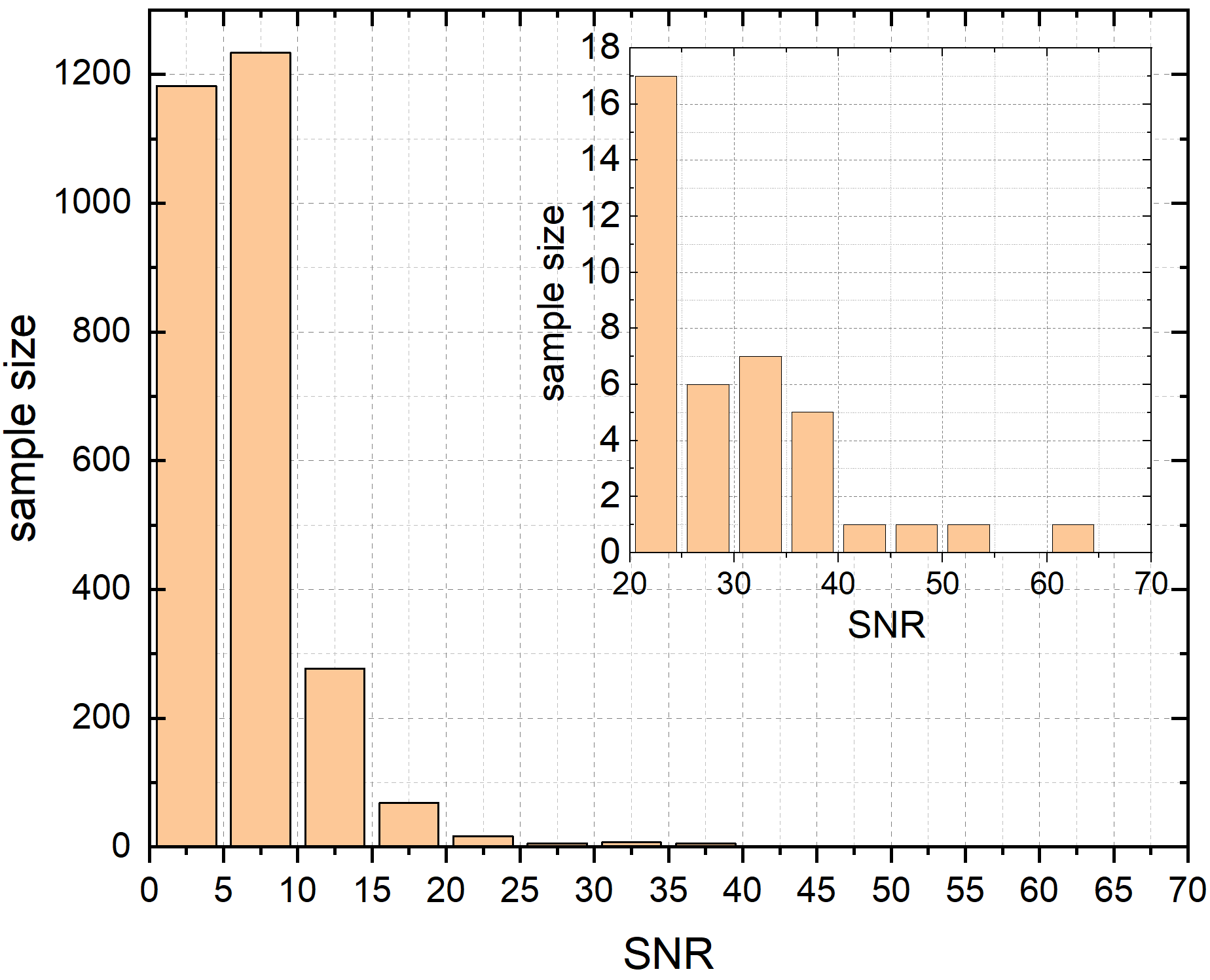}{0.48\textwidth}{(b)}
          }
\caption{
The SNR distribution of the samples
generated in the redshift range of
$z<0.1$ detected by LHV.
According to the lowest and highest event rates given in Ref.\citep{AbbottAbbottAbbott2019},
we generate a total of 250  samples in (a) and a total of 2800 samples in (b).
With the threshold of  SNR$>10$,
there are 30 detectable samples as shown in (a);
and there are 384 detectable samples  in (b).
}
     \label{SNRn}
\end{figure*}

\begin{figure*}[htbp]
\gridline{\fig{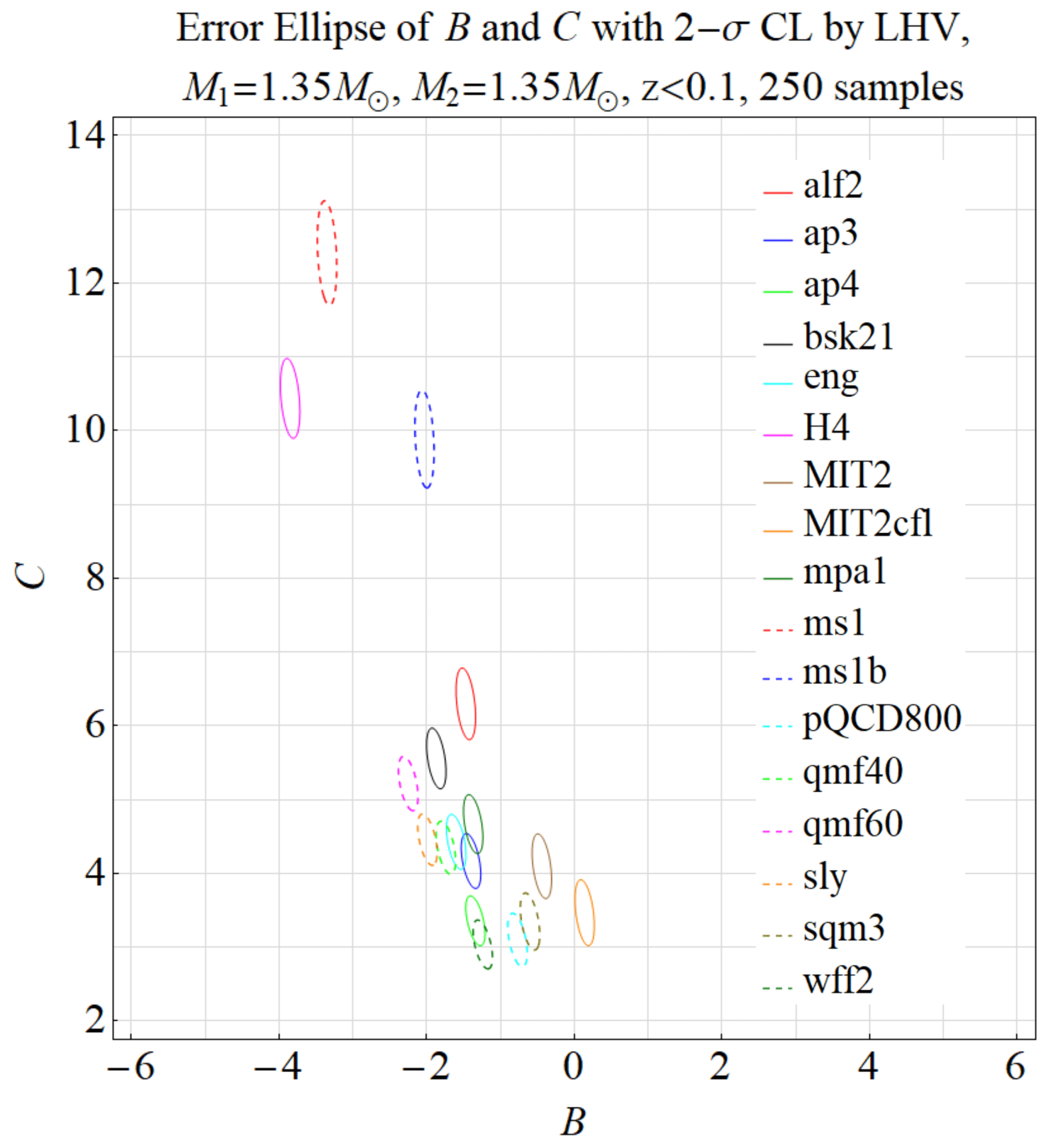}{0.48\textwidth}{(a)}
          \fig{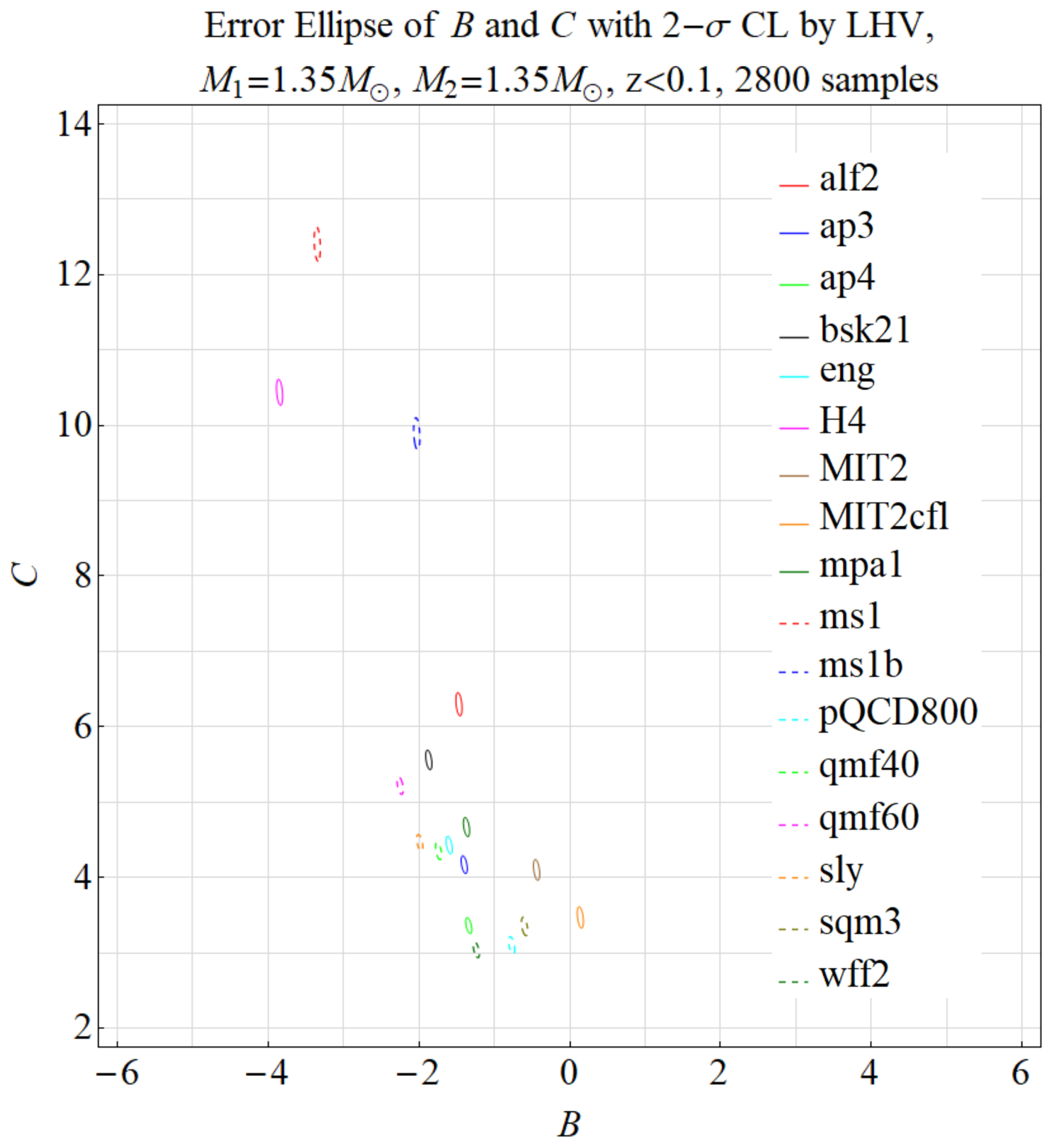}{0.48\textwidth}{(b)}
          }
  \caption{
  The error ellipses of $B$ and $C$  at $2$-$\sigma$ CL for different EOS models  under the detection of LHV.
  We have assumed the EM counterparts for every GW sample are detectable.
  (a) is plotted with  the total 250 samples
  and  (b) is plotted with the total 2800 samples.
  These samples are generated at $z<0.1$.
  The corresponding SNR distributions are  illustrated in Fig.\ref{SNRn}.
  It is observed that in (a), the error ellipses of ms1, H4, ms1b, alf2, bsk21, qmf60, MIT2, MIT2cfl do not overlap with others,
  which implies that they are distinguishable by this GW observation.
  In (b), there is no overlap between any two ellipses,
  thus in this case all the EOS models are distinguishable.
  }\label{LHVlowBCError}
\end{figure*}

We consider the case with LHV network to determine the EOS model in low redshift $z<0.1$.
Given a merger rate as 250--2810 Gpc$^{-3}$  y$^{-1}$ at 90\% confidence
 level  (CL) \citep{AbbottAbbottAbbott2019},
and assuming that BNSs are uniformly distributed in comoving volume,
there will be
$2.5\times10^2\sim 2.8 \times10^3$
BNS merger events
for three-year observation.
In this paper, SNR$>10$ is taken as the criterion for detectable GW.
As an illustration,
we perform a numerical simulation to generate GW samples from BNS merger.
In Fig. \ref{SNRn},
the SNR distribution of the generated samples is presented.
It is observed that there are 30 detectable samples among the total 250 merger events,
and 384 detectable samples among the total 2800 merger events.
For every detectable GW,
we assume that the corresponding electromagnetic counterparts are observable
at this low-redshift region,
which provides the precise redshift $z$
and location $(\theta_s,\phi_s)$ of the source.
The remaining 9 parameters
$(\mathcal M_c,$ $\eta,$ $t_c,$ $\psi_c,$ $\iota,$
$\psi_s,$ $d_{\rm L},$ $B,$ $C)$
need to be determined by GW observation.
To explore the detection uncertainties of $B$ and $C$,
one needs to marginalize the remaining 7 parameters,
which is complicated through a Bayesian approach \citep{PozzoLiMessenger2017}.
We adopt an easy-to-handle process by using Fisher matrix \citep{Coe2009,AmendolaSellentin2016}.
For a detectable GW sample $k$,
we calculate the 9-parameter Fisher matrix
and inverse it to get the 9-parameter covariance matrix.
The removal of the rows and columns of
the 7  parameters except $B$ and $C$ is equivalent to the marginalization of these 7 parameters,
and inverting the remaining $2\times2$ submatrix yields
the Fisher matrix $ (F_{ij})_k$ of $B$ and $C$.
The total Fisher matrix is obtained by
summing  the Fisher matrixes of $(B,C)$
over all detectable GW samples,
$F_{ij}=\sum_{k}(F_{ij})_k$.
Then,
it is straightforward to get the covariance matrix of $B$ and $C$ as the inverse of $F_{ij}$,
with which we can plot the  error ellipse of $B$ and $C$
as illustrated in
Fig.\ref{LHVlowBCError} at $2$-$\sigma$ CL.
If two error ellipses overlap,
the two corresponding models of EOS cannot be distinguished from each other.
We find that when taking 250 samples of merger events,
under $2$-$\sigma$ CL,
some models can be distinguished from the others,
such as 
ms1, H4, ms1b, alf2, bsk21, qmf60, MIT2, MIT2cfl.
The remaining models cannot be distinguished due to the low sensitivity of LHV.
When sampling 2800 merger events,
all models can be distinguished from each other.

In the above, we only consider the BNS coalescences with the same individual NS masses as $1.35 M_\odot$.
In reality, the actual individual masses are more likely to be different.
In order to investigate the effect of different NS masses on constraining the EOS parameters, we set all the merging BNS masses as the medium numbers of the NS masses derived from GW190425, which is $m_1=1.75 M_\odot$ and $m_2=1.57 M_\odot$ \citep{GW190425ligo}, and repeat the same simulation process as before.
We find that the error ellipses of $B$ and $C$ are similar as in Fig.\ref{LHVlowBCError}, which leads to the same conclusion for distinguishing EOS models.
So different NS masses does not significantly affect the results.
However,
since the premised $\lambda$-$m$ relation in (\ref{lambdaMlinear})
is only valid within $1$-$\sigma$ range of the NS mass-distribution adopted in this paper,
the differentiation of EOS models may be modified according to the true merging NS mass-distribution,
which currently remains unknown unfortunately.

\subsection{The LHVIK network}
\label{subsec:LHVIK}

\begin{figure*}[htbp]
\gridline{\fig{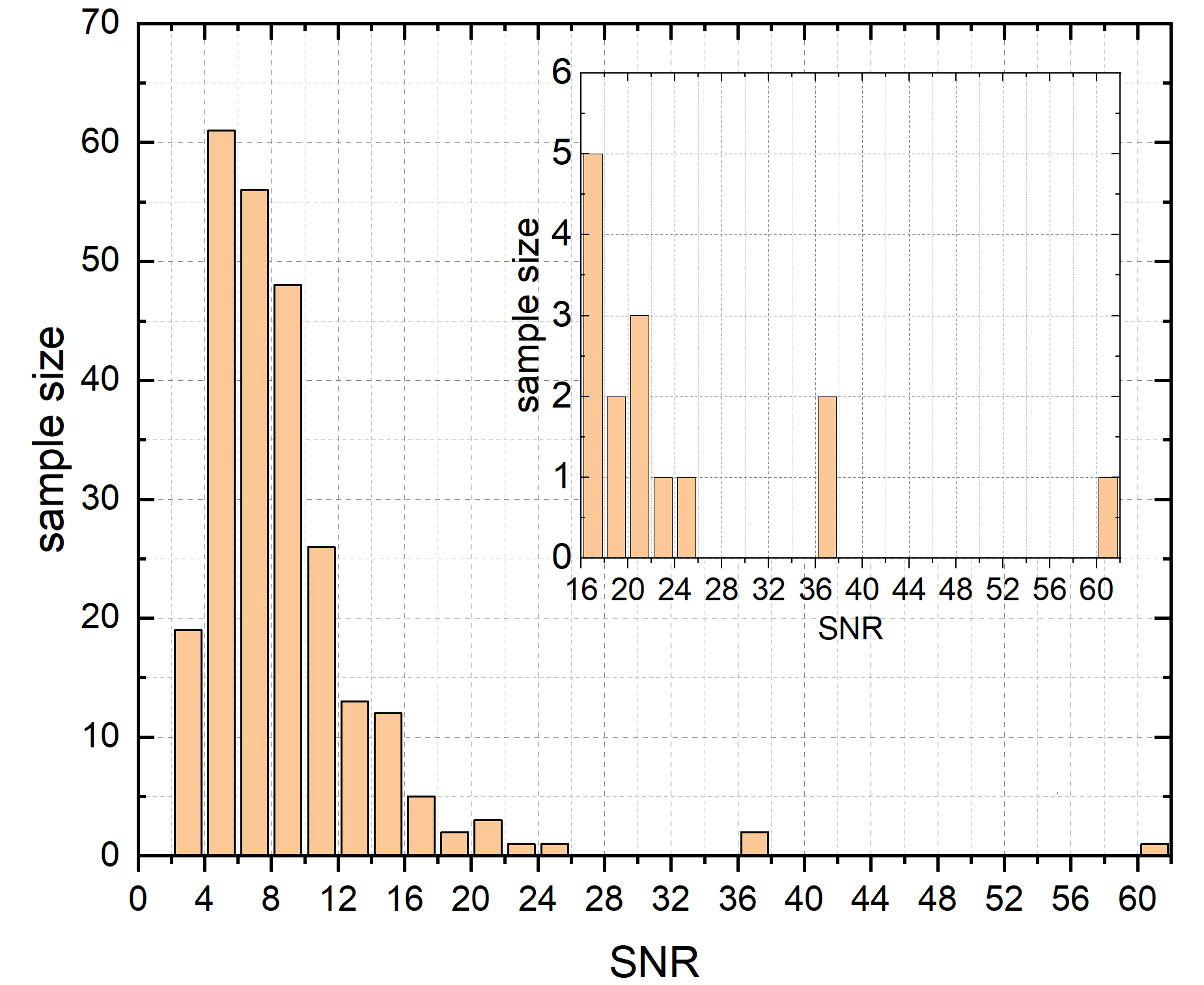}{0.48\textwidth}{(a)}
          \fig{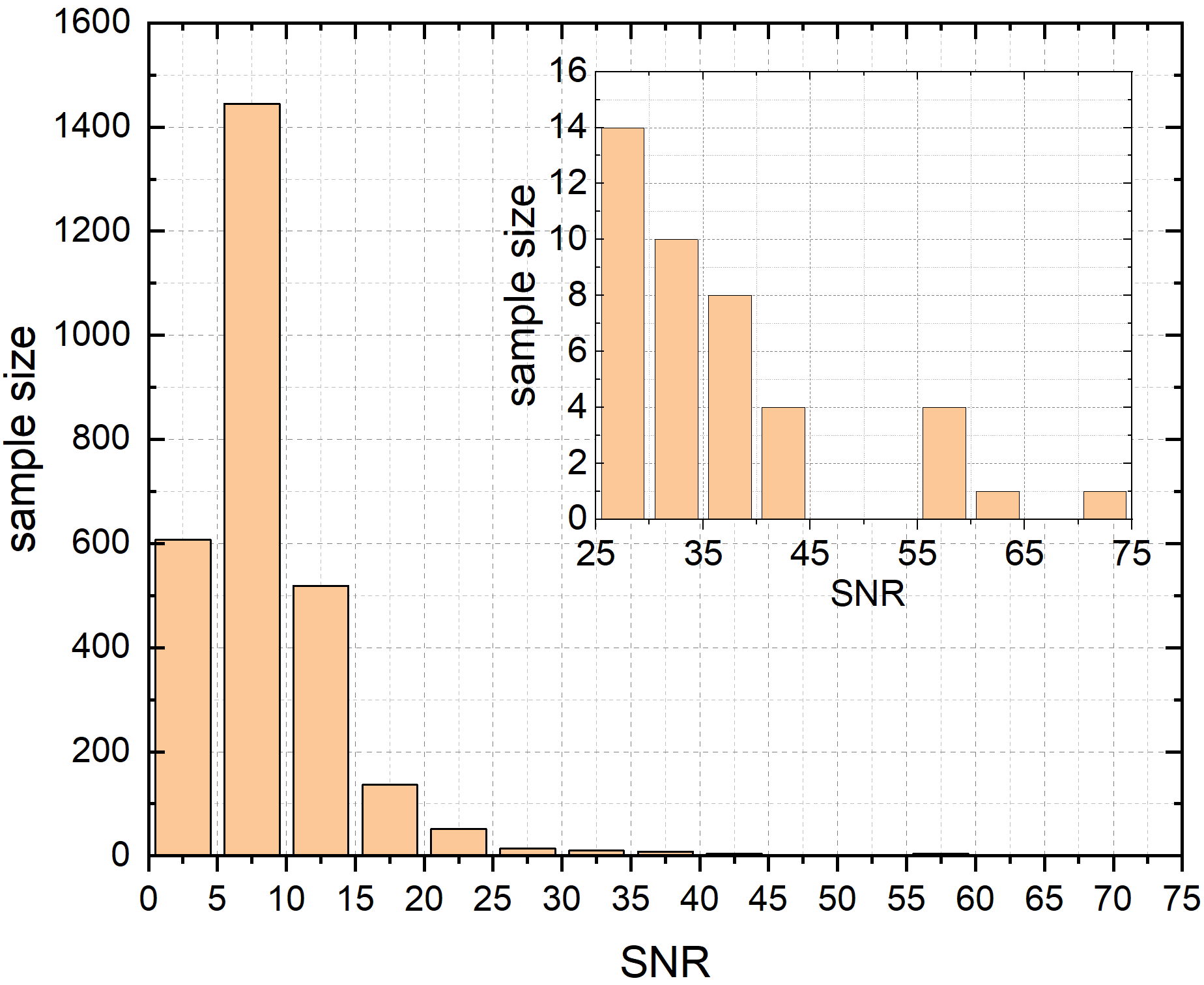}{0.48\textwidth}{(b)}
          }
\caption{ 
By the detection of LHVIK,
the SNR distribution of the samples
generated in the redshift range of
$z<0.1$ is plotted.
In (a),
there are  totally 250 samples and  66 samples  are detectable 
with the threshold of  SNR$>10$.
In (b),
there are totally 2800 samples,
and 748 samples are detectable.
    }
     \label{SNRnLHVIK}
\end{figure*}

\begin{figure*}[htbp]
\gridline{\fig{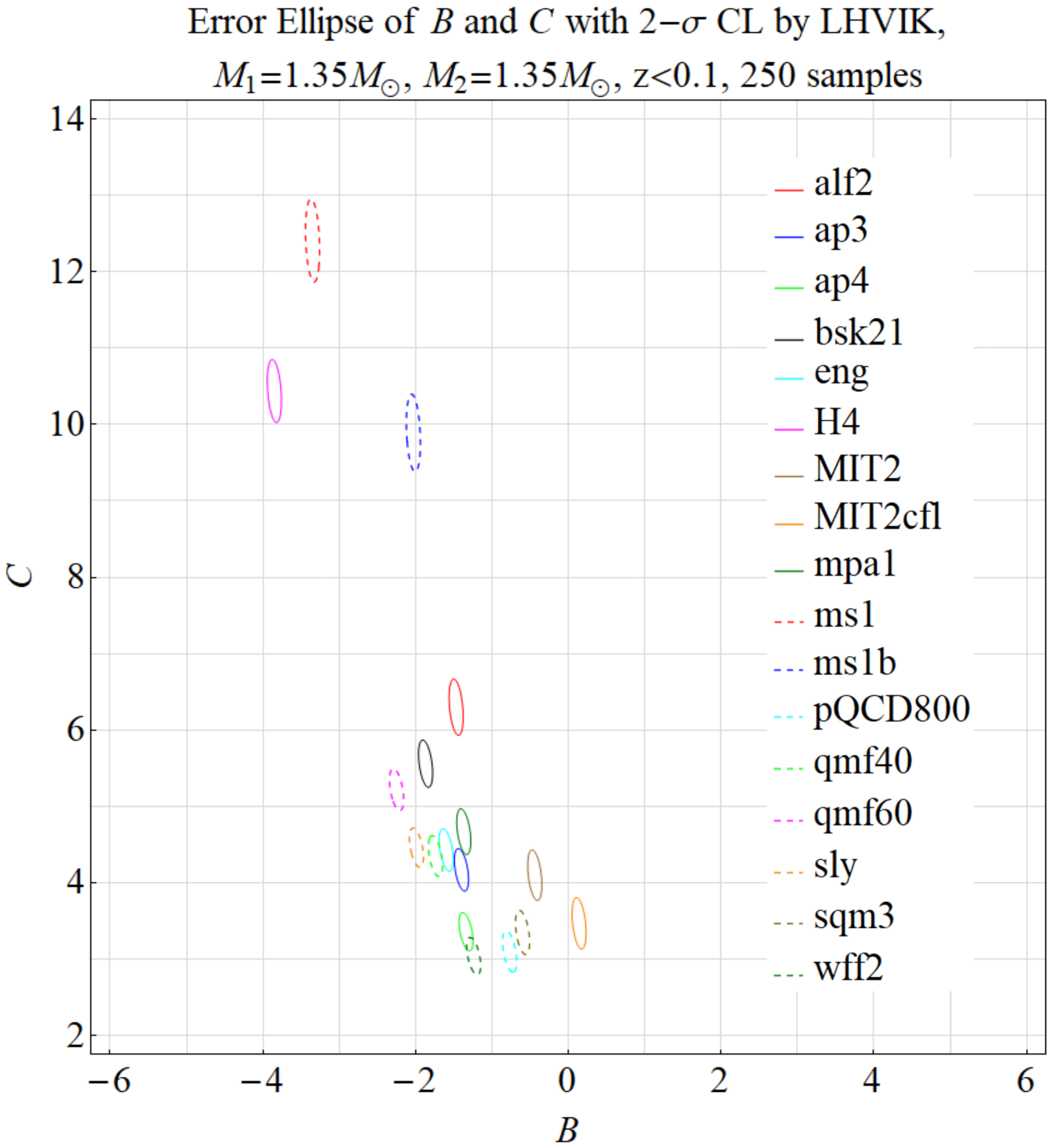}{0.48\textwidth}{(a)}
          \fig{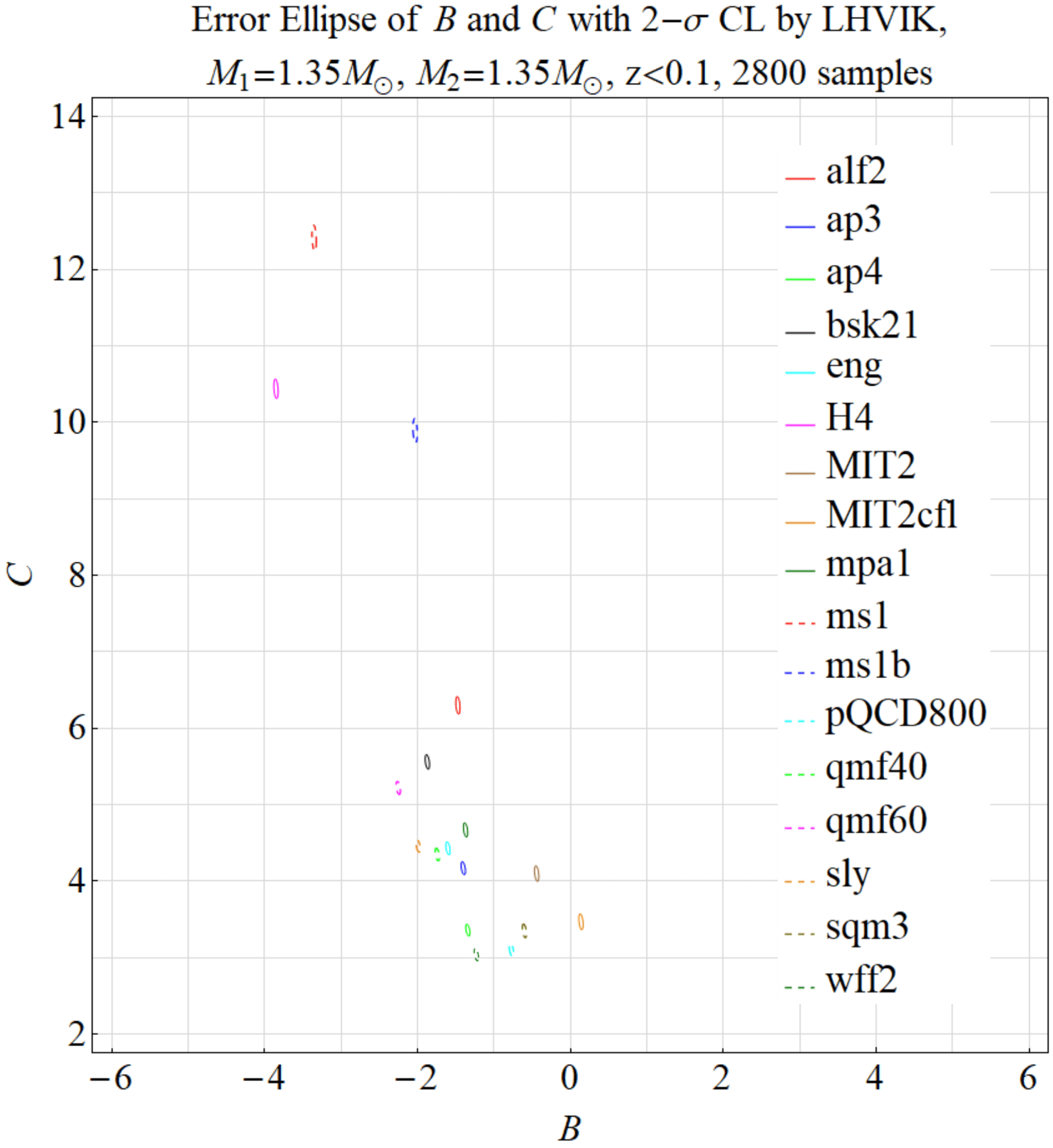}{0.48\textwidth}{(b)}
          }
\caption{
  The error ellipses of $B$ and $C$  at $2$-$\sigma$ CL for different EOS models under the GW detection of LHVIK.
  The EM counterparts of all the GW samples are assumed to be detectable.
  These samples are generated at $z<0.1$.
  (a) is plotted with totally 250 samples
  and  (b) is plotted with totally 2800 samples.
  The corresponding SNR distributions are illustrated in Fig.\ref{SNRnLHVIK}.
  It is observed that in (a), the error ellipses of qmf40, eng, ap4, wff2 have certain overlaps with other models,
  so that they are indistinguishable by this GW observation.
  In (b), there is no overlap between any two ellipses,
  thus in this case all the EOS models are distinguishable.
  }\label{LHVIKlowBCError}
\end{figure*}

Next, we consider the LHVIK network.
The SNR distribution is shown in Fig.\ref{SNRnLHVIK}.
It can be seen that when a total of 250 samples are taken,
there are 66 detectable samples,
more than that in LHV case.
The corresponding error ellipses of $B$ and $C$ for different EOS models are
shown in Fig.\ref{LHVIKlowBCError} (a)
under $2$-$\sigma$ CL.
Compared with Fig.\ref{LHVlowBCError} (a) by LHV,
LHVIK can distinguish 5 additional  EOS models, namely, sly, mpa1, ap3, pQCD800 and sqm3.
When a total of 2800 samples are taken,
there are 748 detectable samples,
and the error ellipses are presented in Fig.\ref{LHVIKlowBCError} (b).
We find that all the EOS models are distinguishable.
It can be seen that the error ellipses by LHVIK are slightly smaller than that by LHV.
This is because the SNR by LHVIK is greater than by LHV and the number of detectable samples by LHVIK is also larger.

\subsection{The 3LA+ network}
\label{subsec:LIGOA+}

\begin{figure*}[htbp]
\gridline{\fig{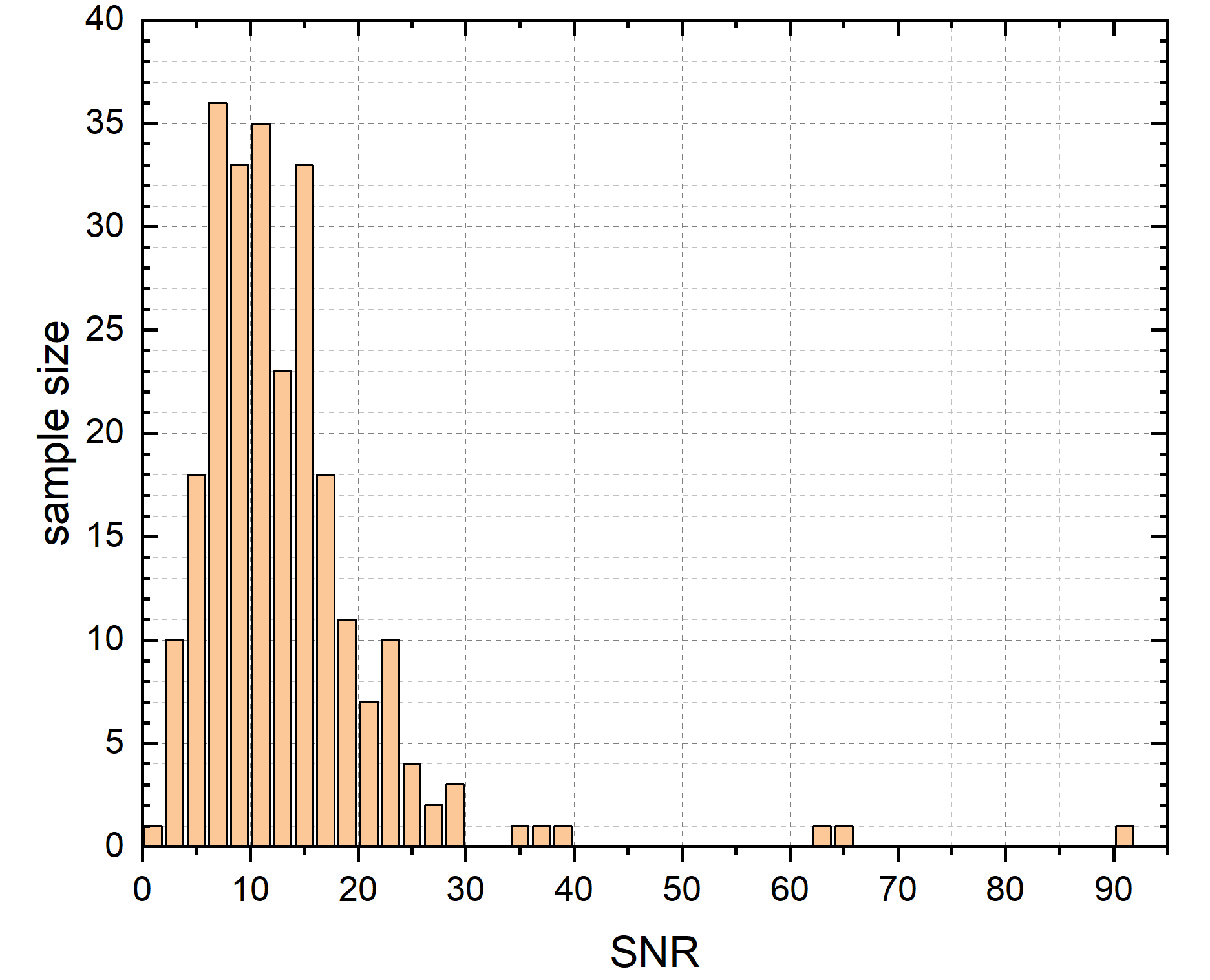}{0.48\textwidth}{(a)}
          \fig{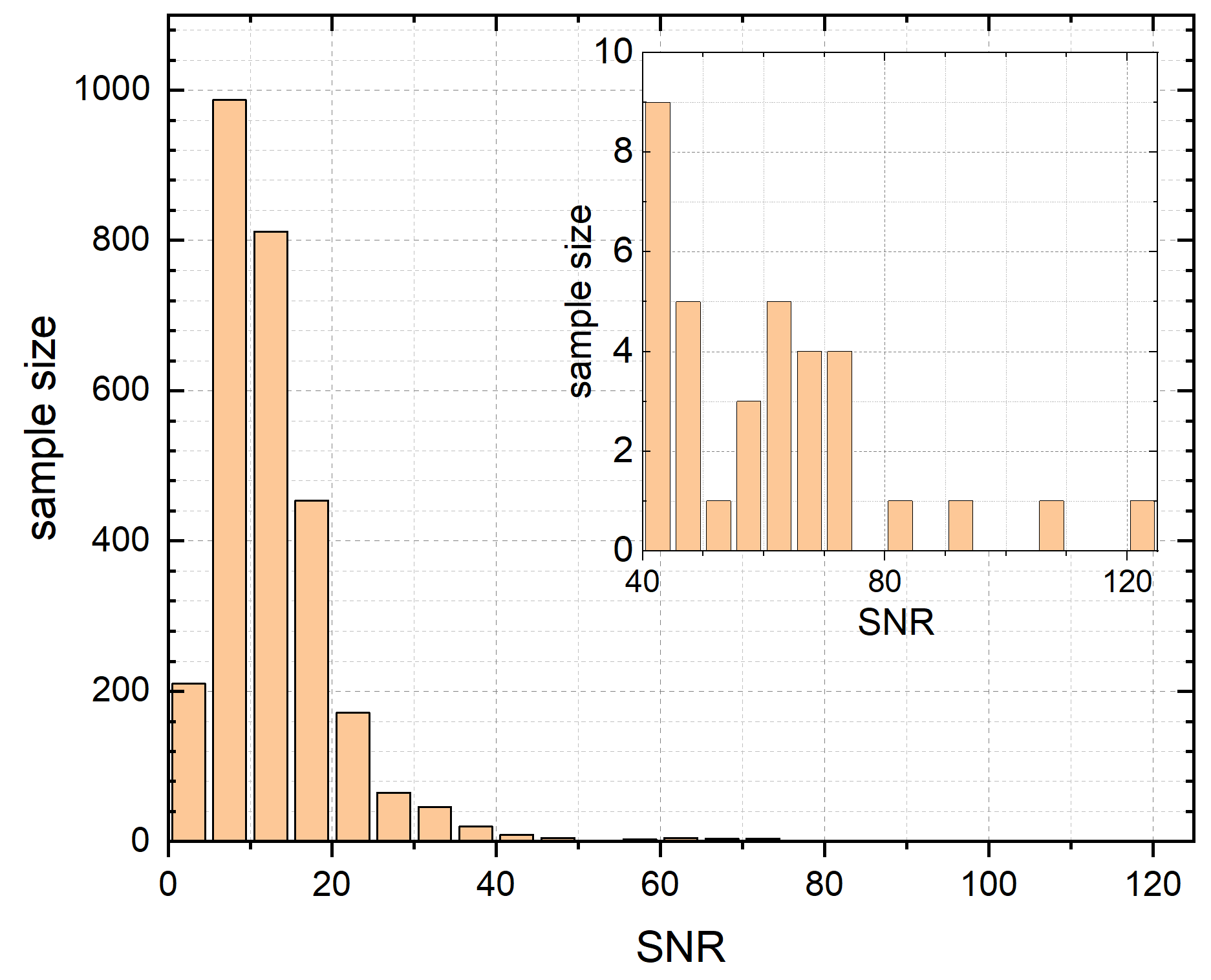}{0.48\textwidth}{(b)}
          }
\caption{
The SNR distribution of the samples
generated in the redshift range of
$z<0.1$ detected by 3LA+.
With the threshold of  SNR$>10$,
there are 152 detectable samples in the total 250 samples as shown in (a);
and there are 1603 detectable samples in the total 2800 samples as shown in (b).
    }
     \label{SNRnLIGOaPlus250}
\end{figure*}

\begin{figure*}[htbp]
\gridline{\fig{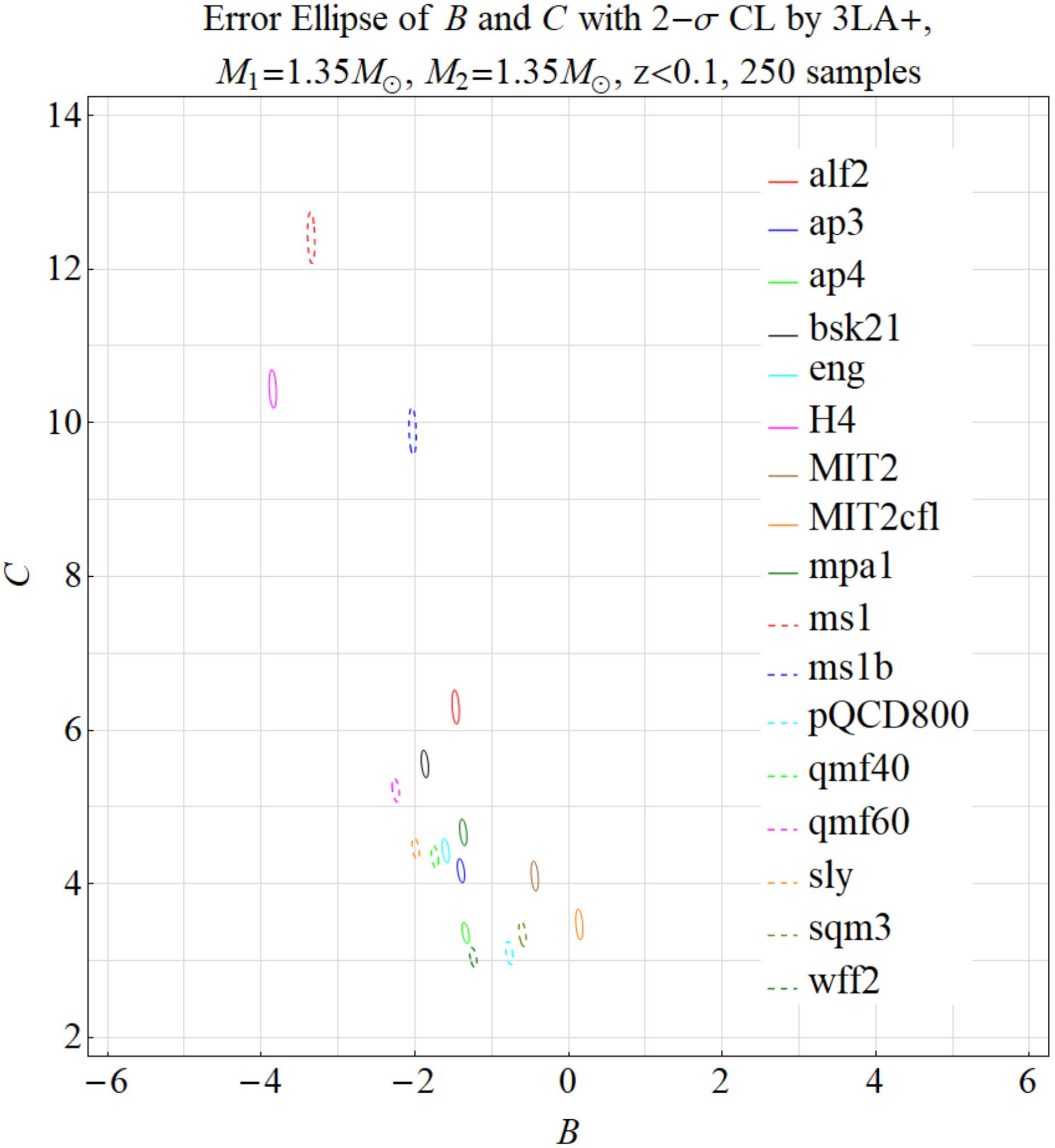}{0.48\textwidth}{(a)}
          \fig{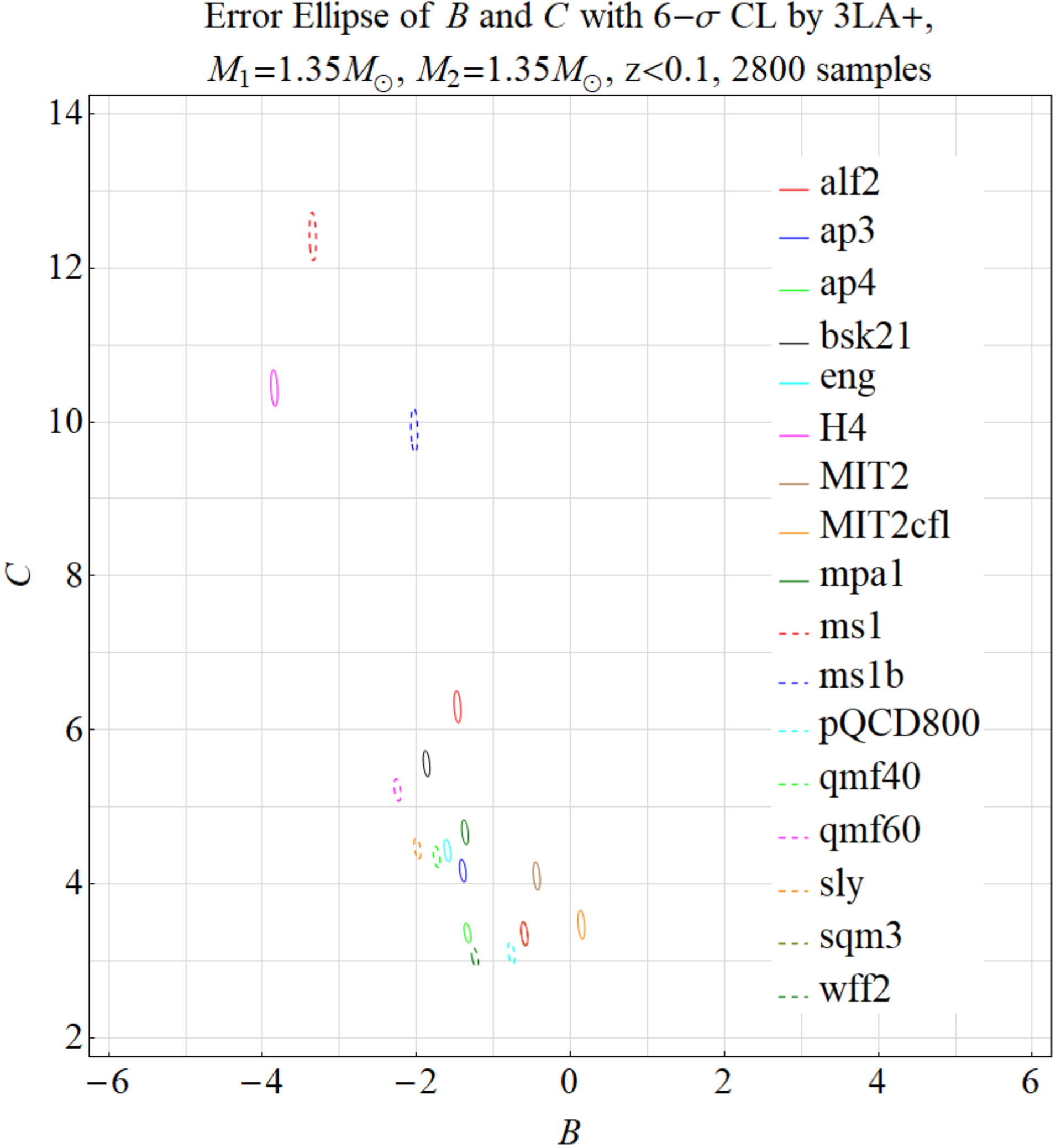}{0.48\textwidth}{(b)}
          }
  \caption{
    The error ellipses of $B$ and $C$  for different EOS models  under the detection of 3LA+.
  We have assumed the EM counterparts for every GW sample are detectable.
  (a) is plotted with the  total 250 samples
  and  (b) is plotted with the total 2800 samples.
  These samples are generated at $z<0.1$,
  and the corresponding SNR distributions are  illustrated in Fig.\ref{SNRnLIGOaPlus250}.
  It is observed that in (a) and (b),
  all the ellipses have no overlap region with the others,
  so that all the EOS models can be distinguished from each other by 3LA+.
  }\label{LIGOaPlus250}
\end{figure*}

Similar to the previous process,
Fig.\ref{SNRnLIGOaPlus250} plots the SNR distribution by 3LA+
in the redshift range of $z<0.1$.
Among a total of 250 samples, 152 samples are detectable,
and the $2$-$\sigma$ error ellipses of $B$ and $C$ are
plotted in Fig.\ref{LIGOaPlus250} (a).
Among a total of 2800 samples, 603 samples can be detected,
and the $6$-$\sigma$ error ellipses are plotted
 in Fig.\ref{LIGOaPlus250} (b).
From the two figures,
it is found that  all the EOS models can be
distinguished by 3LA+ given any merger rate.
Thus, the detection capability of 3LA+ is stronger than both LHV and LHVIK,
and the proportion of samples that can be detected by 3LA+ is higher.

\subsection{The ET2CE network}
\label{subsec:ET2CE}

\begin{figure*}[htbp]
\gridline{\fig{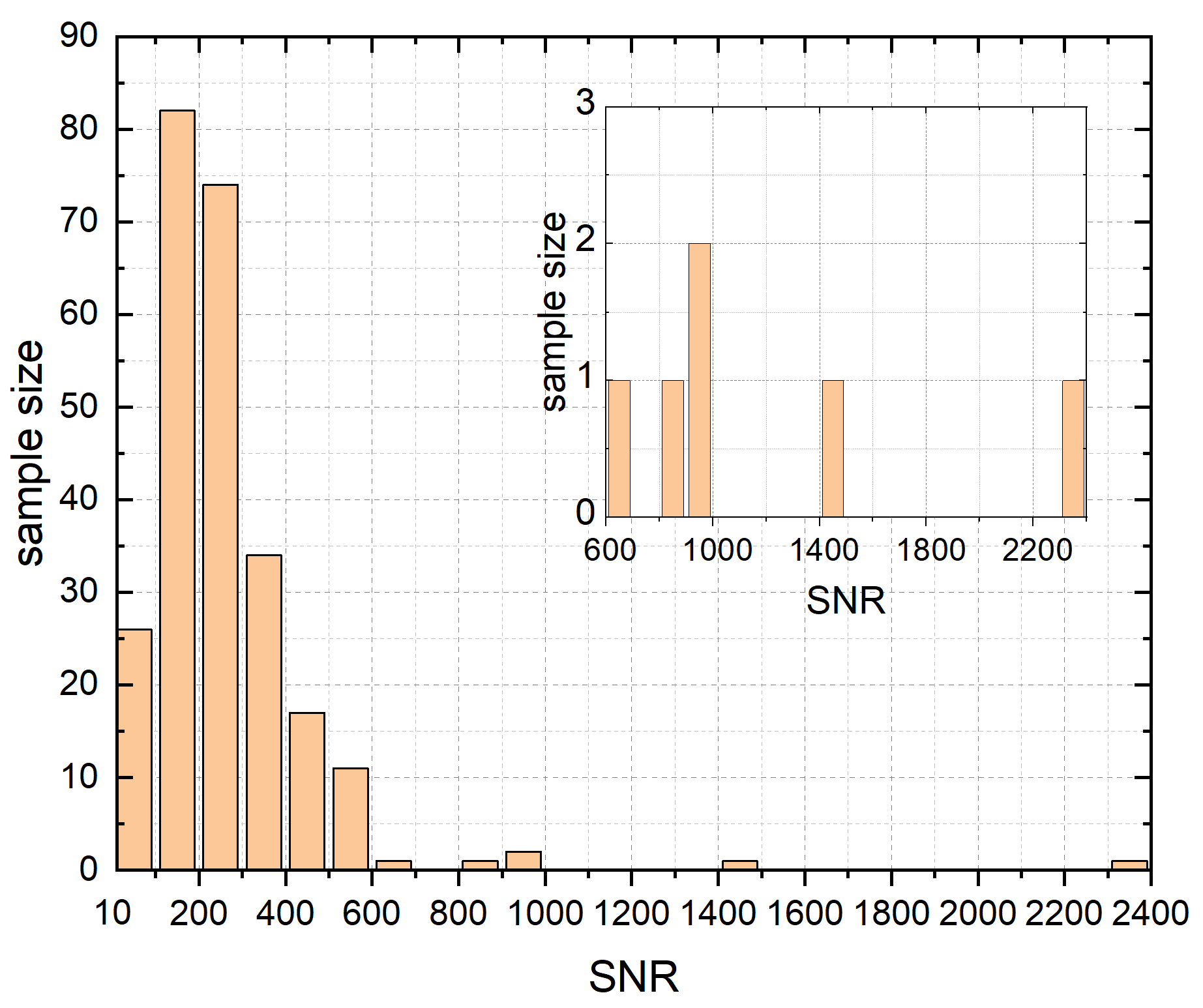}{0.48\textwidth}{(a)}
          \fig{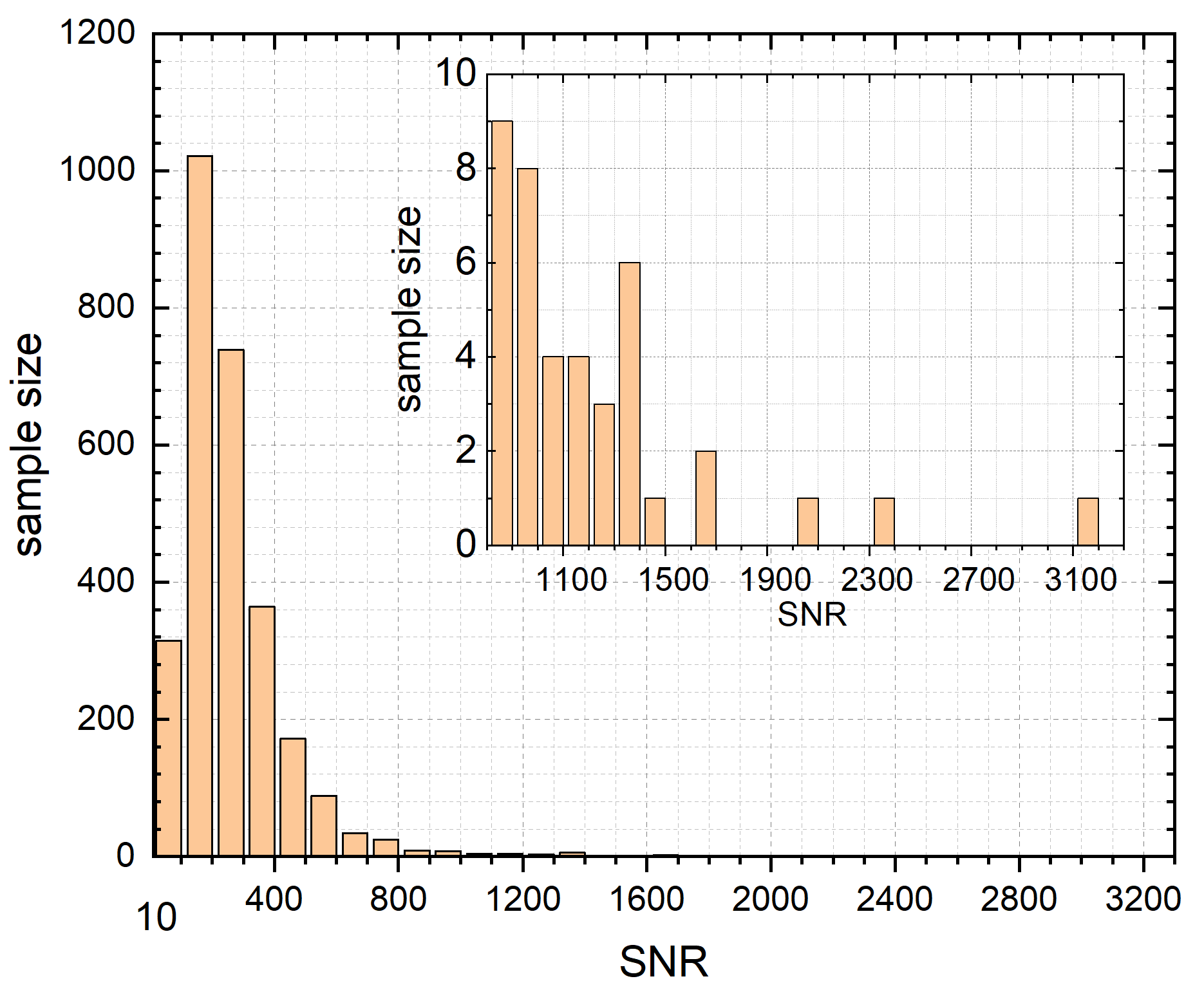}{0.48\textwidth}{(b)}
          }
\caption{The SNR distribution of the samples
generated in the redshift range of
$z<0.1$ detected by ET2CE.
There are totally 250 samples in (a),
and totally 2800 samples in (b).
It is found that with the threshold of  SNR$>10$,
all the GW samples are detectable by ET2CE.
    }
     \label{SNRnET2CE}
\end{figure*}

\begin{figure*}[htbp]
\gridline{\fig{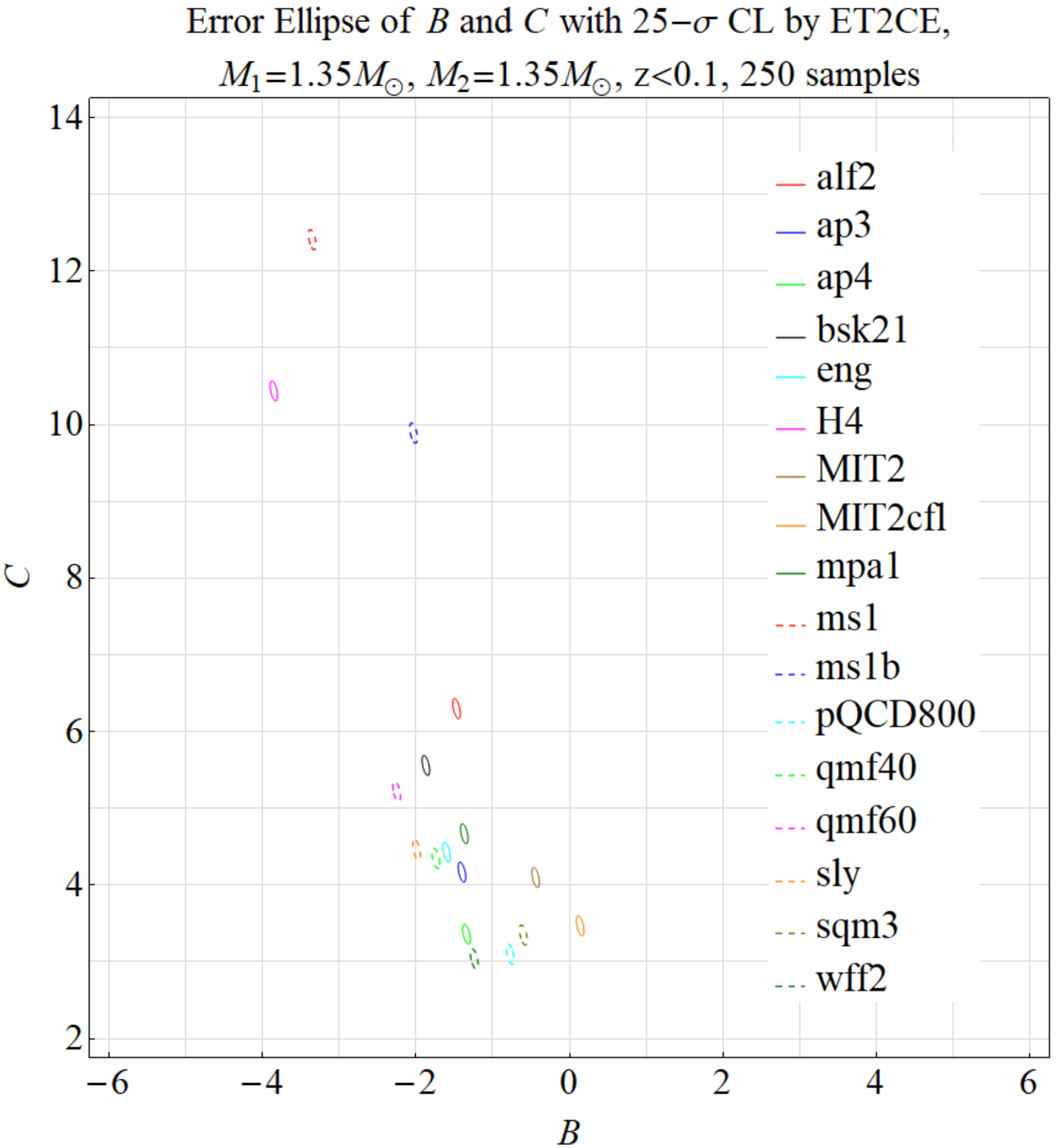}{0.48\textwidth}{(a)}
          \fig{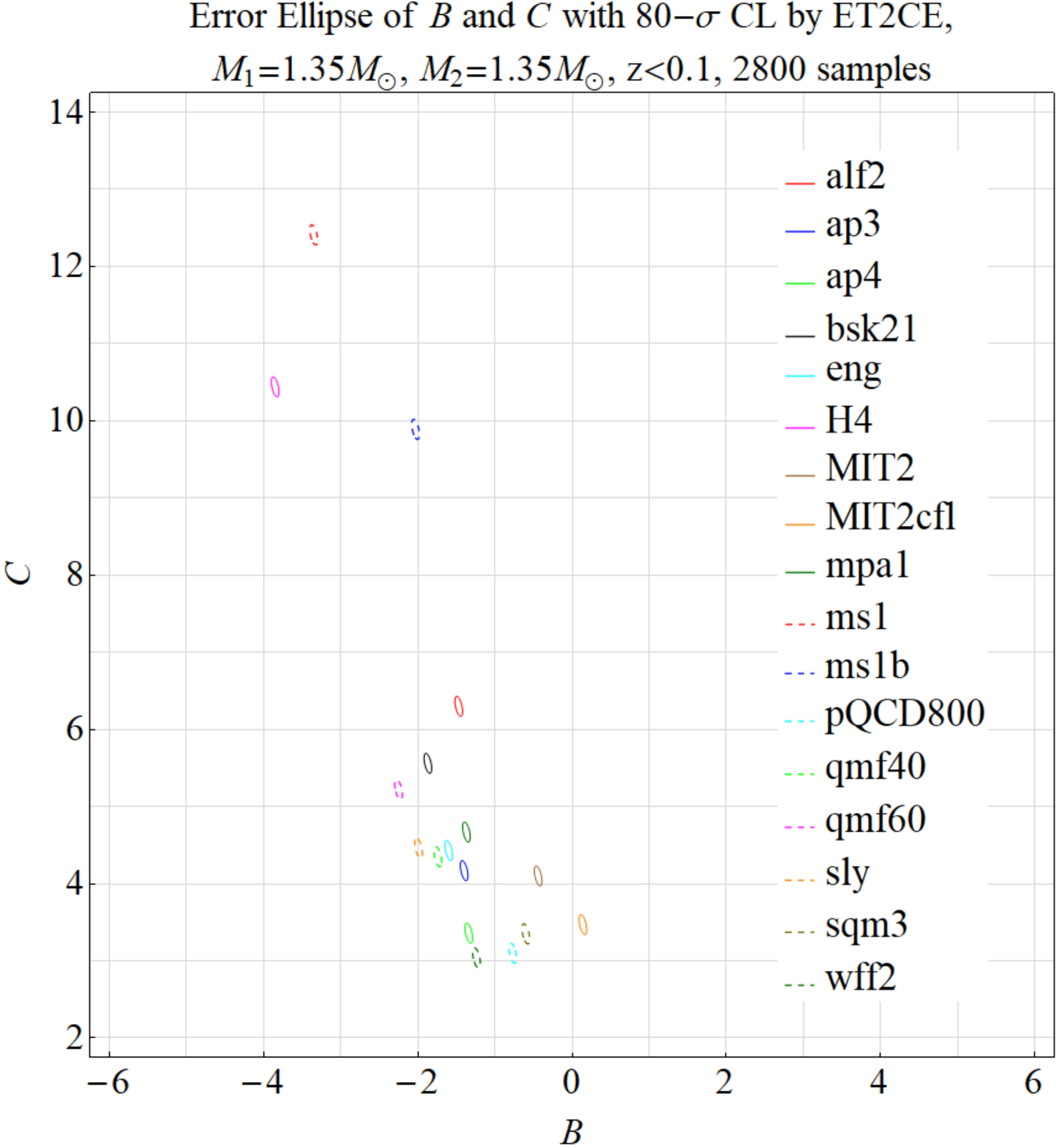}{0.48\textwidth}{(b)}
          }
  \caption{
    The error ellipses of $B$ and $C$   for different EOS models  under the detection of ET2CE.
  The EM counterparts are assumed to be detectable for every GW sample.
  (a) is plotted with the total 250 samples,
  and the error ellipses in (a) are at 25-$\sigma$ CL;
  (b) is plotted with the total 2800 samples,
  and the error ellipses in (b) are at 80-$\sigma$ CL.
  These samples are generated at $z<0.1$,
  and the corresponding SNR distributions are  illustrated in Fig.\ref{SNRnET2CE}.
  It is found that all the EOS models are distinguishable by ET2CE and the confidence level is higher than the  networks of 
  2G and 2.5G detectors.
  These improvements are due to the great sensitivity of the 3G detectors.
  }\label{ET2CElowBCError}
\end{figure*}

Similarly,
the SNR distribution by the ET2CE network is plotted in Fig.\ref{SNRnET2CE} in the redshift $z<0.1$.
Under criterion SNR$>10$,
all GW samples are detectable in the given merger rates.
When the merge events are taken as the lower limit of 250,
the $25$-$\sigma$ error ellipses of $B$ and $C$ are
plotted in Fig.\ref{ET2CElowBCError} (a).
When the merger events is taken
to be its upper limit as 2800 samples,
the $80$-$\sigma$ error ellipses are
 in Fig.\ref{ET2CElowBCError} (b).
It is observed that  all the EOS models can be distinguished  for  every possible merger rate.
This improvement is due to the great sensitivity of the ET2CE network.

\begin{figure*}[htbp]
\gridline{\fig{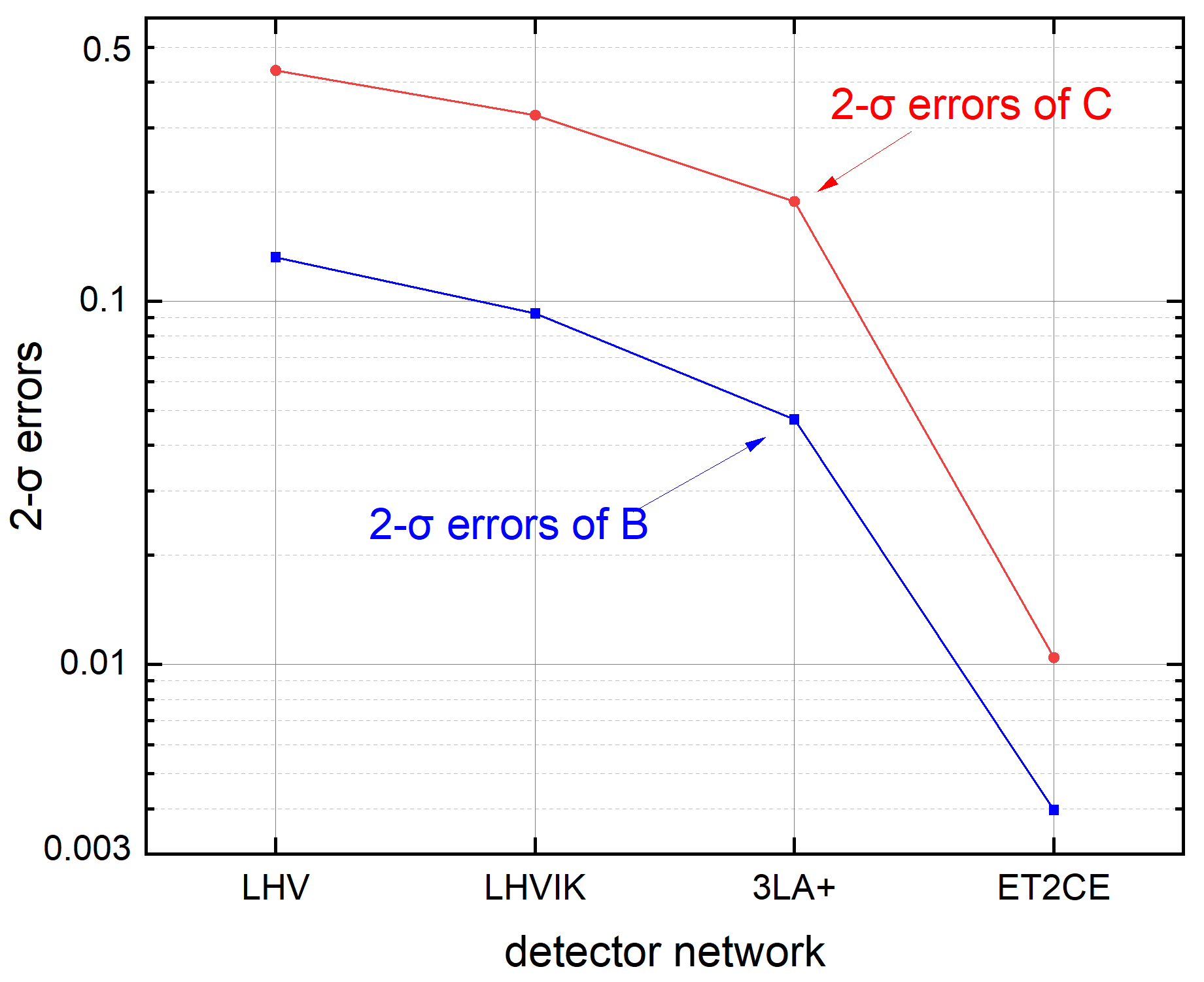}{0.48\textwidth}{(a)}
          \fig{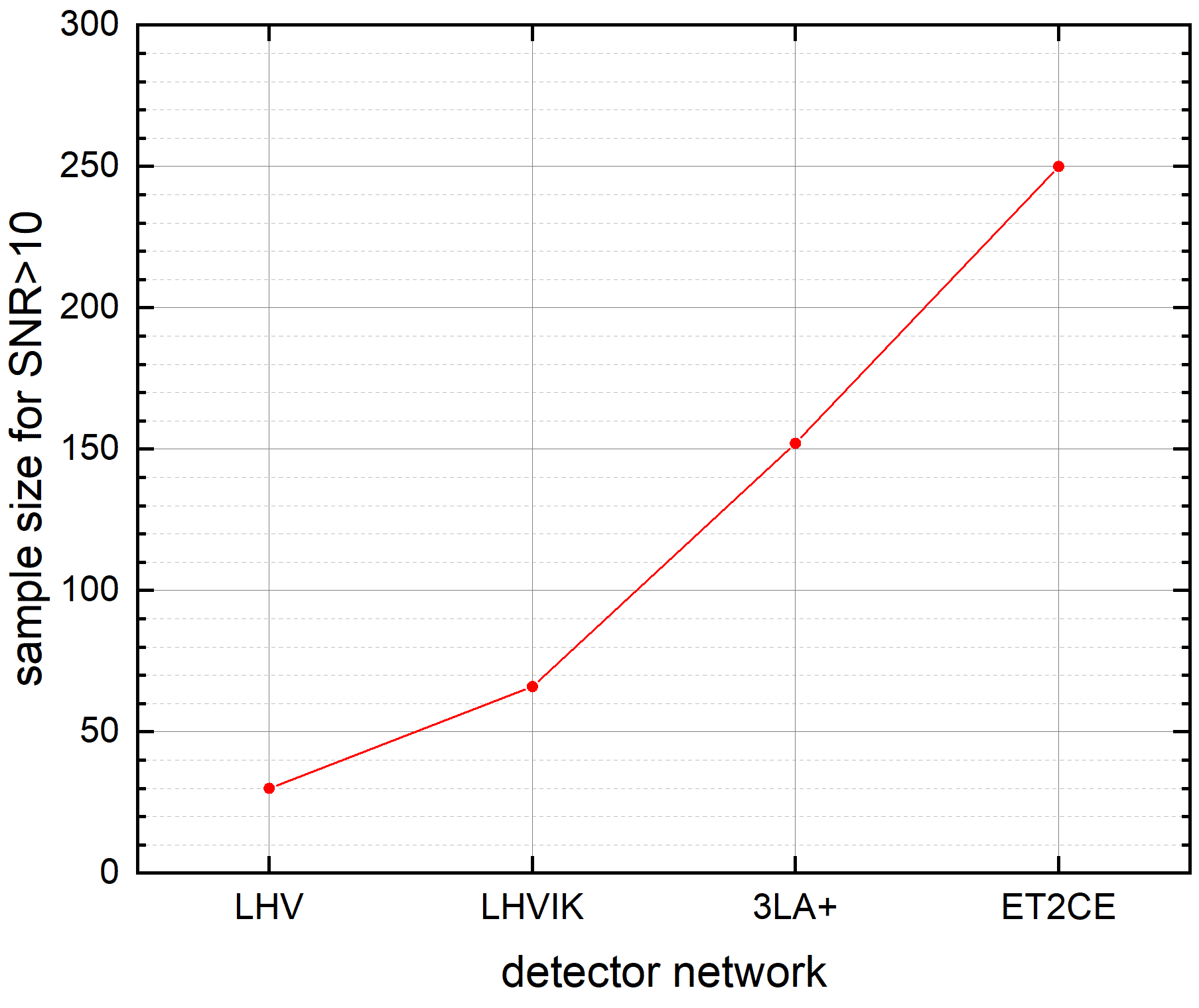}{0.48\textwidth}{(b)}
          }
  \caption{
  (a) 
   illustrates the average of the 2-$\sigma$ errors of $B$ and $C$  over all the EOS models for each network,
   which is calculated with a total of 250 samples at $z<0.1$.
    We have assumed the EM counterparts for every GW sample are detectable.
   (b) illustrates
   the number of the detectable samples in the redshift range of $z<0.1$ for each network. It is seen that the detection capability of the 2.5G and 3G detector networks is greater than the 2G ones.
  }\label{BCerror250net}
\end{figure*}

In order to compare the detection capabilities of different detector networks, at $z<0.1$ with a total of 250 samples, 
we take the average of the 2-$\sigma$ errors of $B$ and $C$ given by each network over all the EOS models.
The result is shown in Fig.\ref{BCerror250net} (a).
The number of the detectable samples at $z<0.1$ With the threshold of SNR$>10$ for each network is also plotted in Fig.\ref{BCerror250net} (b).
It is found that the errors of $B$ and $C$ detected by the 2.5G or 3G detector network are smaller than the 2G one.
This is due to the greater sensitivity of the 2.5G and 3G detectors which leads to more detectable samples as shown in Fig.\ref{BCerror250net} (b).
And a network with more detectors such as LHVIK can increase the detectable sample size and reduce the errors of $B$ and $C$ but not very significantly.

\subsection{Detectable EM counterpart in $z<0.05$}
\label{subsec:z0105}

\begin{figure*}[htbp]
\gridline{\fig{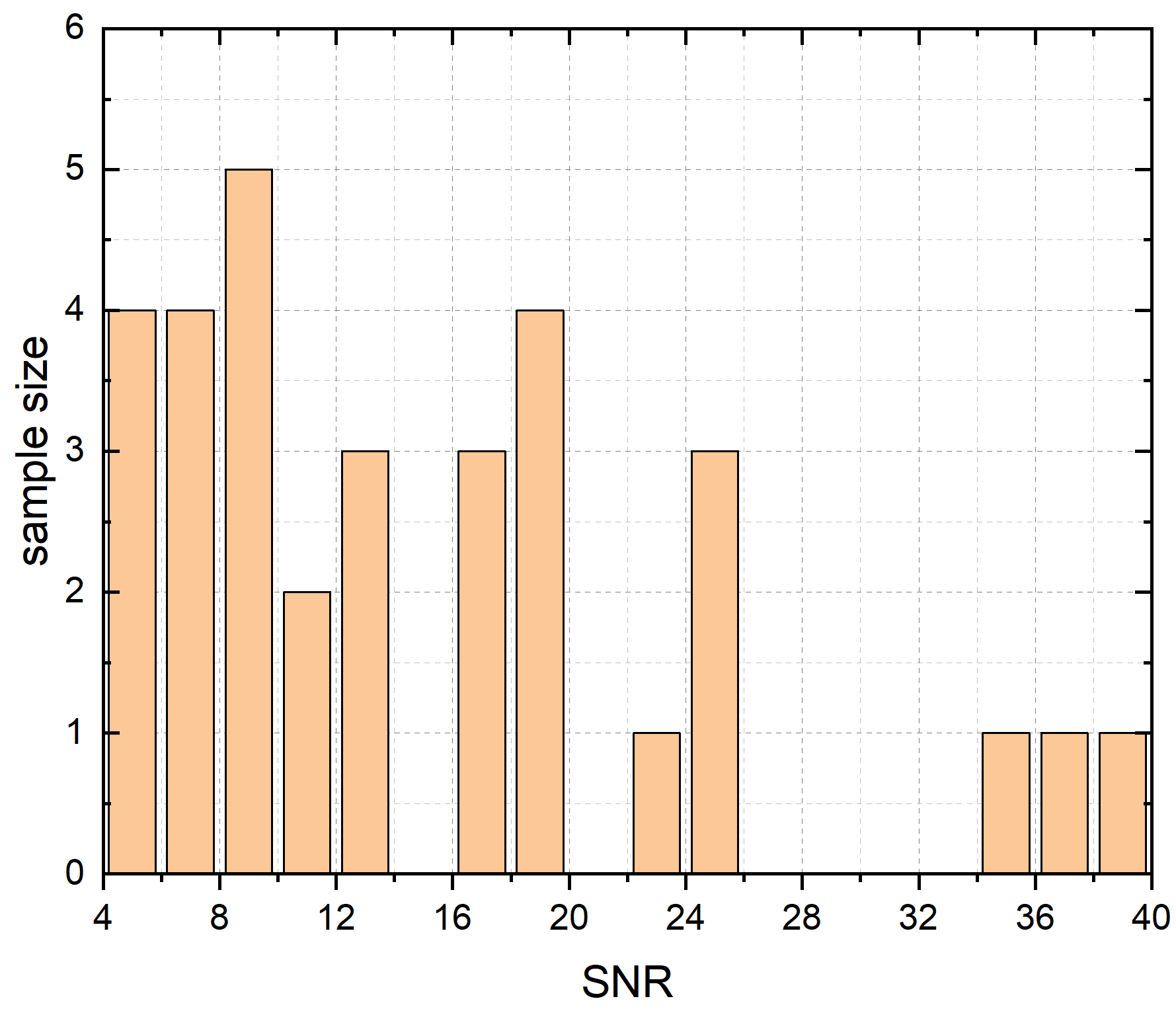}{0.48\textwidth}{(a)}
          \fig{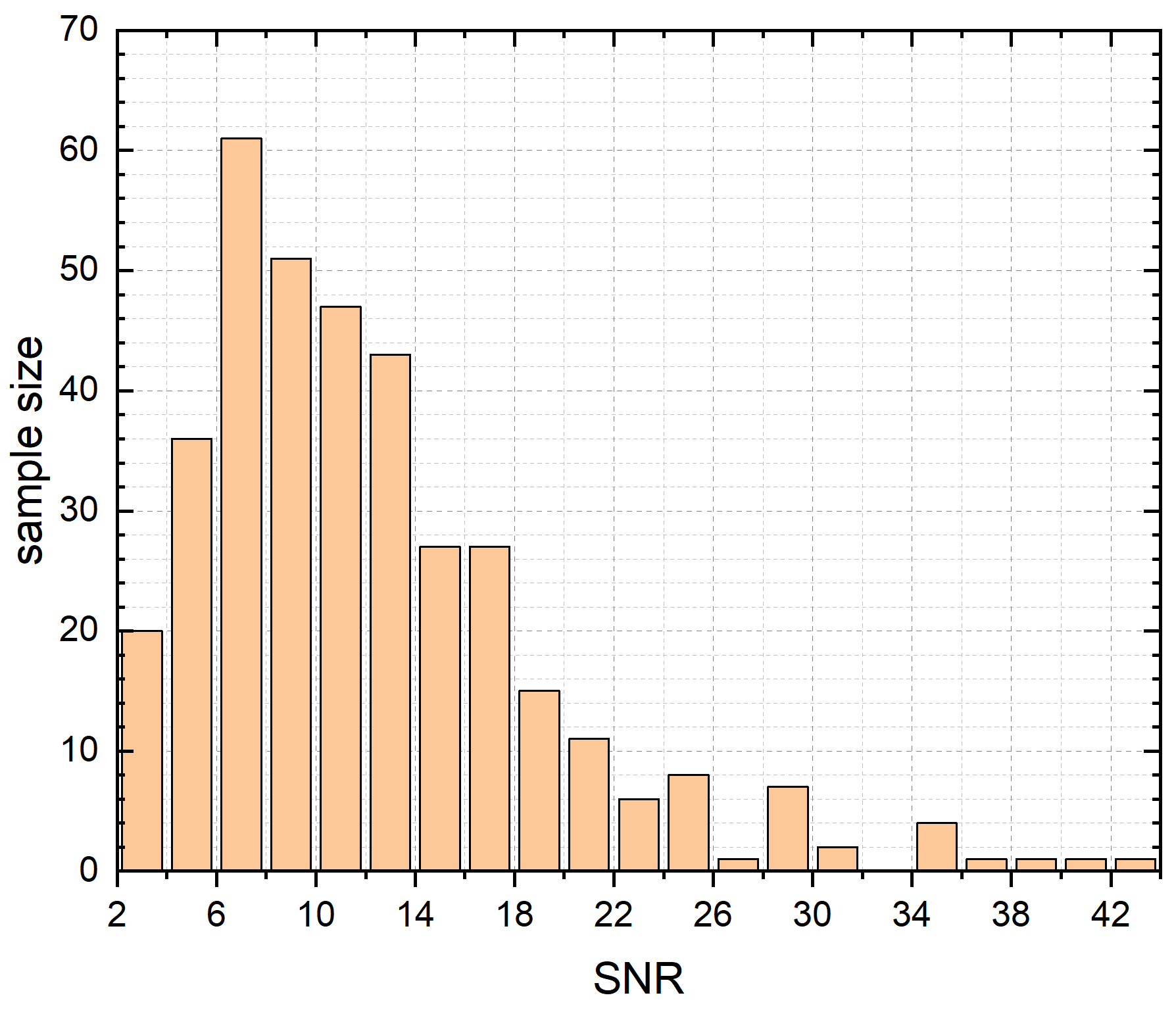}{0.48\textwidth}{(b)}
          }
\caption{
The SNR distribution of the samples
generated in the redshift range of
$z<0.05$ detected by LHV.
There are totally 32 samples in (a),
and totally 370 samples in (b),
according to the lowest and highest event rates given in Ref.\citep{AbbottAbbottAbbott2019} respectively.
It is found that with the threshold of  SNR$>10$,
there are 19 detectable samples in (a),
and 202 detectable samples in (b).
    }
     \label{SNRnLHV32370}
\end{figure*}

\begin{figure*}[htbp]
\gridline{\fig{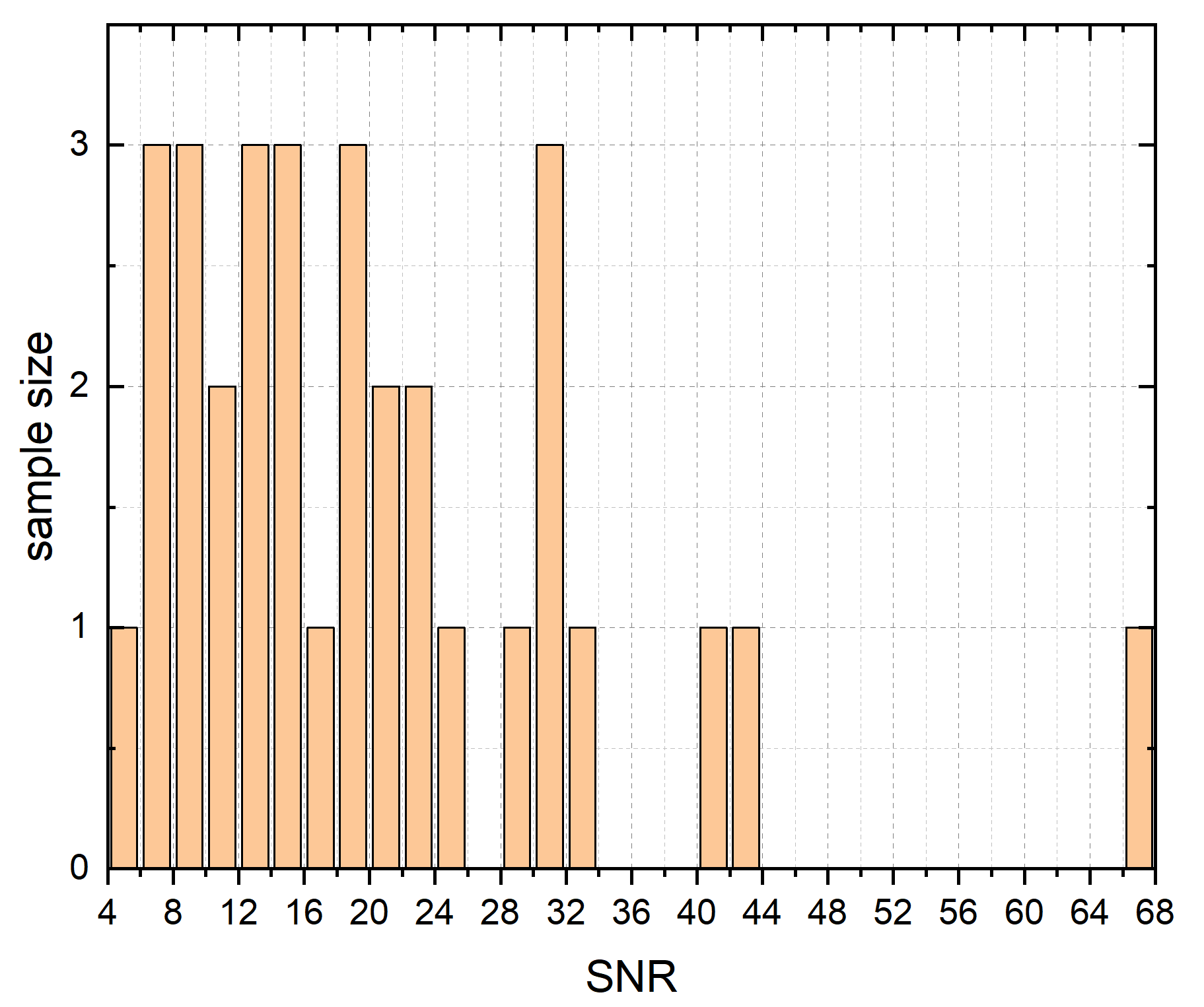}{0.48\textwidth}{(a)}
          \fig{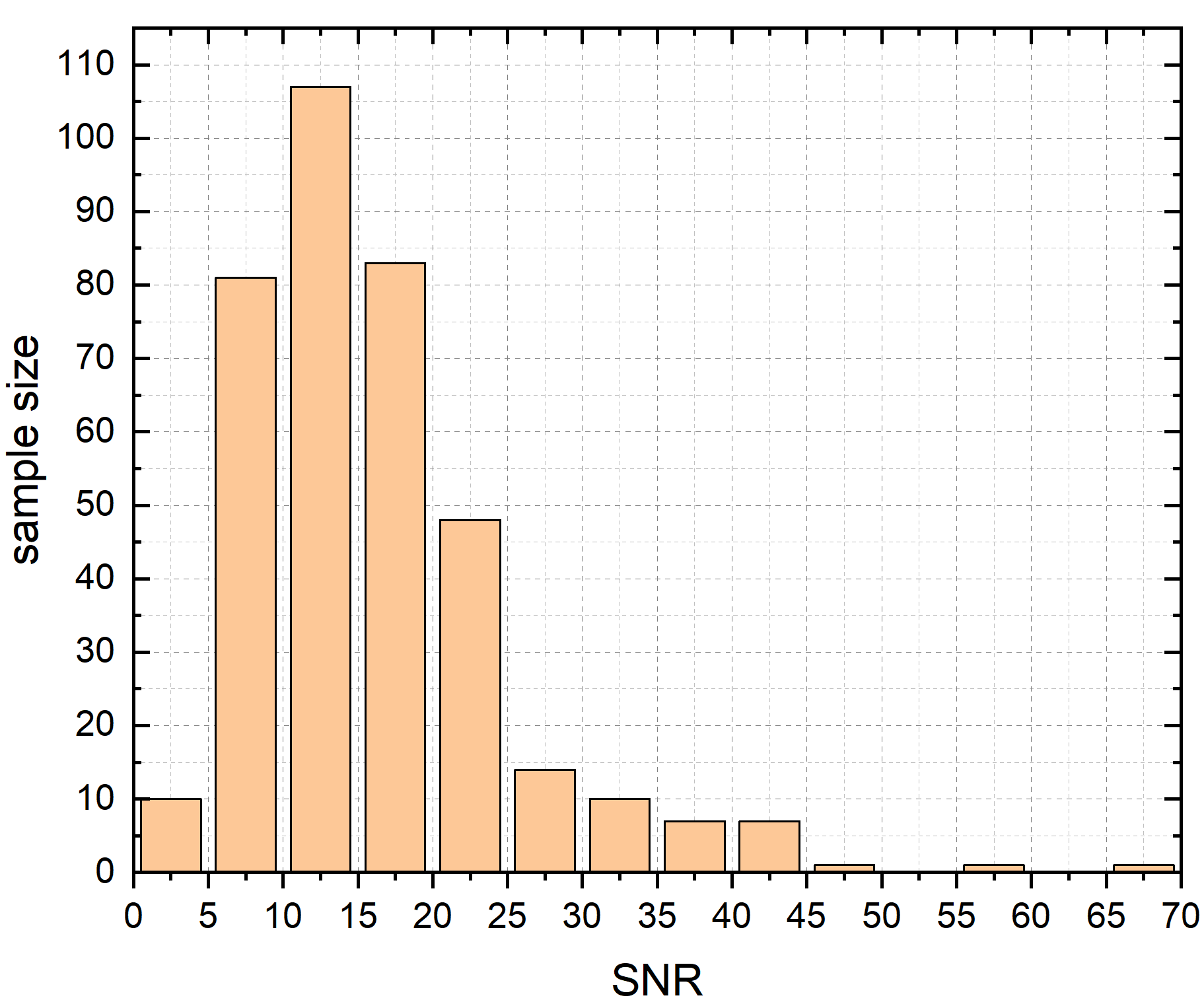}{0.48\textwidth}{(b)}
          }
\caption{The SNR distribution of the samples
generated in the redshift range of
$z<0.05$ detected by LHVIK.
With the threshold of SNR$>10$,
there are 25 detectable samples in the total 32 samples as shown in (a);
and there are 279 detectable samples in the total 370 samples as shown in (b).
    }
     \label{SNRnLHVIK32370}
\end{figure*}

\begin{figure*}[htbp]
\gridline{\fig{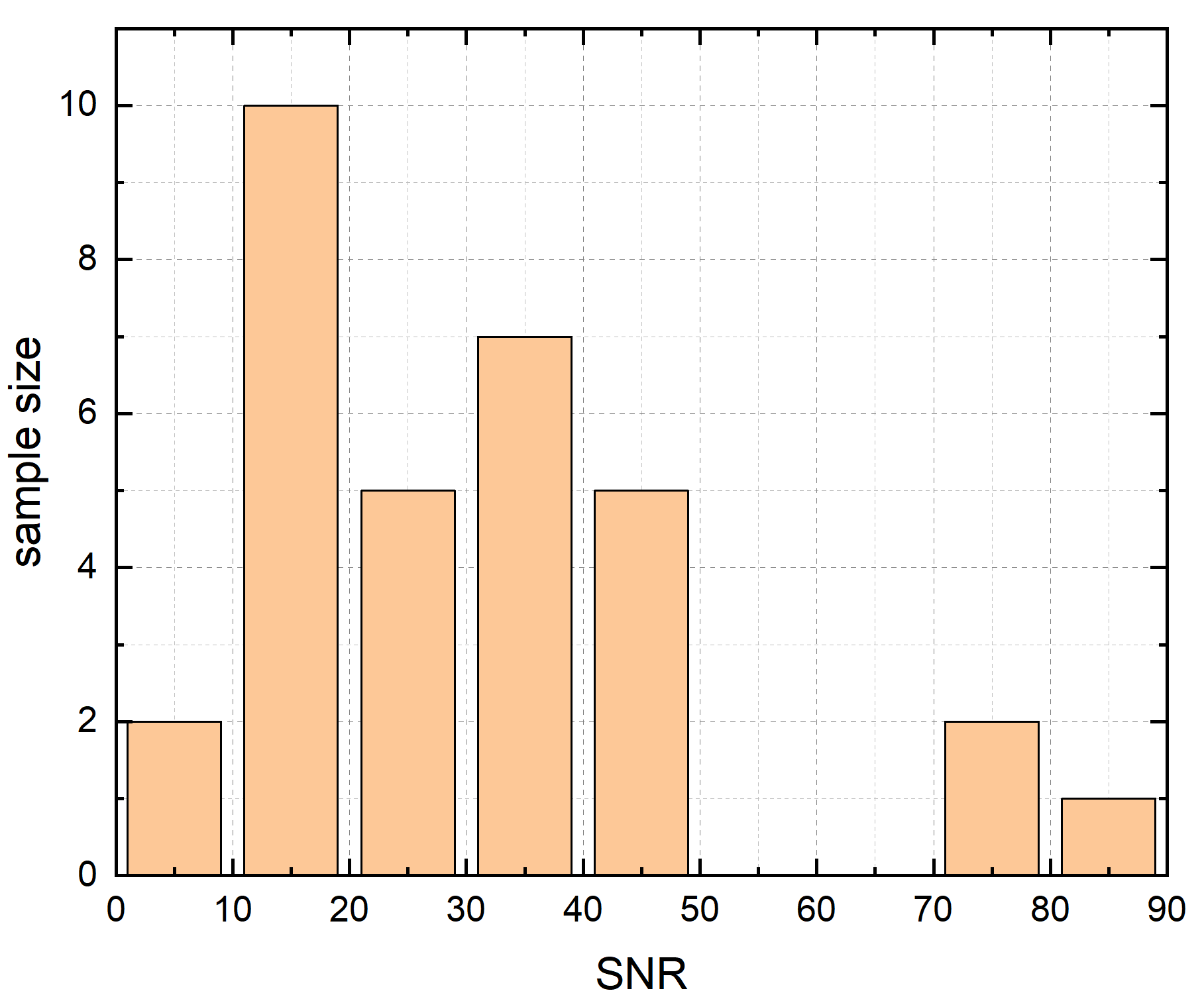}{0.48\textwidth}{(a)}
          \fig{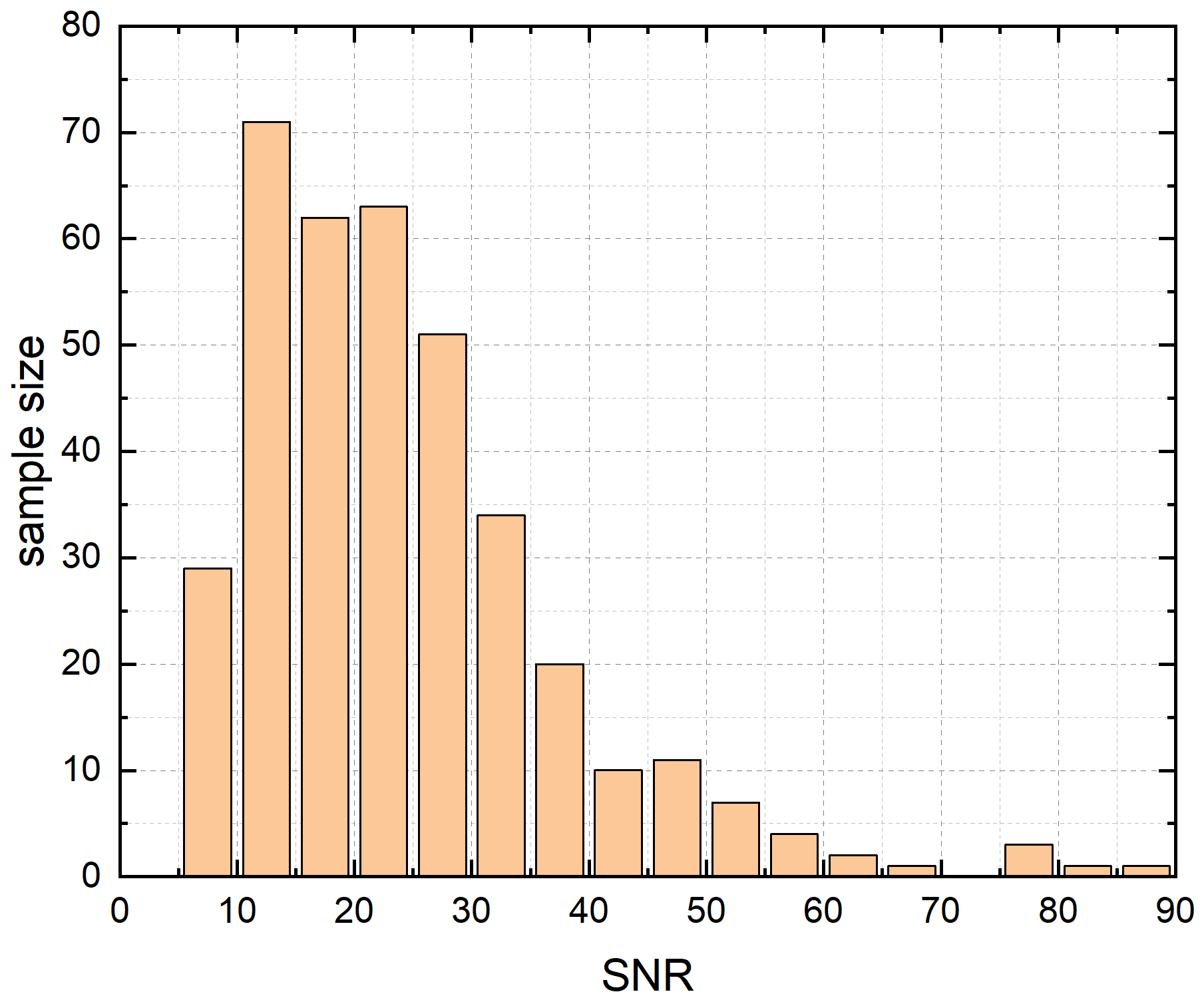}{0.48\textwidth}{(b)}
          }
\caption{In the redshift range of
$z<0.05$ and detected by 3LA+,
the SNR distribution of the generated samples is illustrated.
With the threshold of SNR$>10$,
there are 30 detectable samples in the total 32 samples as shown in (a);
and there are 341 detectable samples in the total 370 samples as shown in (b).
    }
     \label{SNRn3LAplus32370}
\end{figure*}

\begin{figure*}[htbp]
\gridline{\fig{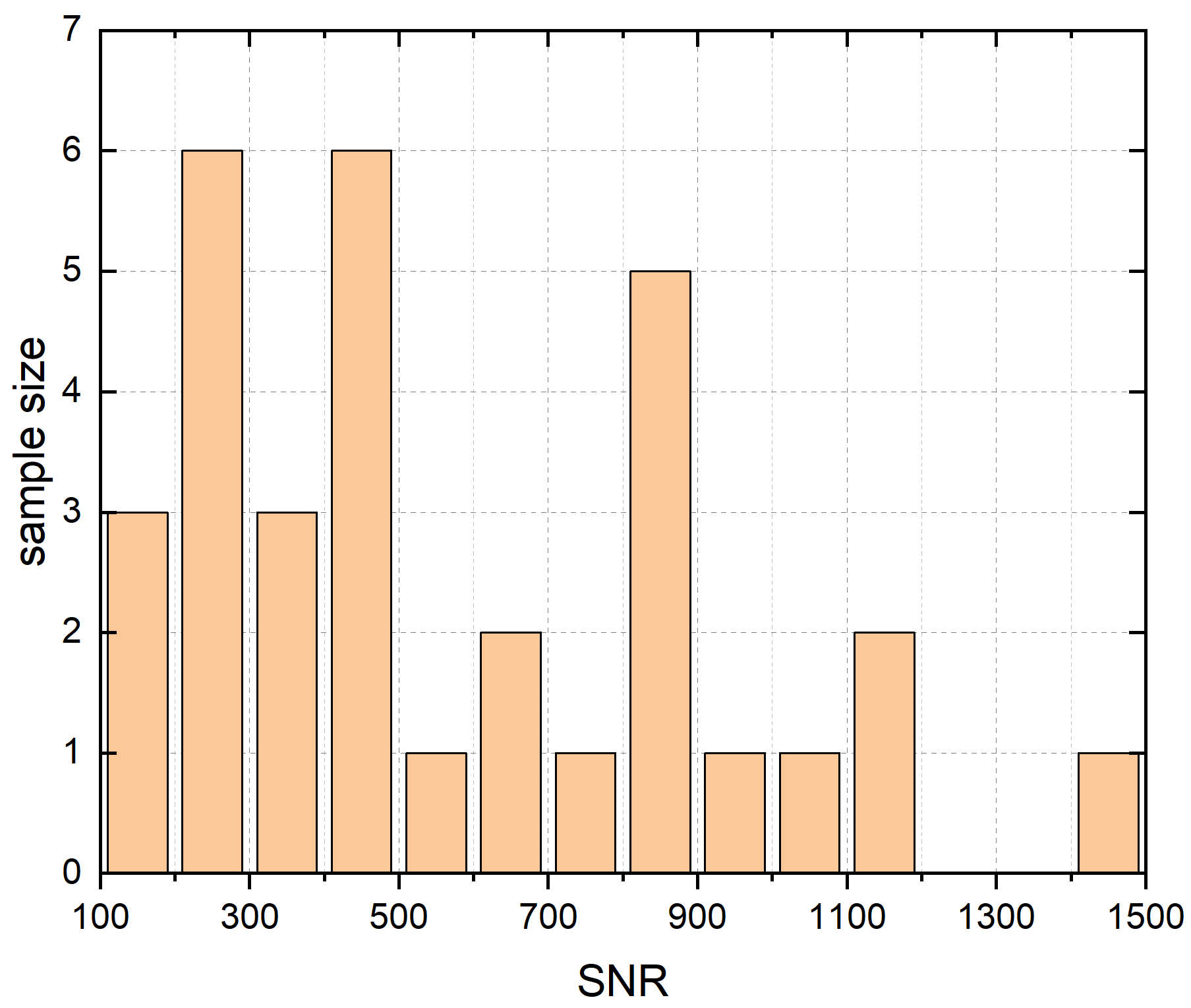}{0.48\textwidth}{(a)}
          \fig{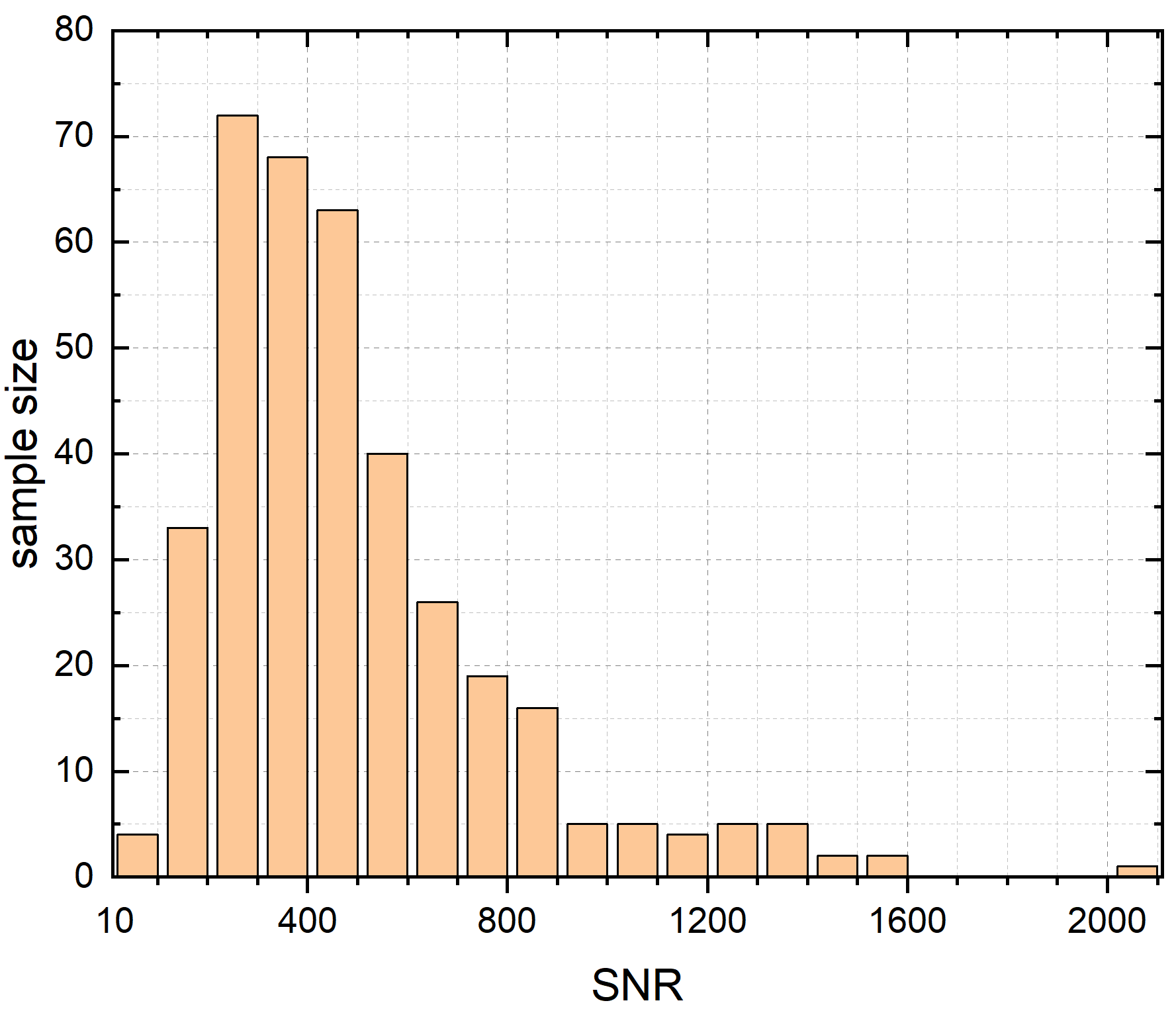}{0.48\textwidth}{(b)}
          }
\caption{The SNR distribution of the  samples generated
in the redshift range of
$z<0.05$ and detected by ET2CE.
There are totally 32 samples as shown in (a);
and there are totally 370 samples as shown in (b).
With the threshold of SNR$>10$, all the samples are detectable by ET2CE.
    }
     \label{SNRnET2CE32370}
\end{figure*}

\begin{figure*}[htbp]
\gridline{\fig{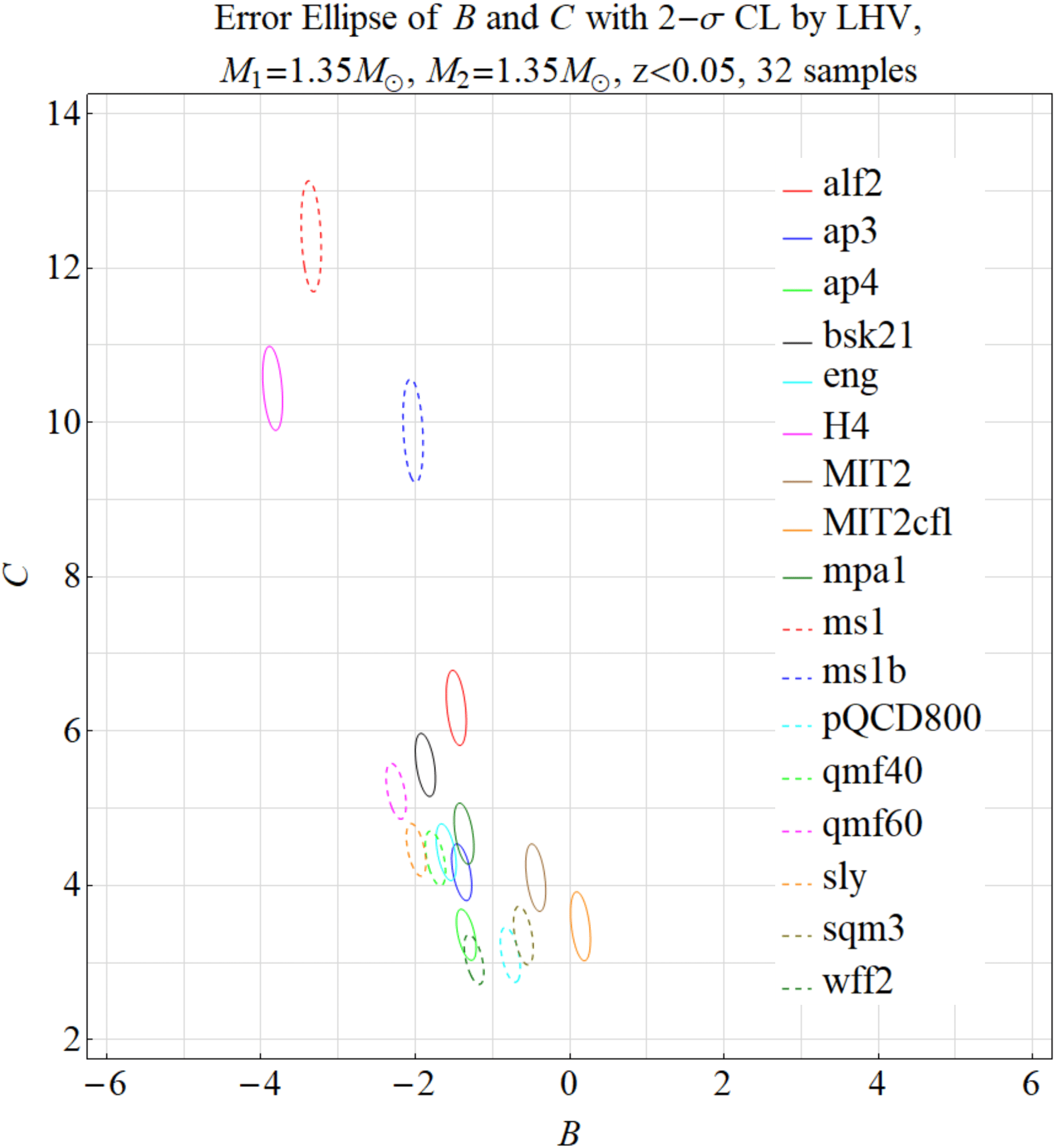}{0.48\textwidth}{(a)}
          \fig{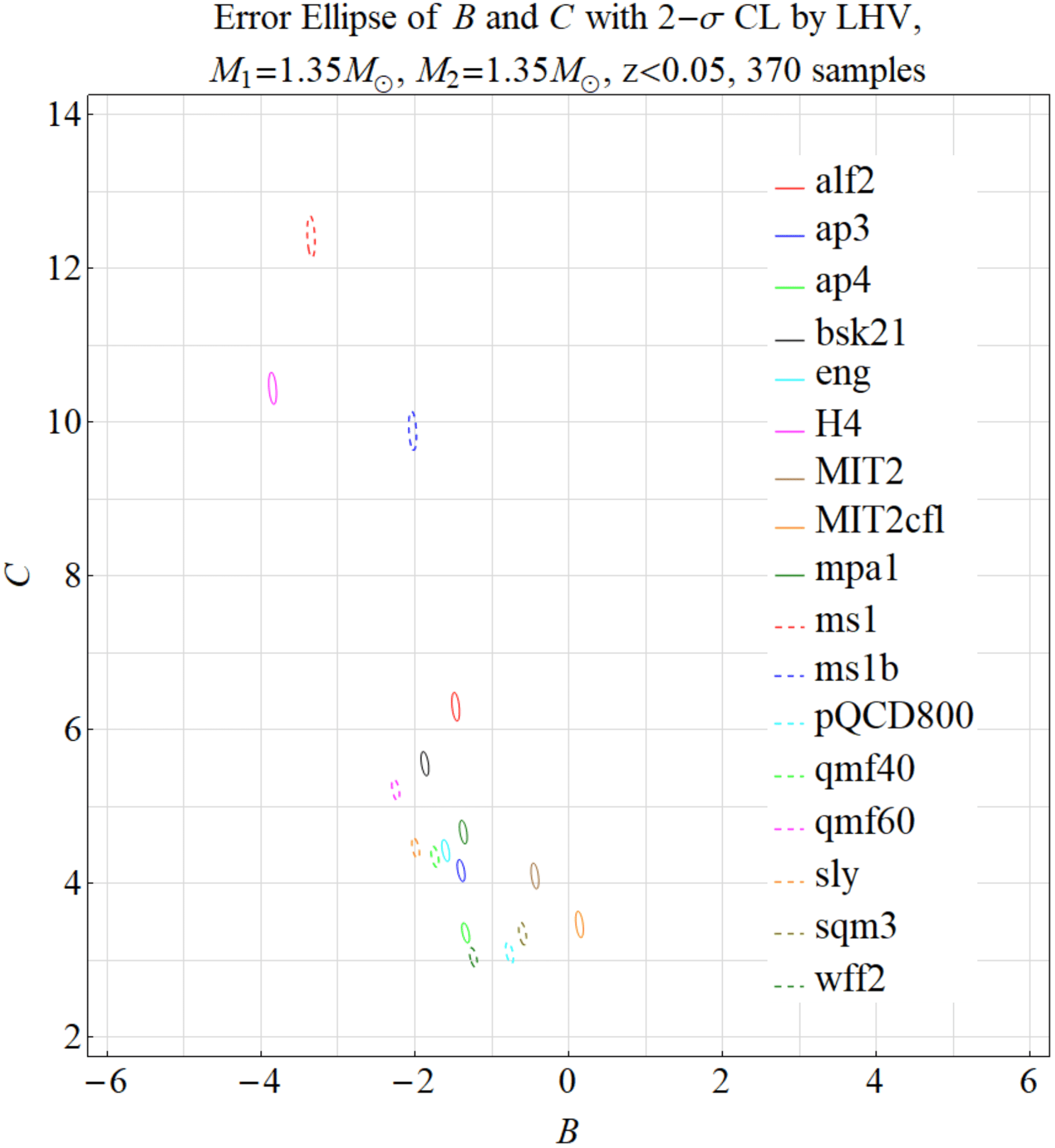}{0.48\textwidth}{(b)}
          }
  \caption{
   The error ellipses of $B$ and $C$  at $2$-$\sigma$ CL for different EOS models  under the detection of LHV.
  (a) is plotted with totally 32 samples
  and  (b) is plotted with totally 370 samples.
  These samples are generated at $z<0.05$,
  and the corresponding SNR distributions are  illustrated in Fig.\ref{SNRnLHV32370}.
  We have assumed the EM counterparts for every GW sample are detectable.
  It is observed that the overlaps between the ellipses as well as the distinguishable EOS models are the same as in Fig.\ref{LHVlowBCError}.  
  }\label{LHV32}
\end{figure*}

\begin{figure*}[htbp]
\gridline{\fig{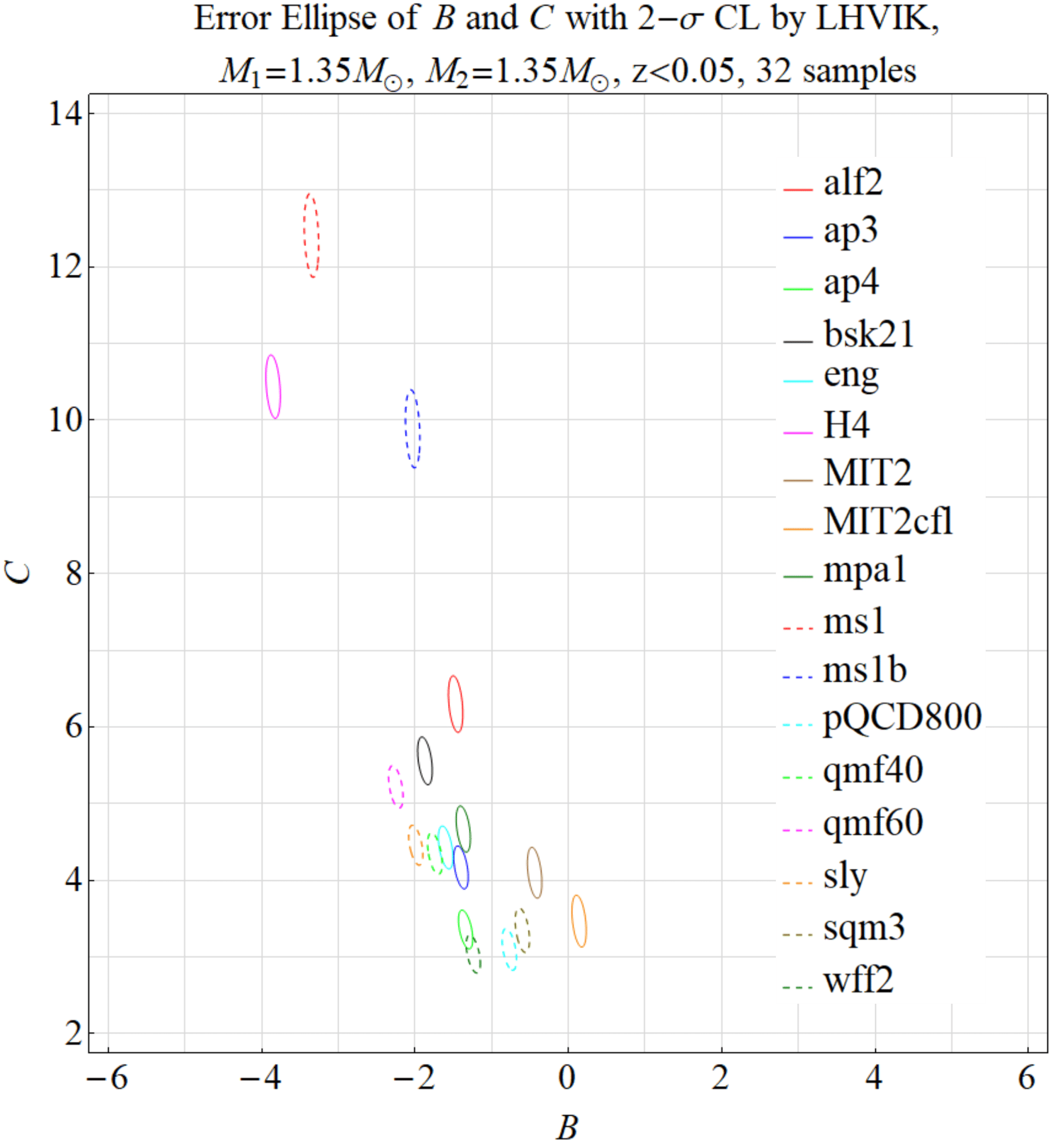}{0.48\textwidth}{(a)}
          \fig{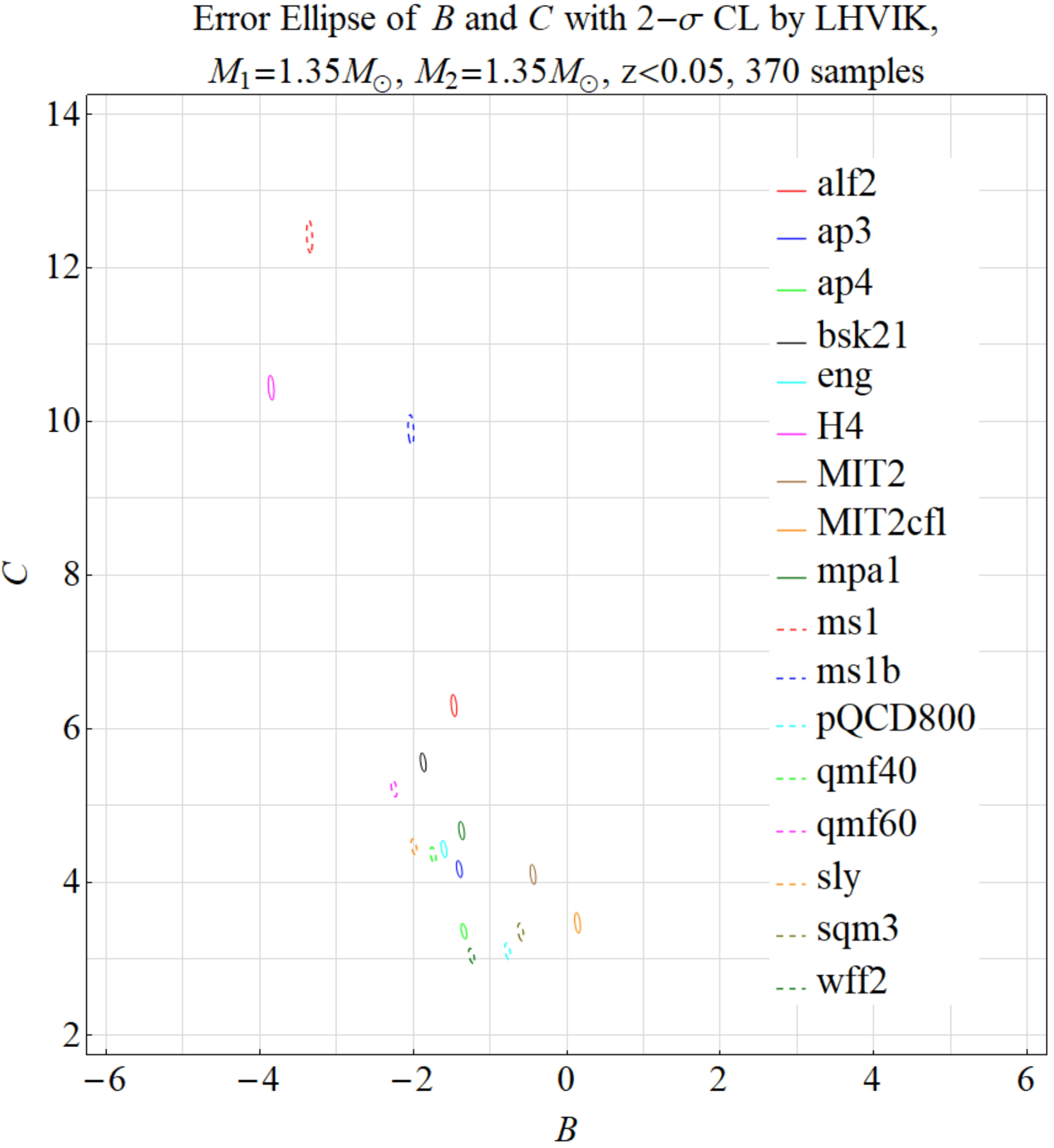}{0.48\textwidth}{(b)}
          }
  \caption{
      The error ellipses of $B$ and $C$  at $2$-$\sigma$ CL for different EOS models  under the detection of LHVIK.
  (a) is plotted with totally 32 samples
  and  (b) is plotted with totally 370 samples.
  These samples are generated at $z<0.05$,
  and the corresponding SNR distributions are  shown in Fig.\ref{SNRnLHVIK32370}.
  We have assumed the EM counterparts for every GW sample are detectable.
  It is observed that the overlaps between the ellipses as well as the distinguishable EOS models are the same as in Fig.\ref{LHVIKlowBCError}.
  }\label{LHVIK32}
\end{figure*}

\begin{figure*}[htbp]
\gridline{\fig{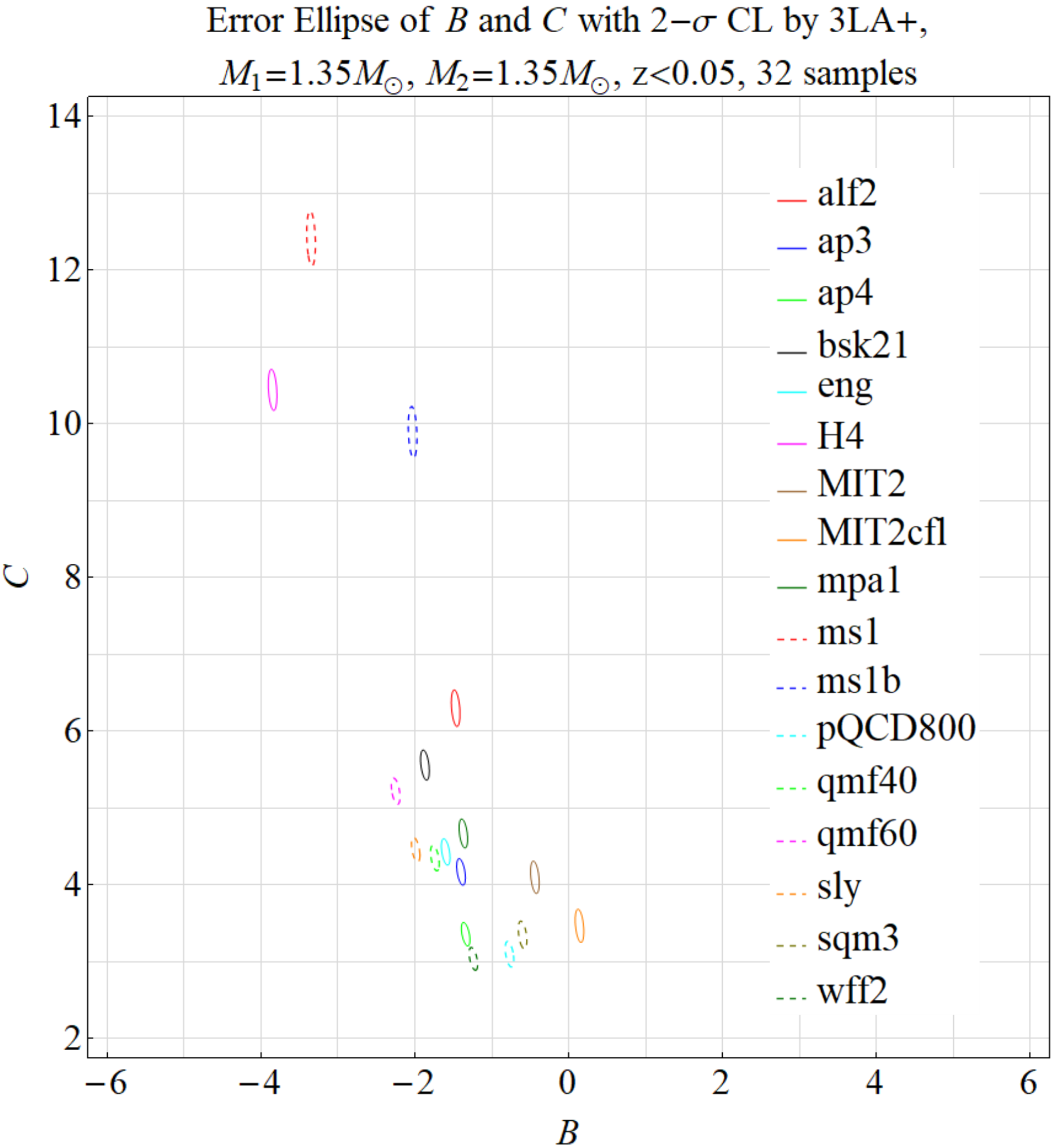}{0.48\textwidth}{(a)}
          \fig{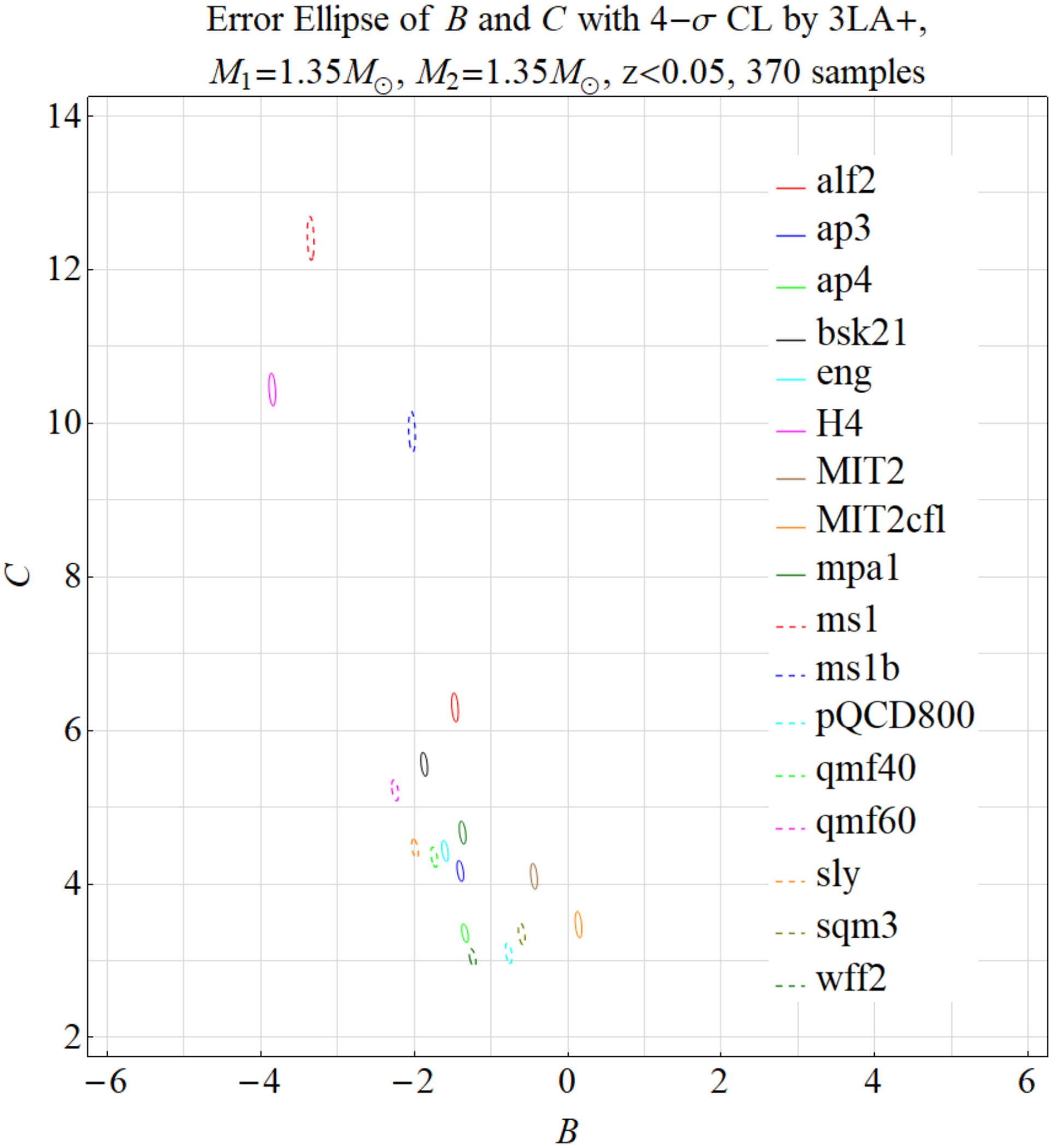}{0.48\textwidth}{(b)}
          }
  \caption{
      The error ellipses of $B$ and $C$ for different EOS models  under the detection of 3LA+.
  (a) is plotted with totally 32 samples  at $2$-$\sigma$ CL 
  and  (b) is plotted with totally 370 samples  at $4$-$\sigma$ CL .
  These samples are generated at $z<0.05$,
  and the corresponding SNR distributions are  shown in Fig.\ref{SNRn3LAplus32370}.
  We have assumed the EM counterparts for every GW sample are detectable.
  It is observed that all the EOS models are distinguishable,
  but the confidence level in (b) of this figure is lower than that in Fig.\ref{LIGOaPlus250} (b).
  }\label{LIGOaPlus32}
\end{figure*}

\begin{figure*}[htbp]
\gridline{\fig{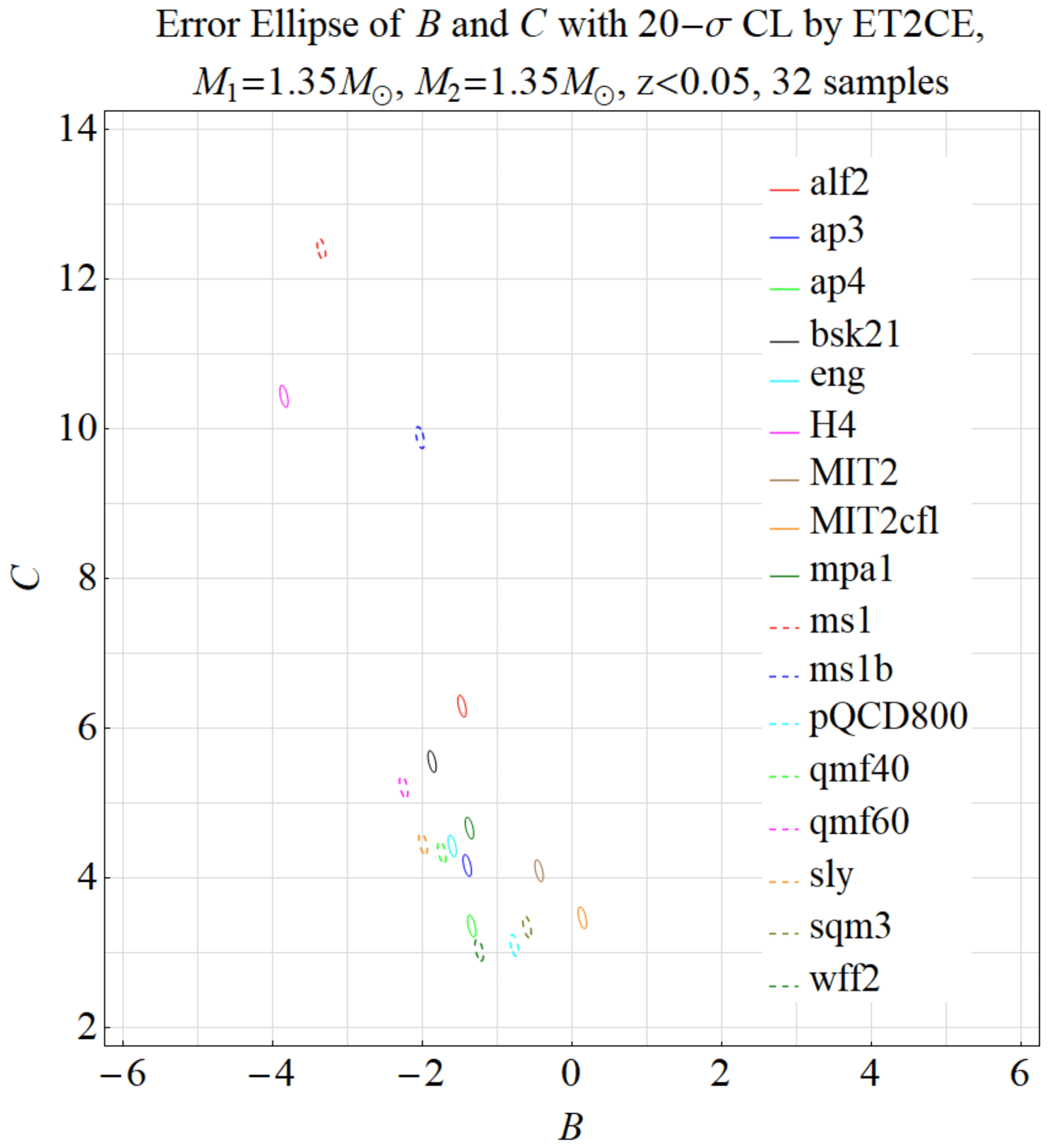}{0.48\textwidth}{(a)}
          \fig{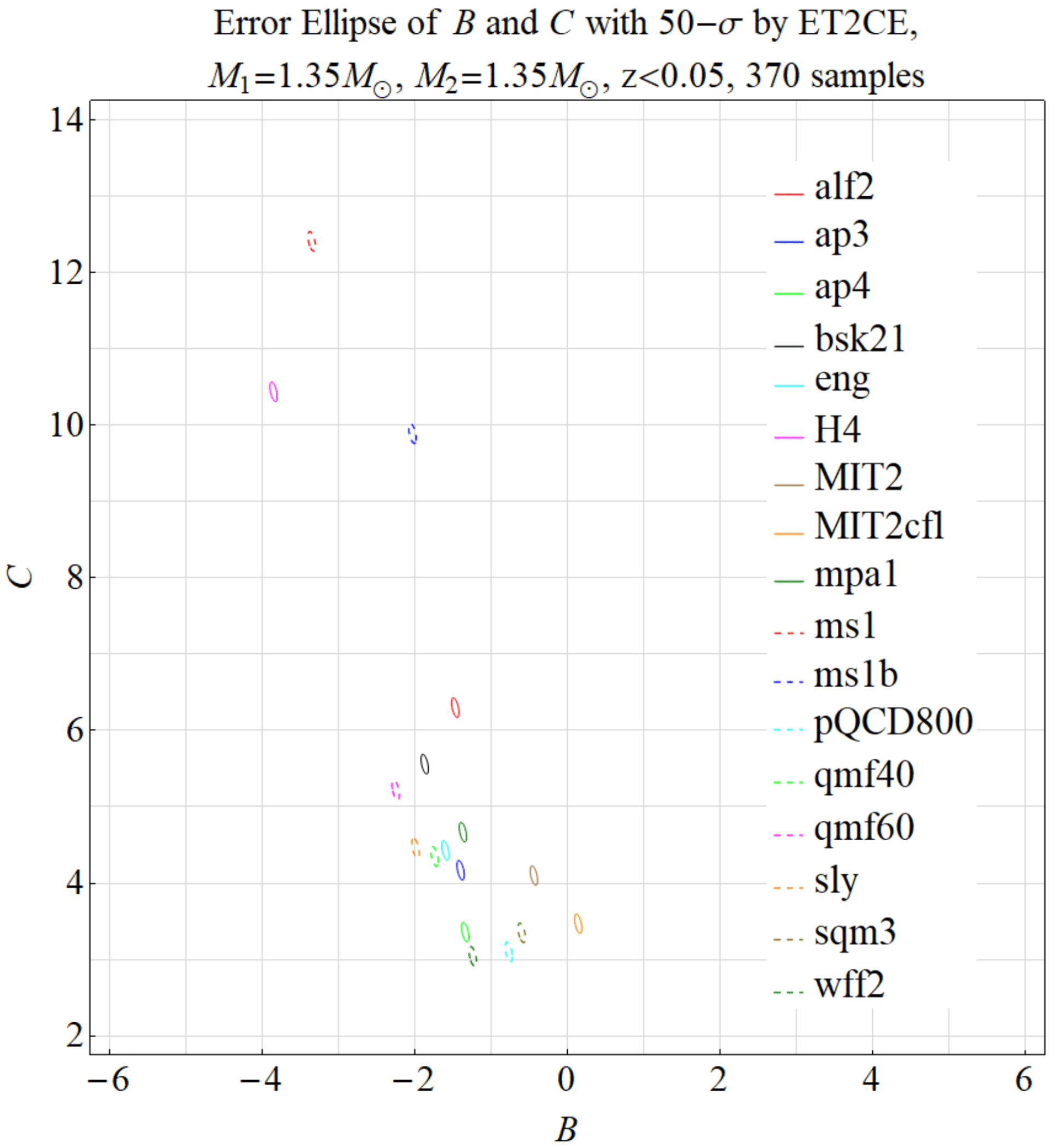}{0.48\textwidth}{(b)}
          }
  \caption{
     The error ellipses of $B$ and $C$ for different EOS models  under the detection of ET2CE.
  (a) is plotted with totally 32 samples  at $20$-$\sigma$ CL 
  and  (b) is plotted with totally 370 samples  at $50$-$\sigma$ CL .
  These samples are generated at $z<0.05$,
  and the corresponding SNR distributions are  shown in Fig.\ref{SNRnET2CE32370}.
  We have assumed the EM counterparts for every GW sample are detectable.
  It is observed that all the EOS models are distinguishable,
  but the confidence levels are lower than that in Fig.\ref{ET2CElowBCError}.
  }\label{ET2CE32}
\end{figure*}

It is noticed that the second BNS-merger GW event GW190425 is at $ z \sim0.03$, and no EM counterpart is reported \citep{GW190425ligo}. 
Thus, as a more conservative consideration, we assume that in the future, the detectable EM counterpart
of the GWs generated by the BNS merger is within the redshift range $ z < 0.05$.
According to Ref.\citep{AbbottAbbottAbbott2019},
in a three-year observation,
the number of BNS merger events will be 32--370.
We repeat a simulation similar to the previous subsections to estimate the error ellipses of the tidal parameters of each EOS model.
The SNR distributions by the LHV, LHVIK, 3LA+ and ET2CE networks are plotted in Fig.\ref{SNRnLHV32370},
Fig.\ref{SNRnLHVIK32370},
Fig.\ref{SNRn3LAplus32370}  and
Fig.\ref{SNRnET2CE32370}, respectively.
Comparing Fig.\ref{SNRnLHV32370}
and Fig.\ref{SNRn},
it can be seen that by LHV,
the rate of the detectable samples
in $z<0.05$ and $z<0.1$
is quite similar.
Comparing Fig.\ref{SNRnLHVIK32370}
and Fig.\ref{SNRnLHVIK},
it can be seen that
by LHVIK,
the number of the detectable samples
in $z<0.05$
is greater than that in $z<0.1$,
the same is true for 3LA+ and ET2CE.
This is because the detection capability of LHV is weaker than LHVIK, 3LA+ or ET2CE,
so the main part of the detectable objects by LHV comes from low redshift.
In particular,
the number of the detectable samples by LHV is almost unchanged
when the redshift is reduced to a lower level.
Using these samples,
the error ellipses of $ B $ and $ C $ 
for different EOS models
by using  LHV, LHVIK, 3LA+ and ET2CE are presented in
Fig.\ref{LHV32}, Fig.\ref{LHVIK32}, Fig.\ref{LIGOaPlus32} and Fig.\ref{ET2CE32}, respectively.
Comparing Fig.\ref{LHV32} and
Fig.\ref{LHVlowBCError},
it is observed that the error ellipses obtained in $z<0.05$ and $z<0.1$ are similar in size,
and the distinguishable EOS models for the two redshift ranges are the same.
The error ellipses in Fig.\ref{LHVIK32} by LHVIK
are smaller than in Fig.\ref{LHVIKlowBCError},
but not significantly,
which leads to the same distinguishable EOS models in these two figures.
Due to the great sensitivity of ET2CE,
in $z<0.05$
all the EOS models are distinguishable by ET2CE
at $20$-$\sigma$ level with totally 32 samples,
and  at $50$-$\sigma$ level with totally 370 samples.
The result for 3LA+ is similar to ET2CE,
that 
using the network of 3LA+,
all the EOS models are distinguishable at $2$-$\sigma$ level
 with totally 32 samples
as well as at $4$-$\sigma$ level with totally 370 samples.
Therefore, we can conclude that if the redshift range of the detectable EM counterparts  is reduced to a smaller but reasonable value,
the outcome of the EOS determination still stands.

\subsection{No EM counterpart}
\label{subsec:noEMcounterpart}

For comparison, we also consider the case where there is no detectable EM counterpart for each GW signal
in a low-redshift range.
Under this assumption,
the position parameters $(\theta_s,\phi_s)$ and redshift $z$ of the source remain unknown.
Subsequently,
for each GW event,
we need to calculate the Fisher matrix of all 12 parameters,
$(\mathcal M_c,$ $\eta,$ $t_c,$ $\psi_c,$ $\iota,$ $\theta_s,$ $\phi_s,$
$\psi_s,$ $d_{\rm L},$ $z,$ $B,$ $C)$,
and then marginalize it into a Fisher matrix of two tidal-effect parameters $(B, C)$
as the processing adopted in Sec. \ref{subsec:LHV}.
Then, we sum the Fisher matrices of $(B,C)$
over all detectable GW samples
to obtain a total Fisher matrix
whose inverse is the covariance matrix of $(B,C)$.

\begin{figure*}[htbp]
\gridline{\fig{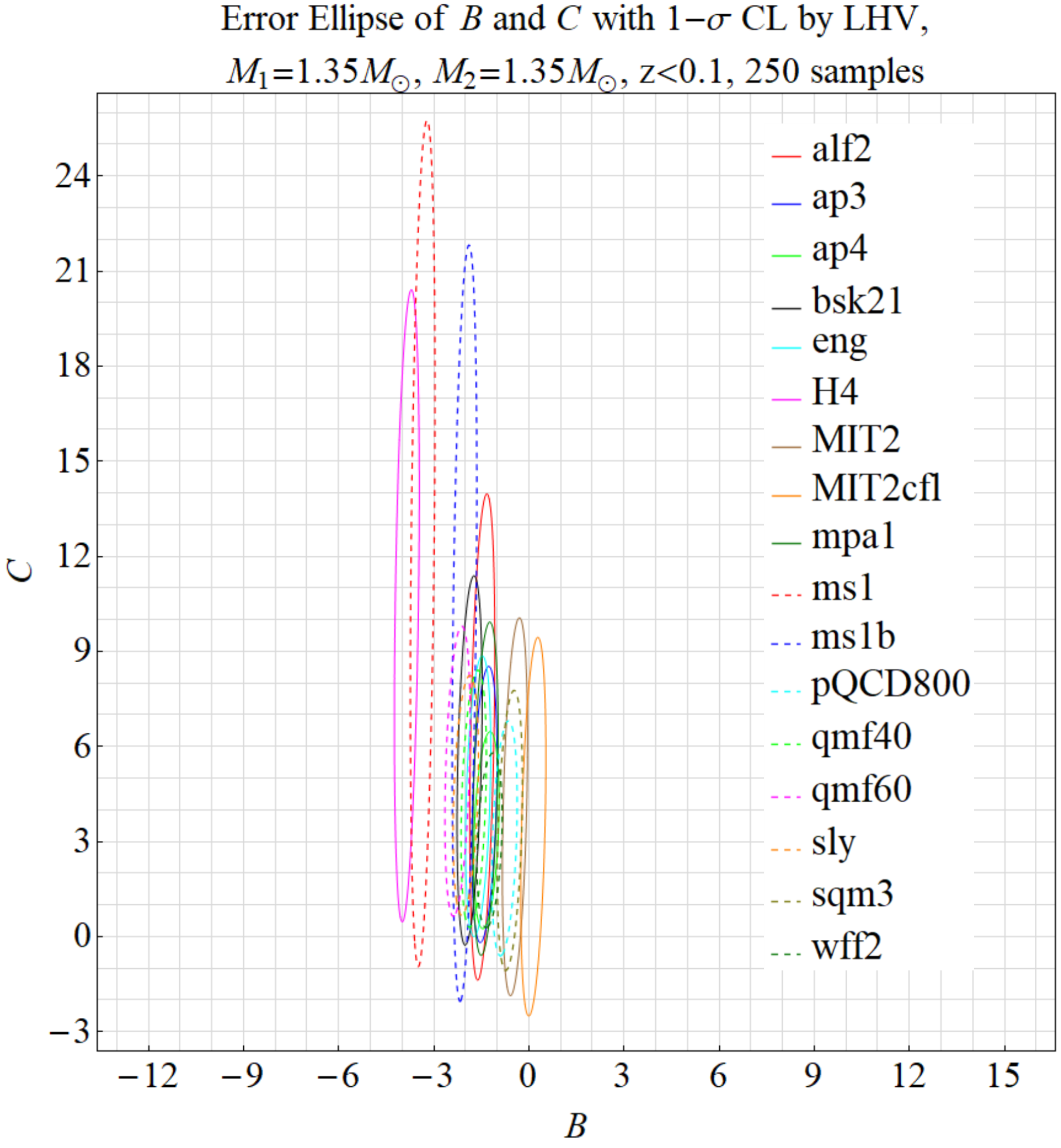}{0.48\textwidth}{(a)}
          \fig{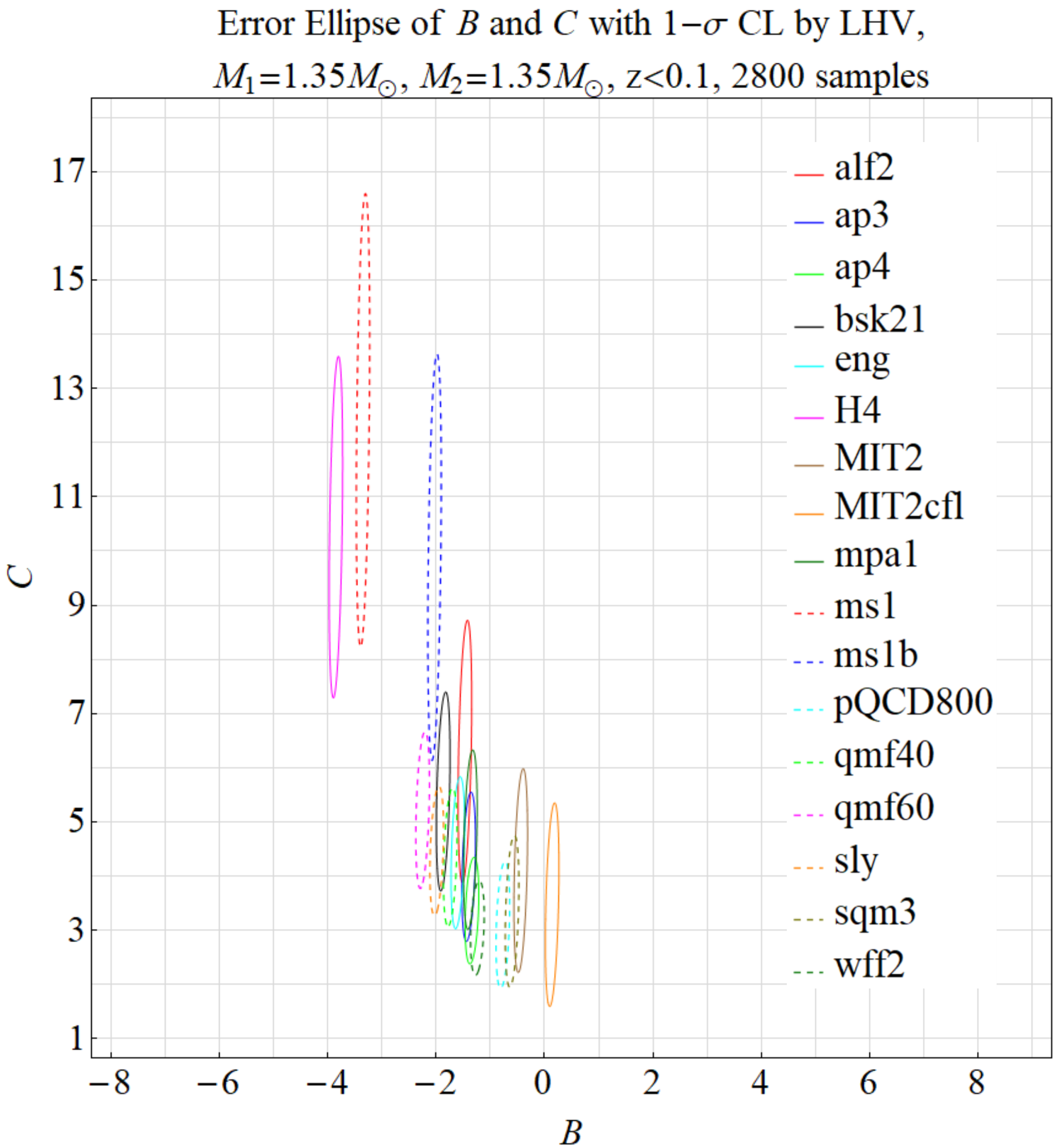}{0.48\textwidth}{(b)}
          }
  \caption{
    The error ellipses of $B$ and $C$ for different EOS models at $1$-$\sigma$ CL  under the detection of LHV.
  (a) is plotted with totally 250 samples 
  and  (b) is plotted with totally 2800 samples,
  which  are generated at $z<0.1$ and correspond to the lowest and highest merger rates of BNS
  \citep{AbbottAbbottAbbott2019}.
  The corresponding SNR distributions are  the same as that in Fig.\ref{SNRn}.
  We have assumed the EM counterparts cannot be detected in these two figures.
  It is observed that these EOS models are indistinguishable in (a),
  and there are only the models of ms1, H4, ms1b, qmf40, qmf60 can be distinguished in (b).
  }\label{LHVno250}
\end{figure*}

\begin{figure*}[htbp]
\gridline{\fig{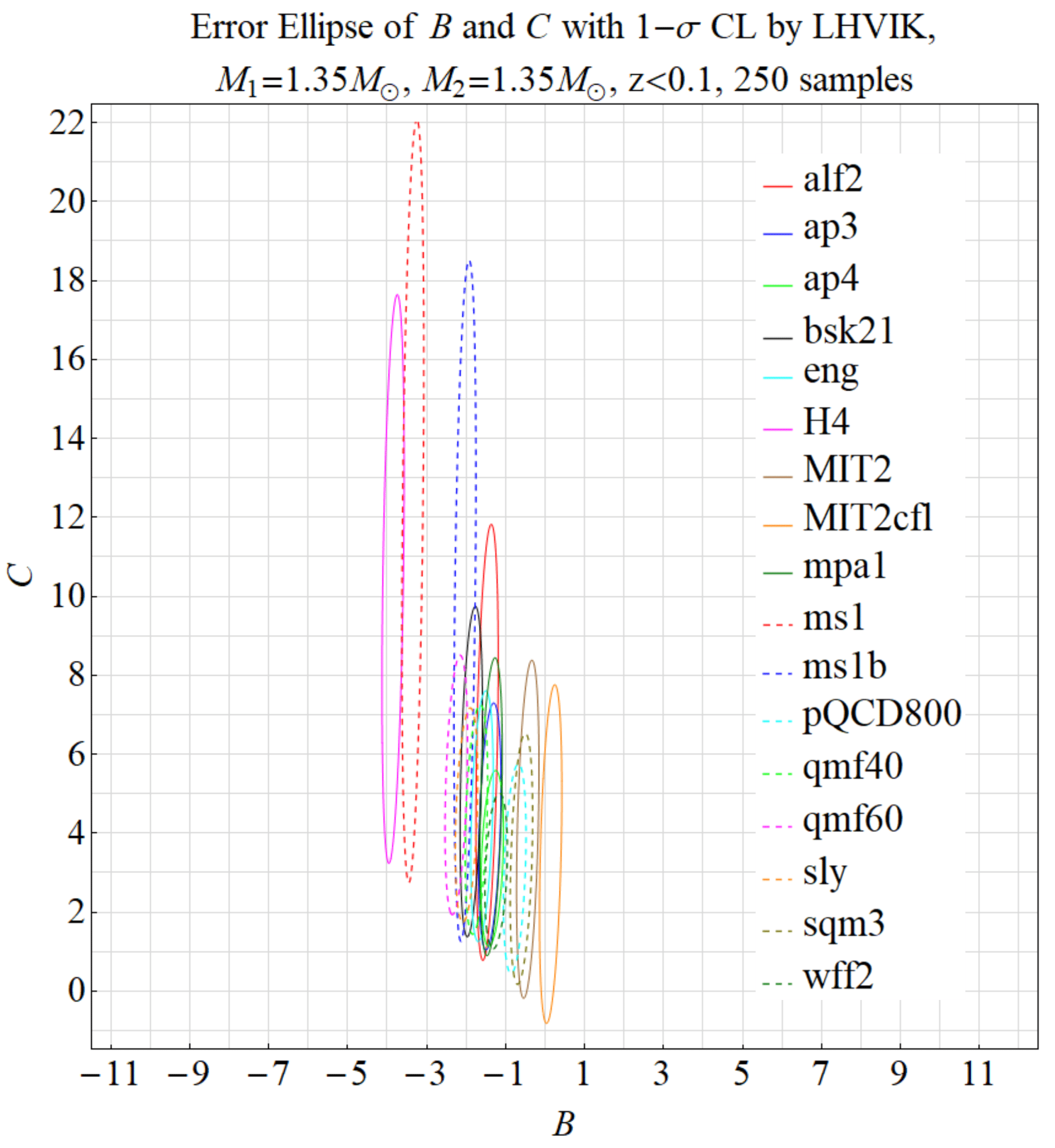}{0.48\textwidth}{(a)}
          \fig{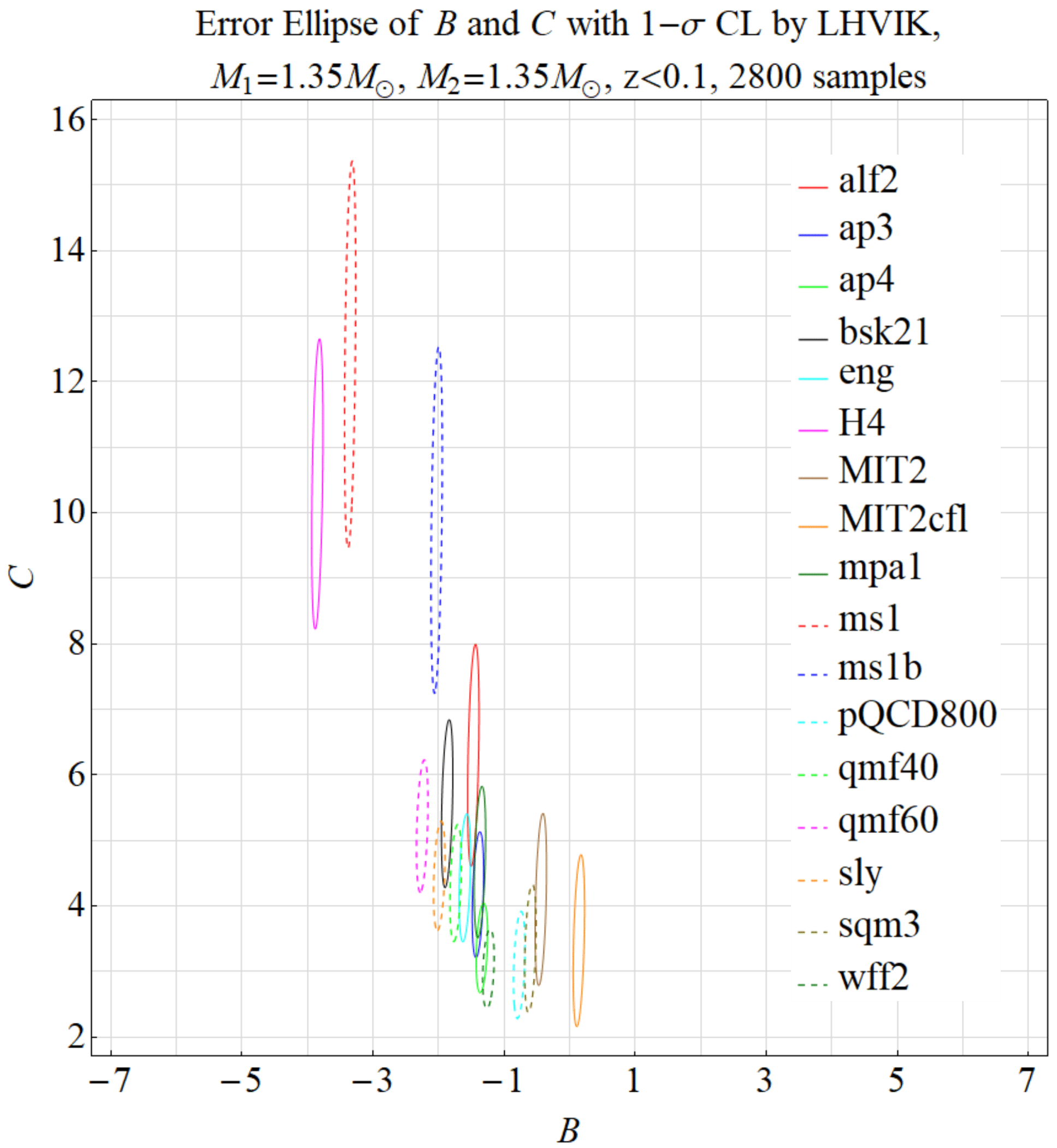}{0.48\textwidth}{(b)}
          }
  \caption{
  The error ellipses of $B$ and $C$ for different EOS models  at $1$-$\sigma$ CL  under the detection of LHVIK.
  (a) is plotted with totally 250 samples 
  and  (b) is plotted with totally 2800 samples.
  The samples  are generated at $z<0.1$, and the corresponding SNR distributions are  the same as that in Fig.\ref{SNRnLHVIK}.
  We have assumed the EM counterparts cannot be detected in these two figures.
    It is observed that MIT2cfl is distinguishable in (a),
  and though the error ellipses in (b) are smaller than that in Fig.\ref{LHVno250} (b), the distinguishable models in (b) are the same as Fig.\ref{LHVno250} (b).
  }\label{LHVIKno250}
\end{figure*}

\begin{figure*}[htbp]
\gridline{\fig{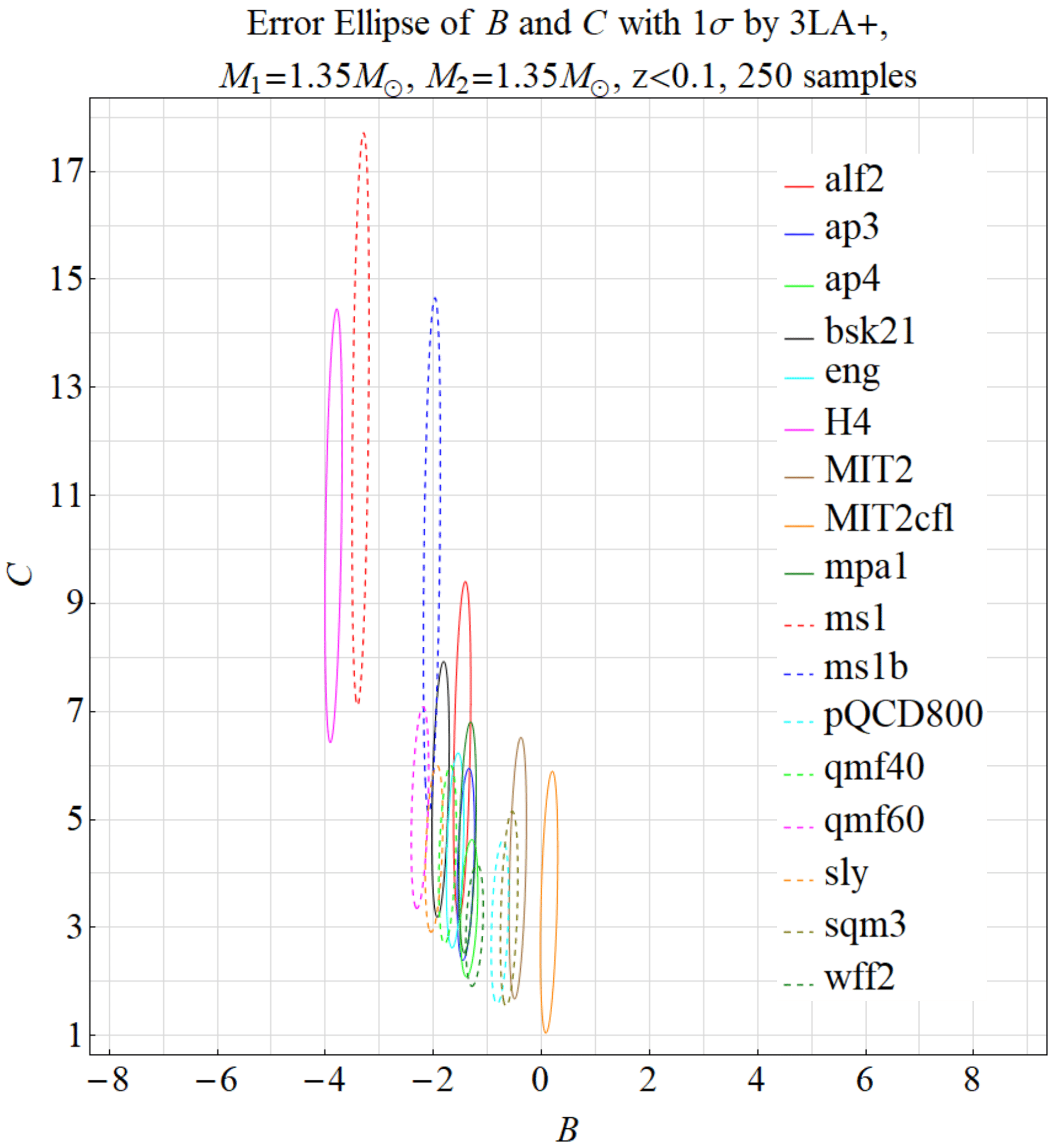}{0.48\textwidth}{(a)}
          \fig{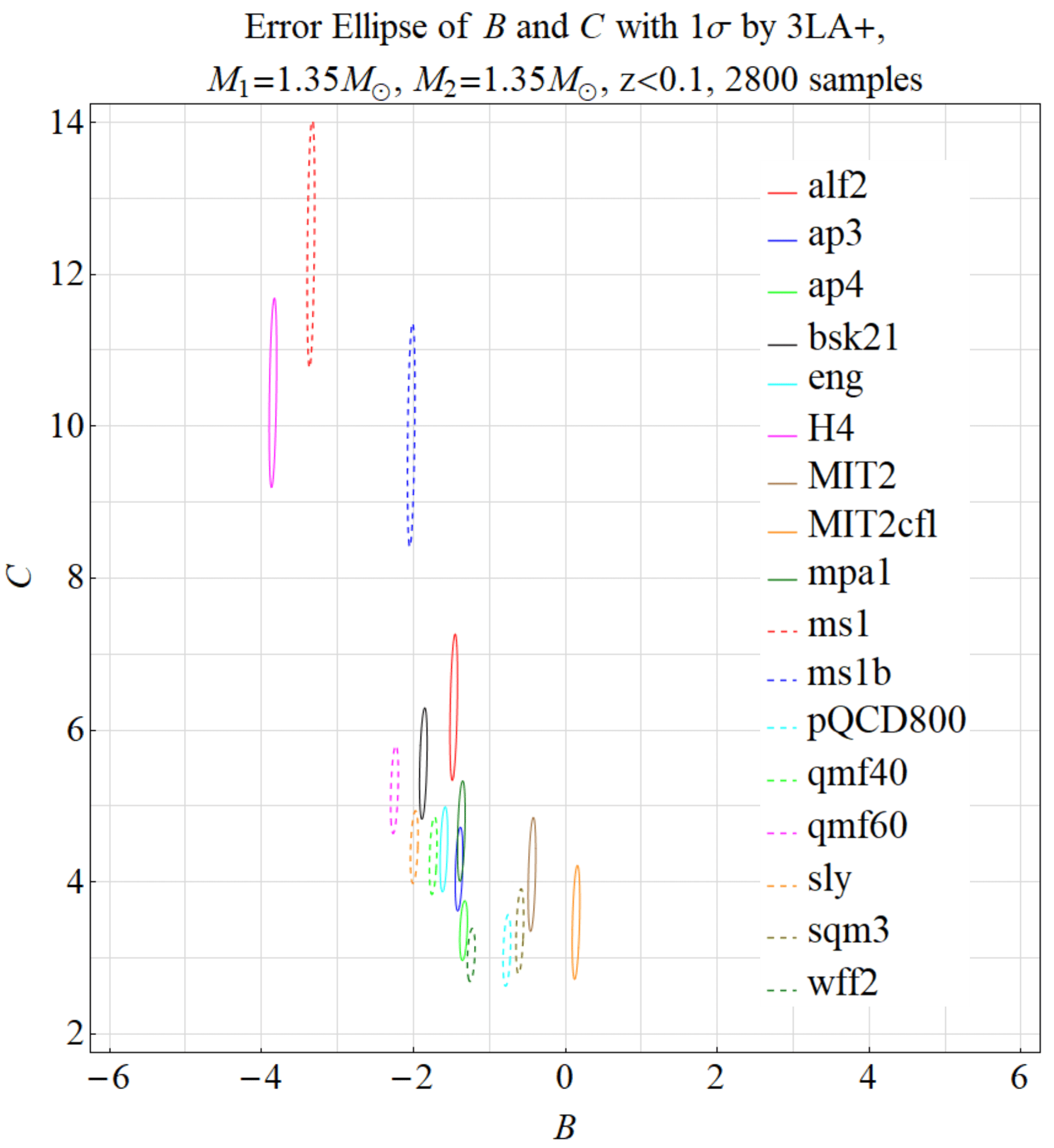}{0.48\textwidth}{(b)}
          }
  \caption{
 The error ellipses of $B$ and $C$ for different EOS models  at $1$-$\sigma$ CL  under the detection of 3LA+.
  (a) is plotted with totally 250 samples 
  and  (b) is plotted with totally 2800 samples.
  The samples  are generated at $z<0.1$, and the corresponding SNR distributions are  the same as that in Fig.\ref{SNRnLIGOaPlus250}.
  We have assumed the EM counterparts cannot be detected in these two figures.
    It is observed that ms1, H4, MIT2cfl can be distinguished in (a),
  and except for mpa1 and ap3,
the other EOS models can be distinguished in (b).
  }\label{3LAplusNo250}
\end{figure*}

\begin{figure*}[htbp]
\gridline{\fig{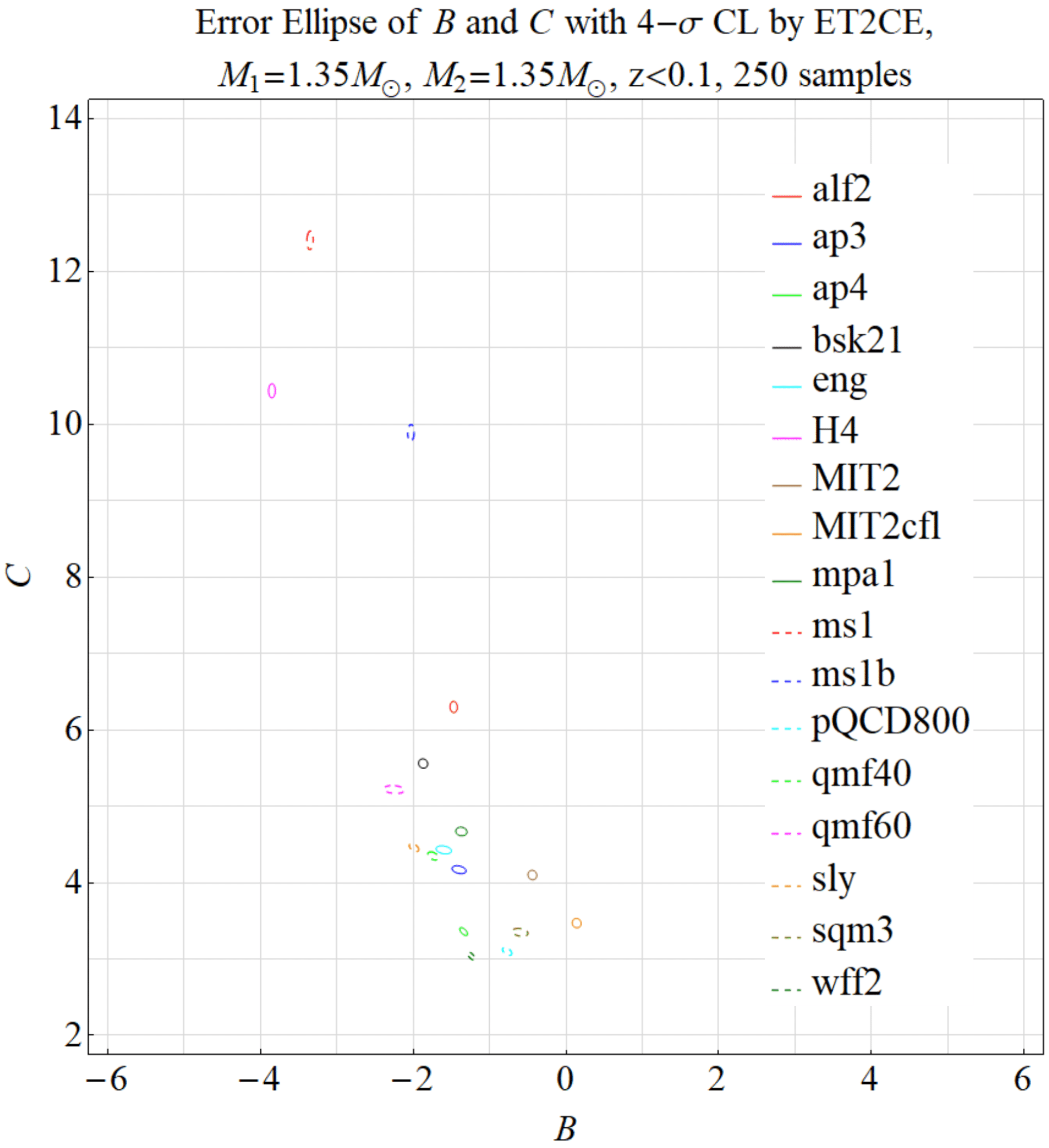}{0.48\textwidth}{(a)}
          \fig{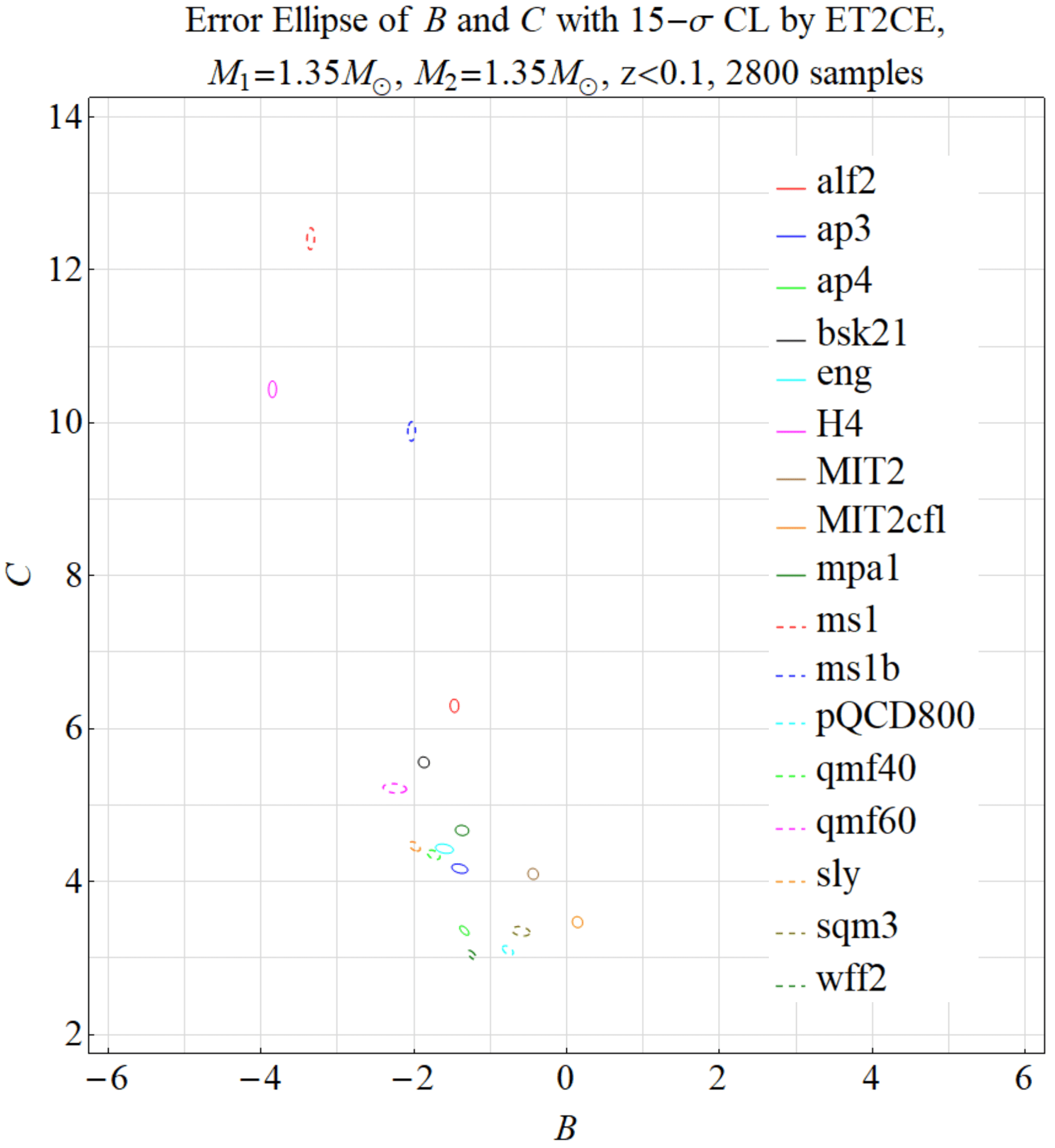}{0.48\textwidth}{(b)}
          }
  \caption{
   The error ellipses of $B$ and $C$ for different EOS models   under the detection of ET2CE.
  (a) is plotted  at $4$-$\sigma$ CL with totally 250 samples 
  and  (b) is plotted  at $15$-$\sigma$ CL with totally 2800 samples.
  The samples  are generated at $z<0.1$, and the corresponding SNR distributions are  the same as that in Fig.\ref{SNRnET2CE}.
  We have assumed the EM counterparts cannot be detected in these two figures.
    It is observed that all the EOS models are distinguishable.
    However, the confidence levels are smaller than that in Fig.\ref{ET2CElowBCError}.
  }\label{ET2CEno250}
\end{figure*}

In this subsection,
we still consider the redshift range of $z<0.1$,
and the number of BNS-merger events for three-year observation ranges from 250 to 2800.
With these setups,
we repeat a simulation similar to Sec.\ref{subsec:LHV}, \ref{subsec:LHVIK}, \ref{subsec:LIGOA+} and \ref{subsec:ET2CE}.
The distributions for SNR  by using LHV, LHVIK, 3LA+ and ET2CE
are given in Fig.\ref{SNRn}, Fig.\ref{SNRnLHVIK},
 Fig.\ref{SNRnLIGOaPlus250},
and Fig.\ref{SNRnET2CE}, respectively.
The corresponding error ellipses of $B$ and $C$
 by using LHV, LHVIK, 3LA+ and ET2CE are plotted in
Fig.\ref{LHVno250},
Fig.\ref{LHVIKno250},
Fig.\ref{3LAplusNo250} and Fig.\ref{ET2CEno250},
respectively.
If there is no electromagnetic counterpart,
in the case of a lower merge rate  with 250 samples, 
when using LHV,all the EOS models are indistinguishable
at 1-$ \sigma $ CL;
but when using LHVIK,
MIT2cfl can be distinguished from other models.
When taking a higher merge rate with 2800 samples, the EOS models 
 ms1, H4, ms1b, qmf40, qmf60  can be distinguished by LHV or LHVIK at $ 1$-$\sigma $ CL,
other models are indistinguishable.
By 3LA+,
at 1-$\sigma$ CL,
with totally 250 samples,
the EOS models 
ms1, H4, MIT2cfl can be distinguished
but the others are indistinguishable.
While with totally 2800 samples,
except for mpa1 and ap3,
{which have similar prediction of radii and $\lambda-m$ slopes for neutron stars,}
other EOS models can be distinguished.
By ET2CE,
all the EOS models are distinguishable,
which is the same as the conclusion of previous subsections.
Comparing these figures with
Fig.\ref{LHVlowBCError},  Fig.\ref{LHVIKlowBCError}
,  Fig.\ref{LIGOaPlus250}
and Fig.\ref{ET2CElowBCError},
it is observed that the errors of $B$ and $C$ are
 much larger than those with detectable EM counterpart.
Thus, EM counterparts are important for the determination of EOS,
especially for the networks of 2G and 2.5G detectors.


\section{Determination of dark energy}
\label{sec:DEdetermination}

Dark energy with positive density but negative pressure
is believed to be an impetus of the
accelerated expansion of the universe
\citep{Albrecht0609591,Amendola2010},
and a large number of dark energy models have been proposed in the literature
\citep{FriemanTurnerHuterer2008}. In order to differentiate these different models,
it is essential to measure the EOS of dark energy.
Currently,
the main methods   include observations of SNIa, CMB,
large-scale structure,
weak gravitational lensing, and so on \citep{Albrecht0609591,Amendola2010},
which are based on the
observations of various electromagnetic waves.
Future GW events of the compact binary coalescences can act as the standard sirens
to explore the physics of dark energy. This issue has been widely discussed in the previous works
\citep{SathyaprakashSchutzBroeck2010,
ZhaoBroeckBaskaran2011,cairg,
ZhaoWen2018},
under the assumption that
for each detectable GW event in the BNS merger,
an EM counterpart can be found,
thereby the position
and the redshift of the source are fixed.
However, the detection ability of this method is limited by the number of available sources, since for most high-redshift ({\it e.g.} $z> 0.1$) events,
it is unlikely that the EM counterparts will be observed.
In this section, we analyze the GW waveforms containing the tidal effects,
from which both redshift and luminosity distance can be obtained.
Furthermore, the evolution of dark matter can be studied.
For the EOS model of BNS,
we will adopt the $ B $ and $ C $
obtained from GW and the corresponding EM counterpart observations
at the low-redshift as a prior condition,
and add it to the process of analyzing the dark energy EOS with only GW observation in the high-redshift range.

In order to quantify the EOS of dark energy, in this article, we adopt the widely-used  Chevallier-Linder-Polarski
parametrization \citep{ChevallierLinder2001,ChevallierLinder2003,Albrecht0609591},
in which
a parameter $w(z)$ is introduced as
the ratio of the pressure $p_{\rm de}$
and energy density $\rho_{\rm de}$
of dark energy as
\be\label{CPL}
w(z)=\frac{p_{\rm de}}{\rho_{\rm de}}
=w_0+w_a\frac{z}{1+z},
\ee
where $w_0$ and $w_a$ are two constants.
$w_0$ represents the a present-day value of EOS and
$w_a$ represents its evolution with redshift.
In the $\Lambda$CDM model,
a cosmological constant
$\Lambda$,
corresponds to $w_0=-1$ and $w_a=0$.
In a general Friedmann-Lemaitre-Robertson-Walker (FLRW) universe,
the luminosity distance of astrophysical sources is a function of redshift $z$ as \citep{Weinberg2008}
\begin{equation}
d_{\rm L}(z)=(1+z)\left\{\begin{array}{ll}
{|k|^{-1 / 2} \sin \left[|k|^{1 / 2} \int_{0}^{z} \frac{d z^{\prime}}{H\left(z^{\prime}\right)}\right]} & {\left(\Omega_{k}<0\right)} \\
{\int_{0}^{z} \frac{d z^{\prime}}{H\left(z^{\prime}\right)}} & {\left(\Omega_{k}=0\right)} \\
{|k|^{-1 / 2} \sinh \left[|k|^{1 / 2} \int_{0}^{z} \frac{d z^{\prime}}{H\left(z^{\prime}\right)}\right]} & {\left(\Omega_{k}>0\right)}
\end{array}\right.
\end{equation}
where $\Omega_k$ is the contribution of spatial curvature density,
and the Hubble parameter $H(z)$ is
\be
H(z)=H_0\l[
\Omega_m(1+z)^3
+\Omega_k(1+z)^2
+(1-\Omega_m)E(z)\r]^{1/2}
,
\ee
with
\be
E(z)=(1+z)^{3(1+w_0+w_a)}e^{-3w_a z/(1+z)}.
\ee
There are five cosmological parameters
to be determined in the above $d_{\rm L}$-$z$ relation, i.e.,
($H_0$,$\Omega_m$, $\Omega_k$,$w_0$,$w_a$).
However,
as Ref.\citep{ZhaoBroeckBaskaran2011} has shown,
the background parameters
$(H_0,\Omega_m,\Omega_k)$
and the dark energy parameters $(w_0,w_a)$
have strong degeneracy,
thus one cannot constrain the full parameters from GW standard sirens alone.
The SNIa and BAO methods for dark energy detection also suffer the same problem.
A general way to break this degeneracy is to
take the CMB constraints for
the parameters $(H_0,\Omega_m,\Omega_k)$ as a prior,
which is nearly equivalent to treat the three parameters
as known in data analysis \citep{ZhaoBroeckBaskaran2011,ZhaoWen2018}.
Therefore,
in the following,
we only study the constraints on parameters $(w_0,w_a)$ under GW observation.
The Fisher matrix of $(w_0,w_a)$
can be obtained by converting the measurement error of $d_{\rm L}$ as following
\be\label{FisherDMpara}
F_{ij}
=\sum_{k}\frac{(\partial{\rm ln}d_{\rm L}(k)/\partial p_i)
    (\partial{\rm ln}d_{\rm L}(k)/\partial p_j)}
    {(\Delta d_{\rm L}/d_{\rm L}(k))^2
    +\left(\Delta d_{\rm L}^{\,\prime} / d_{\rm L}(k)\right)^{2}
    +\sigma_{l}^{2}},
\ee
where both the indices $i$ and $j$ run from 1 to 2,
and $(p_1, p_2)$
indicate two free parameters  $(w_0, w_a)$.
The index $k=1,2,\cdot\cdot\cdot,N$
 labels the event with redshift and position at
$(z_k, \theta_{sk}, \psi_{sk})$
in the total $ N $ sources.
$ \sigma_ {l} = 0.05 z $ is the contribution of the weak-lensing effect to the error of $ d_{\rm L} $ \citep{SathyaprakashSchutzBroeck2010,ZhaoBroeckBaskaran2011},
$\Delta d_{\rm L}^{\,\prime}$ is induced from
the redshift error $\Delta z$ as
\be\label{Deltadlz'}
\Delta d_{\rm L}^{\,\prime}=\frac{\partial d_{\rm L}(z)}{\partial z}\Delta z.
\ee
In (\ref{FisherDMpara}) and (\ref{Deltadlz'}),
$\Delta d_{\rm L}$ and $\Delta z$ are
the $1$-$\sigma$ errors.
To get $\Delta d_{\rm L}$ and $\Delta z$,
we first calculate
a  $12\times12$ Fisher matrix from the $k$-th GW sample
of the full 12 parameters
$(\mathcal M_c,$ $\eta,$ $t_c,$ $\psi_c,$ $\iota,$ $\theta_s,$ $\phi_s,$
$\psi_s,$ $d_{\rm L},$ $z,$ $B,$ $C)$
in the redshift range of $0.1<z<2$.
Then, we add the Fisher matrix of $ B $ and $ C $ obtained
from the low-redshift GW and corresponding EM counterparts
to the $B$- and $C$-related elements
in this $ 12 \times12 $ Fisher matrix as a prior \citep{Coe2009}.
$\Delta d_L$ and $\Delta z$ are the square root
of the $(d_{\rm L},d_{\rm L})$- and $(z,z)$-diagonal elements
of the inverse of this new $12\times12$ matrix.

There is also another quantity,
the figure of merit (FoM) \citep{Albrecht0609591},
indicating the goodness of constraints from
observational data sets,
which is proportional to the inverse area of the error ellipse in the $w_0$-$w_a$ plane as following
\be
{\rm FoM}=\l[
{\rm Det}\,C(w_0,w_a)
\r]^{-1/2},
\ee
where $C(w_0,w_a)$ is the covariance matrix of $w_0$ and $w_a$.
Larger FoM means stronger constraint on the parameters, since it corresponds to a smaller error ellipse.

By adopting the merger rates in Ref.\citep{AbbottAbbottAbbott2019},
there will be
$2.1\times10^5\sim 7.2\times10^6$
merger events for three-year observation.
And when the number of events is large,
the sum of events in Eq.(\ref{FisherDMpara})
can be replaced by the following integral \citep{ZhaoBroeckBaskaran2011}
\be
F_{ij}=N\times\int_0^{z_{\rm max}}
\frac{\partial\ln d_{\rm L}(z)}{\partial p_i}
\frac{\partial\ln d_{\rm L}(z)}{\partial p_j}
f(z)
\l\langle
\frac{1}{{(\Delta d_{\rm L}/d_{\rm L}(k))^2
    +\left(\Delta d_{\rm L}^{\,\prime} / d_{\rm L}(k)\right)^{2}
    +\sigma_{l}^{2}}}
\r\rangle
dz,
\ee
where $f(z)$ is the number distribution of the GW sources over redshift $z$,
and $\langle\cdot\cdot\cdot\rangle$ is the average over the
angles $(\theta_s,$ $\phi_s,$
$\iota,$ $\psi_s)$.
Thus, the $1$-$\sigma$ errors of $\Delta w_0$
and $\Delta w_a$,
and FoM are related to a large  number $N$ of the total events
as following
\be\label{Nrelation}
\Delta w_0=\frac{A_{w_0}}{\sqrt N},
~~~~~~
\Delta w_a=\frac{A_{w_0}}{\sqrt N},
~~~~~~
{\rm FoM}=A_{\rm FoM}N,
\ee
where the values of the coefficients $A_{w_0}$, $A_{w_a}$
and $A_{\rm FoM}$ depend on the type
of detector network.
Eq.(\ref{Nrelation}) makes it easy to convert the estimation errors for one large number of events  to another.

\begin{figure*}[htbp]
\gridline{\fig{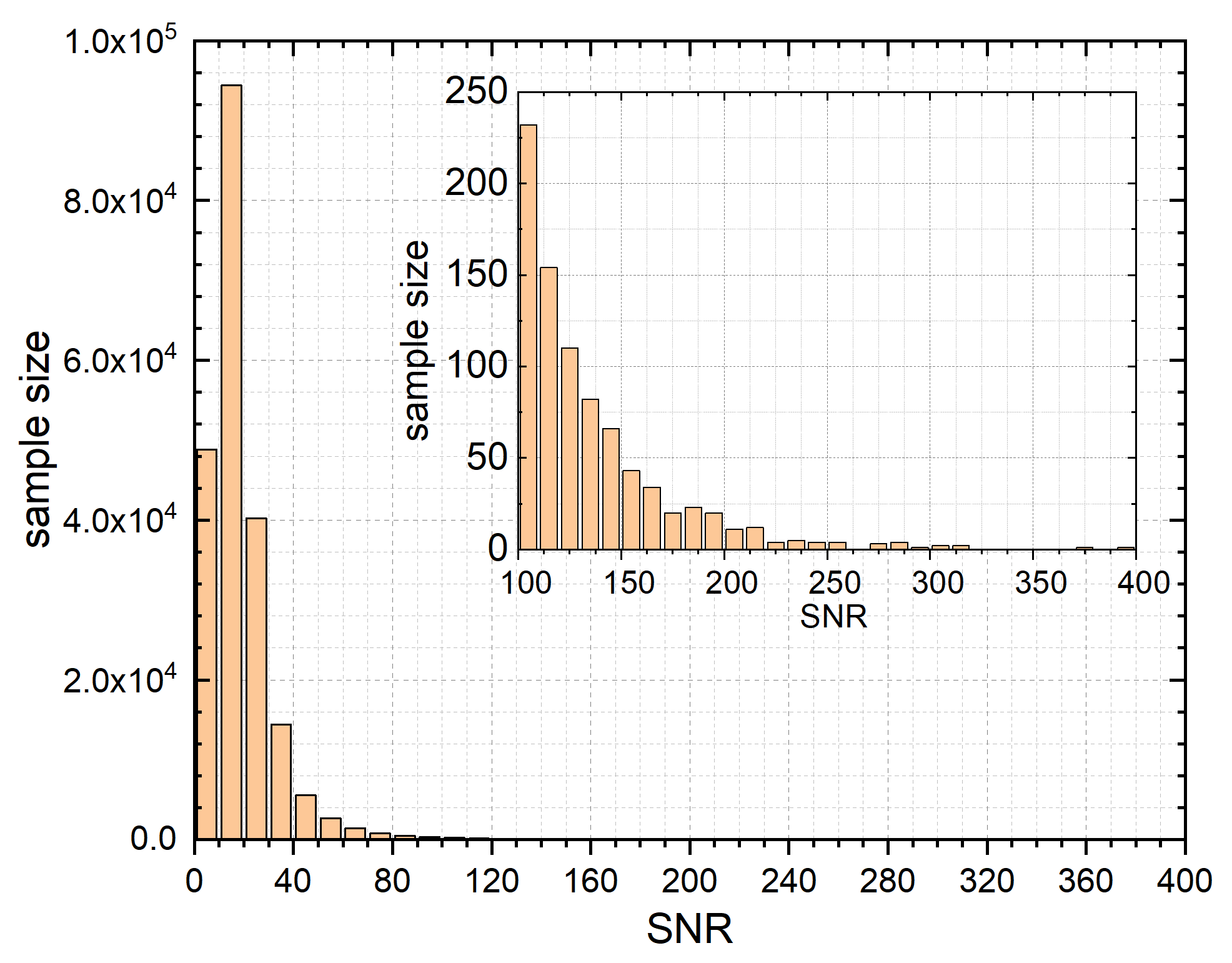}{0.48\textwidth}{(a)}
          \fig{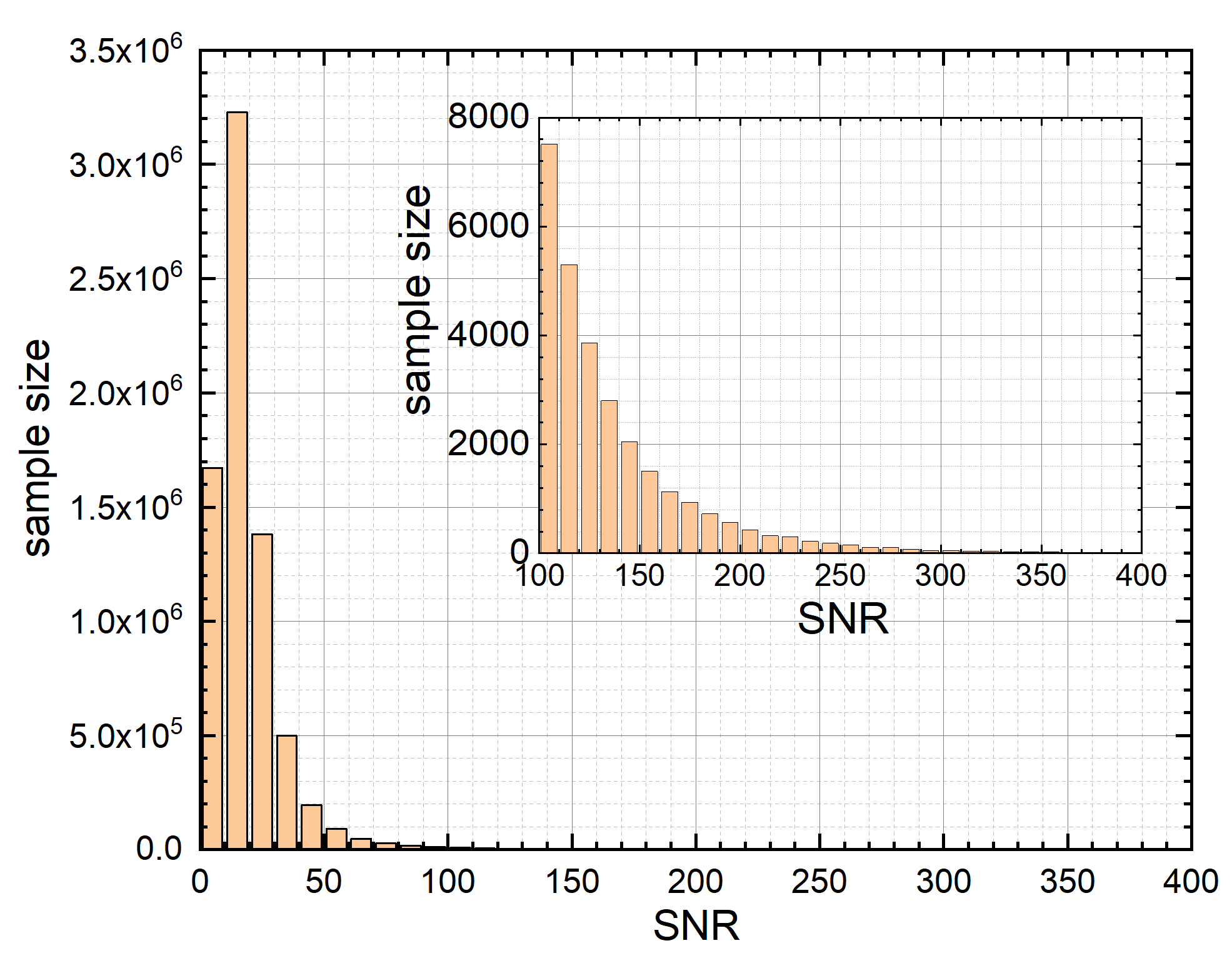}{0.48\textwidth}{(b)}
          }
\caption{The SNR distribution of the samples
generated in the redshift range of
$0.1<z<2$ detected by ET2CE.
According to the lowest and highest event rates given in Ref.\citep{AbbottAbbottAbbott2019},
we generate a total of $2.1\times10^5$  
and $7.2\times10^6$ samples in (a) and (b) respectively.
With the threshold of  SNR$>10$,
$1.6\times10^5$  samples are detectable in (a),
and $5.5\times10^6$ samples are detectable in (b).
The detectable samples account for  $\sim76\%$ of the total samples in both (a) and (b).    }
     \label{SNRhighz}
\end{figure*}

Since only in the 3G era, a larger number of high-redshift BNSs are expected to be detected by GW observations, in this article, we will focus on the 3G detector network for the determination of dark energy. We perform a simulation to estimate the errors of
$w_0$ and $w_a$ by the ET2CE network.
Following Sec.\ref{subsec:ET2CE},
we also take the assumption that
the EM counterparts are detectable only in $z<0.1$.
Thus,
at the lowest merger rate,
we generate $2.1\times10^5$ GW samples
within the redshift range of $0.1<z<2$ numerically,
and its SNR distribution is shown in Fig.\ref{SNRhighz} (a),
where the number of detectable sample is $1.6\times10^5$,
accounting for  $\sim76\%$ of the total samples.
At the highest merger rate,
the SNR distribution with totally $7.2\times10^6$ GW samples is shown in Fig.\ref{SNRhighz} (a),
where the number of detectable sample is $5.5\times10^6$,
accounting for  $\sim76\%$ of the total samples too.
Note that Ref.\citep{PozzoLiMessenger2017} generated 1000 GW samples and plotted the SNR distribution detected by ET,
which showed a $\sim5\%$ detectable sample proportion under SNR$>10$.
The difference between these two SNR distributions is due to ET2CE's better detection capability.
Taking the Fisher matrix of $B$ and $C$ in Sec.\ref{subsec:ET2CE} as a prior,
with the total $2.1\times10^5$ GW samples,
we calculate the Fisher matrix (\ref{FisherDMpara})
for fiducial values $w_0=-1$ and $w_a=0$,
and yield $\Delta w_0$, $\Delta w_a$
at $1$-$\sigma$ CL
for different EOS models,
which are listed in  Table \ref{w0waFoM}.
At the highest merger rate
with $7.2\times10^6$ events,
the corresponding $\Delta w_0$, $\Delta w_a$ and FoM
are listed in Table \ref{w0waFoMUPPER}.
Consistent with the claim in Ref.\citep{PozzoLiMessenger2017}, we find that all the EOSs of NSs yield very similar results. For the case with pessimistic estimation of event rate, GW standard sirens in this method can follow the constraints of $\Delta w_0\sim 0.004$, $\Delta w_a\sim 0.02$ and FOM$\sim 4\times 10^{4}$. While in the case with optimistic estimation of event rate, the constraints are improved to $\Delta w_0\sim 0.0006$, $\Delta w_a\sim 0.004$ and FOM$\sim 2\times 10^{6}$.

As a continuation of the discussion in Sec.\ref{subsec:z0105},
we also assume that the detectable EM counterparts are only distributed within $z<0.05$.
Repeating the similar numerical simulation process
as previous under this assumption,
we find that the errors of $w_0$ and $w_a$,
and the FoM values are totally the same as in Table \ref{w0waFoM} and Table \ref{w0waFoMUPPER}.
This result is conceivable from the results in  Sec.\ref{subsec:z0105},
which shows that there is no significant difference between these two redshift ranges in the determination of the EOS model.

To investigate the importance of the observation of
EM counterparts,
we shall consider a situation that there is no detectable EM counterparts at low redshift.
Similar numerical simulation process
as previous yields the values of $\Delta w_0$, $\Delta w_a$ and FoM
for each EOS model,
which are listed in Table \ref{w0waFoMnoPrior} and Table \ref{w0waFoMnoPriorUPPER}.
Comparing with Table \ref{w0waFoMnoPrior} and Table \ref{w0waFoM},
it is observed that with no EM counterparts, the errors $\Delta w_0$ and $\Delta w_a$ become significantly larger.
Therefore, detectable EM counterparts
at low redshift can reduce the errors of $w_0$ and $w_a$ to a certain extent.

It is important to compare our results with those in previous works. Table III in Ref.\citep{PozzoLiMessenger2017} listed the errors of $w_0$ and $w_a$
which are determined by a single ET through Bayesian approach.
The $1$-$\sigma$ errors of $w_0$ and $w_a$
for $2.1\times10^5$ GW events can be derived from
Table III of Ref.\citep{PozzoLiMessenger2017}
as $\Delta w_0=0.028$ and $\Delta w_a=0.031$,
which are larger than our results in Table \ref{w0waFoM}.
This is mainly due to the better sensitivity of ET2CE network than a single ET. More importantly, Ref.\citep{PozzoLiMessenger2017} assumed
the EOS of NS are known in advance. In this article, we consider a more realistic case, in which we assume the EOS of NSs will be determined by the BNS events at low redshifts with both GW and EM observations.  We find that, in 3G era, this approach is nearly equivalent to the case with known NS's EOS, and the assumption in Ref.\citep{PozzoLiMessenger2017} is justified in our analysis.

In most previous works, the authors investigate the potential constraint of dark energy by GW standard sirens based on the assumption that the high-redshift BNS events have detectable shGRB counterparts. Due to the difficulty of EM observations, the available event number is always to be $\sim 1000$ in the 3G detection era \citep{SathyaprakashSchutzBroeck2010,ZhaoBroeckBaskaran2011,cairg,ZhaoWen2018}, which is justified in our recent simulations \citep{Yu2020b}. In this approach, the potential constraints of dark energy parameters by various 3G detector networks have been explicitly studied in our previous work \citep{ZhaoWen2018}. For instance, the detector network, including 3 CE-like detectors, is expected to give the constraints of $\Delta w_0\sim 0.03$ and $\Delta w_a \sim 0.2$, if assuming 1000 face-on BNSs at $z<2$ have the detected EM counterparts. In comparison with these results, we find that
GW standard sirens with tidal effect have a much stronger detection ability for dark energy, which might guide the research direction in this issue.

The most recent determination by Planck+SNIa+BAO
of the parameters $w_0$ and $w_a$
reports $\Delta w_0\sim0.077$
and $\Delta w_a\sim 0.3$
\citep{Planck2018CosPara},
which are much larger than our results. In the next generation of SNIa and BAO observations, with the CMB priors, the potential constraints on the dark energy parameters are around $\Delta w_0\sim 0.03$ and $\Delta w_a\sim 0.1$ \citep{DE-taskforce}. Therefore, in comparison with the tradition EM methods, we find that in 3G era with GW standard sirens, the constraints on dark energy parameters would be able to improve by 1-3 orders in magnitude. This is mainly attributed to much larger event number of GW sources. So, we conclude that in the near future, the GW standard sirens provide a much more powerful tool for constraining the cosmological parameters.

\begin{table*}[htbp]
\centering
\caption{With the total $2.1\times10^5$ BNS samples in $0.1<z<2$,
$\Delta w_0$, $\Delta w_a$ and FoM 
at $1$-$\sigma$ CL are listed in this table
for each EOS model derived by ET2CE under a detectable criterion SNR$>10$.
We use the Fisher matrix of $B$ and $C$ for a total of 250  samples given in Sec.\ref{subsec:ET2CE} as a prior.}
\label{w0waFoM}
\begin{threeparttable}
\begin{tabular}{l  c c c }
\hline
\hline
\multirow{2}*{EOS}
& \multirow{2}*{$\Delta w_0$ }
& \multirow{2}*{$\Delta w_a$ }
& \multirow{2}*{FoM}
\\
\\
\hline
alf2    & 0.0037 & 0.021  & 59619
\\
\hline
ap3    &  0.0041 & 0.022  & 54936
\\
\hline
ap4    &  0.0044 & 0.024  & 50295
\\
\hline
bsk21  &  0.0039 & 0.021  & 58548
\\
\hline
eng    &  0.0041 & 0.022  & 55503
\\
\hline
H4      &  0.0035 & 0.020  & 63756
\\
\hline
mpa1   &  0.0040 & 0.022  & 56784
\\
\hline
ms1    &  0.0034& 0.019  & 63756
\\
\hline
ms1b    & 0.0035 & 0.020  & 62097
\\
\hline
qmf40    &  0.0041 & 0.022  & 54726
\\
\hline
qmf60    &  0.0040 & 0.022  & 57204
\\
\hline
sly     &  0.0042 & 0.023  & 54390
\\
\hline
wff2       &  0.0046 & 0.024  & 48174
\\
\hline
MIT2\tnote{*}    &  0.0039 & 0.021  & 56280
\\
\hline
MIT2cfl\tnote{*}    &  0.0040 & 0.022  & 55377
\\
\hline
pQCD800\tnote{*}   & 0.0043 & 0.023  & 51681
\\
\hline
sqm3\tnote{*}       &  0.0041 & 0.022  & 53634
\\
\hline\hline
\end{tabular}
 \begin{tablenotes}
        \footnotesize
        \item[*] These are the EOS models of quark stars.
      \end{tablenotes}
    \end{threeparttable}
\end{table*}

\begin{table*}[htbp]
\centering
\caption{Same with Table \ref{w0waFoM}, but here we consider the optimal event rate with totally $7.2\times10^6$ BNS samples in
$0.1<z<2$.}
\label{w0waFoMUPPER}
\begin{threeparttable}
\begin{tabular}{l c c c }
\hline
\hline
\multirow{2}*{EOS}
&  \multirow{2}*{$\Delta w_0$ }
& \multirow{2}*{$\Delta w_a$ }
& \multirow{2}*{FoM}
\\
\\
\hline
alf2    &0.00064 & 0.0035  &$2.04408\times10^6$
\\
\hline
ap3    & 0.00070 & 0.0038  &$1.88352\times10^6$
\\
\hline
ap4    & 0.00076 & 0.0040  &$1.72440\times10^6$
\\
\hline
bsk21  & 0.00066 & 0.0036  &$2.00736\times10^6$
\\
\hline
eng    & 0.00070 & 0.0038  &$ 1.90296\times10^6$
\\
\hline
H4     & 0.00060 & 0.0033  &$2.18592\times10^6$
\\
\hline
mpa1   & 0.00068 & 0.0037  &$1.94688\times10^6$
\\
\hline
ms1    & 0.00058 & 0.0033  &$2.18592\times10^6$
\\
\hline
ms1b    &0.00060 & 0.0033  &$2.12904\times10^6$
\\
\hline
qmf40  & 0.00071 & 0.0038  &$1.87632\times10^6$
\\
\hline
qmf60  & 0.00069 & 0.0037  &$1.96128\times10^6$
\\
\hline
sly   & 0.00072 & 0.0038  &$1.86480\times10^6$
\\
\hline
wff2   & 0.00078 & 0.0041  &$1.65168\times10^6$
\\
\hline
MIT2\tnote{*}  & 0.00067 & 0.0037  &$1.92960\times10^6$
\\
\hline
MIT2cfl\tnote{*}& 0.00068 & 0.0037  &$1.89864\times10^6$
\\
\hline
pQCD800\tnote{*} &0.00073 & 0.0039  &$1.77192\times10^6$
\\
\hline
sqm3\tnote{*}   & 0.00071 & 0.0038  &$1.83888\times10^6$
\\
\hline\hline
\end{tabular}
 \begin{tablenotes}
        \footnotesize
        \item[*] These are the EOS models of quark stars.
      \end{tablenotes}
    \end{threeparttable}
\end{table*}

\begin{table*}[htbp]
\centering
\caption{With totally $2.1\times10^5$ BNS samples in
$0.1<z<2$,
the values of $\Delta w_0$, $\Delta w_a$ and FoM
at $1$-$\sigma$ CL
under the detection of  ET2CE are listed in this table.
We assume there is no detectable EM counterparts at low redshift.}
\label{w0waFoMnoPrior}
\begin{threeparttable}
\begin{tabular}{l  c c c }
\hline
\hline
\multirow{2}*{EOS}
& \multirow{2}*{$\Delta w_0$ }
& \multirow{2}*{$\Delta w_a$ }
& \multirow{2}*{FoM}
\\
\\
\hline
alf2  & 0.0067  & 0.034  & 29715
\\      			
\hline  			
ap3    & 0.0061  & 0.031  & 33369
\\      			
\hline  			
ap4   & 0.0062  & 0.032  & 31269
\\      			
\hline  			
bsk21 & 0.0062  & 0.032  & 36246
\\      			
\hline  			
eng   & 0.0061  & 0.031  & 33495
\\      			
\hline  			
H4    & 0.0070  & 0.037 & 20055
\\      			
\hline  			
mpa1  & 0.0062  & 0.031  & 34986
\\      			
\hline  			
ms1   & 0.0057  & 0.030  & 32571
\\      			
\hline  			
ms1b   & 0.0063  & 0.034  & 25389
\\      			
\hline  			
qmf40 & 0.0062  & 0.031 & 32886
\\      			
\hline  			
qmf60 & 0.0061  & 0.031  & 33705
\\                                			
\hline                            			
sly     & 0.0062  & 0.032  & 32466
\\       			
\hline   			
wff2   & 0.0063  & 0.032  & 30492
\\
\hline
MIT2\tnote{*}        & 0.0063  & 0.032  & 36372
\\
\hline
MIT2cfl\tnote{*}   & 0.0063  & 0.032  & 36372
\\
\hline
pQCD800\tnote{*}   & 0.0062  & 0.032 & 32319
\\
\hline
sqm3\tnote{*}        & 0.0061  & 0.031  & 33474
\\
\hline\hline
\end{tabular}
 \begin{tablenotes}
        \footnotesize
        \item[*] These are the EOS models of quark stars.
      \end{tablenotes}
    \end{threeparttable}
\end{table*}

\begin{table*}[htbp]
\centering
\caption{Same with Table \ref{w0waFoMnoPrior}, but here we consider the optimal event rate with totally $7.2\times10^6$ BNS samples in
$0.1<z<2$.}
\label{w0waFoMnoPriorUPPER}
\begin{threeparttable}
\begin{tabular}{l  c  c c }
\hline
\hline
\multirow{2}*{EOS}
& \multirow{2}*{$\Delta w_0$ }
& \multirow{2}*{$\Delta w_a$ }
& \multirow{2}*{FoM}
\\
\\
\hline
alf2  &0.00115 & 0.0057 & $1.01880\times10^6$
\\
\hline
ap3    &0.00105 & 0.0054 &$ 1.14408\times10^6$
\\
\hline
ap4   &0.00107 & 0.0055 &$ 1.07208\times10^6$
\\
\hline
bsk21 &0.00107 & 0.0054 &$ 1.24272\times10^6$
\\
\hline
eng    &0.00105 & 0.0054 &$ 1.14840\times10^6$
\\
\hline
H4     &0.00119 & 0.0063 &$ 0.68760\times10^6$
\\
\hline
mpa1  &0.00105 & 0.0054 &$ 1.19952\times10^6$
\\
\hline
ms1   &0.00097 & 0.0051 &$ 1.11672\times10^6$
\\
\hline
ms1b   &0.00108 & 0.0058 &$ 0.87048\times10^6$
\\
\hline
qmf40 &0.00105 & 0.0054 &$ 1.12752\times10^6$
\\
\hline
qmf60 &0.00104 & 0.0054 &$ 1.15560\times10^6$
\\
\hline
sly      &0.00105 & 0.0054 &$ 1.11312\times10^6$
\\
\hline
wff2   &0.00108 & 0.0055 &$ 1.04544\times10^6$
\\
\hline
MIT2\tnote{*}        &0.00107 & 0.0054& $1.24704\times10^6$
\\
\hline
MIT2cfl\tnote{*}   &0.00107 & 0.0054& $1.24704\times10^6$
\\
\hline
pQCD800\tnote{*}   &0.00105 & 0.0054 &$ 1.10808\times10^6$
\\
\hline
sqm3\tnote{*}        &0.00105 & 0.0054 &$ 1.14768\times10^6$
\\
\hline\hline
\end{tabular}
 \begin{tablenotes}
        \footnotesize
        \item[*] These are the EOS models of quark stars.
      \end{tablenotes}
    \end{threeparttable}
\end{table*}

\section{Conclusions}
\label{sec:conclusion}

In this paper,
we study the potential detection of the NS's EOS by using the 2G, 2.5G and 3G detectors to detect the GW with corresponding EM in the low redshift
under the framework of Fisher matrix analysis.
We also study the potential of using 3G detector network with GW from BNS merger
as a cosmological probe to determine the dark energy parameters.
Specifically,
we fit the relation between
the tidal deformation $\lambda$ and the NS mass $m$
around $m=1.35 M_\odot$ by a linear approximation
$\lambda=B m+C$,
and choose the parameters $B$ and $C$ to represent different EOS models.
At low redshift, the
detectable EM counterparts can give
the accurate redshift and position of the GW sources.
With this,
we estimate the errors of $B$ and $C$ by using  the tidal effect in GW observation,
and propose to use the overlap of two error ellipses of $B$ and $C$
to study the ability of distinguishing different EOS models,
which is different from other papers in literature that only set constraints on $\lambda$ \citep{HindererLackey2010,GW170817ligo,GW190425ligo}.
Using the EOS model determined by the low redshift as a prior,
the GW events of BNS mergers from high redshift can be used
as a standard siren to determine its luminosity distance and redshift,
which can be used to constrain
the dark energy parameters.

According to the BNS merger rate given by Ref.\citep{AbbottAbbottAbbott2019},
for a 3-year duration of observation,
we simulate and generate 250 and 2800 events within the low redshift $ z <0.1 $, respectively. With the tidal effect in GW from BNS merger,
under the Fisher matrix analysis,
we compare the capability of
distinguishing EOS models
by different 2G, 2.5G, and 3G detectors.
The main result of the study is that,
at the 2-$ \sigma $ CL,
if EM counterparts are detectable at low redshift,
when the merge rate of BNS is taken as the upper limit of the theoretical value,
the 2G detector networks can distinguish all the 17 EOS models
adopted in this paper. While,
if the merge rate is taken as its lower limit,
most EOS models can be distinguished by the 2G detector networks,
but there are still some indistinguishable EOS models.
For any merger rate in the theoretical range,
the 2.5G and 3G detector networks
can distinguish all these EOS models
due to their great sensitivity.
We also check that if the detectable EMs lie in the redshift range of $ z<0.05$,
the above results still hold.
To emphasize the importance of detectable EM,
we have taken the assumption that there is no EM
detectable at the low redshift
as shown in Sec. \ref{subsec:noEMcounterpart}.
The result shows that in the given range of merger rate,
none of the  17 EOS models can be distinguished by
the 2G detector networks,
some of the EOS models can be distinguished by the 2.5G detector networks,
and only the 3G detector networks will be able to distinguish all the 17 EOS models.
Thus, EM counterparts are important for the determination of EOS,
especially for the networks of 2G and 2.5G detectors.

It needs to be noticed that 
the above conclusions are based on the assumption that all merging NSs have the same mass as $1.35 M_\odot$.
The actual merging NS masses are more likely to be different.
As discussed at the end of Sec. \ref{subsec:LHV},
different NS masses will not significantly affect the differentiation of the EOS models under the approximated NS mass-distribution adopted in this paper.
However,
the true mass-distribution remains unknown.
Therefore,
to distinguish EOS models with more realistic mass-distribution requires more careful analysis in the future.

With the EOS model determined in low redshift by ET2CE as a prior,
and considering tidal effect,
we calculate the errors of redshift and luminosity distance of GW sources
by only using the GW observation from the high-redshift range $0.1<z<2$ by ET2CE.
Then, we convert these errors into the errors of parameters of dark matter $w_0$ and $w_a$ according to Eq.(\ref{FisherDMpara}).
The results are listed in Table \ref{w0waFoM}
and Table \ref{w0waFoMUPPER}
for the lowest and highest merger rate respectively.
We have shown that our results are consistent with Ref.\citep{ZhaoWen2018,PozzoLiMessenger2017},
and the future GW observation shall give better constraints on $w_0$ and $w_a$ than the CMB observation \citep{Planck2018CosPara}.
We also calculate the errors  of $w_0$ and $w_a$ with no EM counterparts at low redshift,
which are listed in Table \ref{w0waFoMnoPrior} and Table \ref{w0waFoMnoPriorUPPER}.
It is seen that without EM counterparts,
the errors of $w_0$ and $w_a$ become significantly larger.

In our analysis,
for simplicity,
we neglect the NS spins,
which is believed to be small in BNS systems \citep{ShaughnessyBelczynski2008}. In addition, we only consider the post-Newtonian expansion
at the inspiral stage of BNS merger,
which has been proved to be in good agreement with observations \citep{GW170817ligo,GW190425ligo}.
We have also made a strong assumption in our analysis
that the EOS of all neutron stars is the same,
which allows us to refer to the EOS parameters obtained from the low redshift to the high redshift for constraining parameters of dark energy.
But in reality, neutron stars may have different EOS depending on the internal material composition.
For example, the EOS of a quark star and a neutron star are likely to be different \citep{ArroyCruz2020,OrsariaRodrigues2012}.
We shall leave this problem in our future research.


\

\

\textbf{Acknowledgements}

We appreciate the helpful discussions with Yong Gao and Lijing Shao. This work is supported by the China Postdoctoral Science Foundation grant
No. 2019M662168, NSFC
Grants Nos. 11773028, 11603020, 11633001, 11873040, and the Strategic Priority Research Program of
the Chinese Academy of Sciences Grant No. XDB23010200.

\bibliography{ApJ_Tidal_1}{}
\bibliographystyle{aasjournal}



\end{document}